\documentclass{aa} 
\usepackage{graphicx}
\usepackage{txfonts}
\begin{document} 
	\hyphenation{MontAGN}
	\title{Exploring the inner parsecs of Active Galactic Nuclei using near-infrared high resolution polarimetric simulations}
	\titlerunning{Exploring the inner parsecs of AGN using NIR high resolution polarimetric simulations}
	
	\subtitle{with MontAGN}
	
	\author{L. Grosset\inst{1}\thanks{\email{lucas.grosset@obspm.fr}}
		\and 
			D. Rouan\inst{1} 
		\and 
			D. Gratadour\inst{1} 
		\and 
			D. Pelat\inst{2}
		\and 
			J. Orkisz\inst{3,4}
		\and
			F. Marin\inst{5} 
		\and
			R. Goosmann\inst{5}
			}

   \institute{
   			  LESIA, Observatoire de Paris, PSL Research University, CNRS, Sorbonne Universit\'es, UPMC Univ. Paris 06, Univ. Paris Diderot, Sorbonne Paris Cit\'e, 5 place Jules Janssen, 92190 Meudon, France
          \and
              LUTH, Observatoire de Paris, CNRS, Universit\'e Paris Diderot, Sorbonne Paris Cit\'e, 5 place Jules Janssen, 92190 Meudon, France
          \and
           	  Univ. Grenoble Alpes, IRAM, 38000 Grenoble, France
          \and
          	  LERMA, Observatoire de Paris, PSL Research University, CNRS, Sorbonne Universit\'es, UPMC Univ. Paris 06, \'Ecole normale sup\'erieure, 75005 Paris, France
          \and
         	  Observatoire Astronomique de Strasbourg, Universit\'e de Strasbourg, CNRS, UMR 7550, 11 rue de l’Universit\'e, 67000 Strasbourg, France
         	  }

   \date{Received June 26, 2017; accepted November 30, 2017}

 
  \abstract
   {}
   {In this paper, we aim to constrain the properties of dust structures in the central first parsecs of Active Galactic Nuclei (AGN). Our goal is to study the required optical depth and composition of different dusty and ionised structures.}
   {We developed a radiative transfer code, MontAGN, optimised for polarimetric observations in the infrared. With both this code and STOKES, designed to be relevant from the hard X-ray band to near-infrared wavelengths, we investigate the polarisation emerging from a characteristic model of the AGN environment. For that purpose, we compare predictions of our models with previous infrared observations of NGC 1068, and try to reproduce several key polarisation patterns revealed by polarisation mapping.}
   {We constrain the required dust structures as well as their densities. More precisely, we find out that the electron density inside the ionisation cone is about $2.0 \times 10^9$ m$^{-3}$. With structures constituted of spherical grains of constant density, we also highlight that the torus should be thicker than 20 in term of K band optical depth to block direct light from the centre. It should also have a stratification in density, with a lesser dense outer rim with an optical depth at 2.2 $\mu$m typically between 0.8 and 4 for observing the double scattering effect previously proposed.}
   {We bring constraints on the dust structures in the inner parsecs of an AGN model supposed to describe NGC 1068.  When compared to observations, this leads to optical depth of at least 20 in Ks band for the torus of NGC 1068, corresponding to $\tau_V\approx 170$, which is within the range of current estimation based on observations. In the future, we will improve our study by including non uniform dust structures and aligned elongated grains to constrain other possible interpretations of the observations.}


   \maketitle
%


\section{Introduction}

Polarimetric observations are one of the key methodologies to investigate the nature of Active Galactic Nuclei (AGN), as demonstrated 30 years ago by the study of \cite{Antonucci1984} on optical polarisation of radio-loud AGN and \cite{Antonucci1985} who discovered broad Balmer lines and Fe II emission in the polarised light of NGC 1068, the archetypal Seyfert 2 galaxy. 

Based on previous observations of the zoology of AGN (see, e.g., \citealt{Rowan-Robinson1977,Neugebauer1980,Lawrence1982,Antonucci1984}), it became suspected that a circumnuclear obscuring region was responsible for the disappearance of the broad line region in Seyfert 2, only revealed in AGN seen from directions around the pole (type-1 view). Following this idea, \cite{Antonucci1993} proposed the unified model for radio-quiet AGN, stating that Seyfert 1 or 2 were the same type of object harbouring a luminous accretion disk surrounded by a thick torus, but seen under different viewing angles. This model has been thoroughly tested for many years by several observers, like \cite{Gratadour2003}, \cite{Das2006} or \cite{Lopez2015} and compared to many simulations (see for example \citealt{Wolf1999,Marin2014,Marin2016b}), who found a zeroth-order agreement with empirical measures. But as regards the torus, one of the most important pieces of the model, we still lack informations due to its limited extension and to the high contrast required to observe it. Constraints were recently brought thanks to Adaptive Optics (AO), polarisation and interferometry. \cite{Raban2009} detected two components in the centre of NGC 1068 using mid-infrared interferometry, while \cite{Gratadour2015} used high angular resolution polarimetry to detect the likely signature of the core of the same AGN in near infrared (NIR). More recently, \cite{Garcia-Burillo2016} used ALMA to map the submillimeter counterpart of the torus, finding a molecular disk of about 10 pc diameter.

With the addition of polarimetry, we have access to an information less affected by the intensity contrast between the torus signal and the extremely bright central core than it would be with spectroscopy, interferometry or imaging alone. In addition to intensity, we can measure informations about the oscillation directions of the electromagnetic field, translated into three additional and independent parameters: the linear polarisation degree, the polarisation position angle and the circular polarisation (rarely measured).

Polarisation gives clues on the history of the successive interactions of light with matter. Photons can be emitted already polarised if the source is for instance composed of elongated and aligned grains (see e.g. \citealt{Efstathiou1997}). But light can also become polarised if scattered or if there is unbalanced absorption by aligned grains. Polarisation is therefore a powerful tool to study the properties of scatterers such as dust grains or electrons in AGN environments. Polarimetric measurements can also help to disentangle the origin of polarisation between spherical and oblate grains, as discussed in \cite{Lopez2015} and put constraints on the properties of the magnetic fields around the source, as it is likely to align non spherical dust grains. This would be seen as polarised light through dichroic emission and absorption (see \citealt{Packham2007}). The downside of this method is that interpretation of polarimetric data, with multiple possible origins, is not straightforward and requires a proper modelling.

Following the NIR observations of NGC 1068 by \cite{Gratadour2015}, we developed a numerical simulation code to interpret the observed polarisation features, namely a central region with low degree of polarisation at a constant position angle with a ridge of higher polarisation degree at the very centre. In this paper, the general context and reference observations are presented in section 2 and the numerical code is described in section 3. We then present two sets of parameters: a toy AGN model, used to validate the output of the simulation code and a more realistic set of parameters used to reproduce our previous observations of NGC 1068 (section 4). We analyse and discuss the simulation results in section 5 before concluding.

\section{Observational constraints}

\begin{figure}[ht!]
 \centering    
 \includegraphics[width=0.40\textwidth,clip]{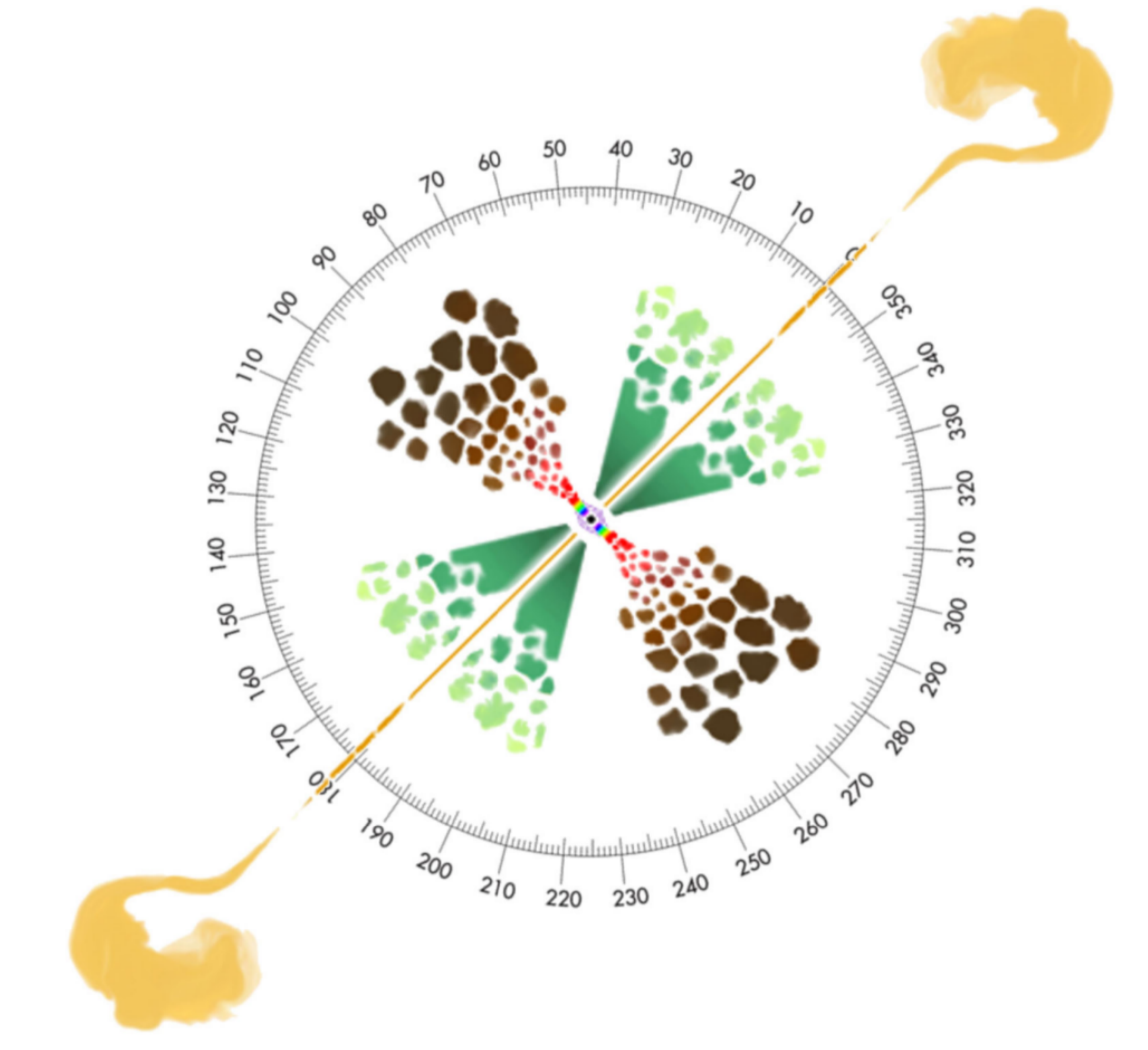} 
  \caption{Unscaled sketch of the AGN unification theory. A type 1 AGN is seen at inclinations 0$^\circ$ - 60$^\circ$  while a type 2 AGN is seen at 60$^\circ$ - 90$^\circ$, approximately. Colour code: the central super massive black hole is in black, the surrounding X-ray corona is in violet, the multi-temperature accretion disc is shown with the colour pattern of a rainbow, the broad line region is in red and light brown, the circumnuclear dust is in dark brown, the polar ionized winds are in dark green and the final extension of the narrow line region is in yellow-green. A double-sided, kilo-parsec jet is added to account for radio-loud AGN. From \cite{Marin2016}.}
  \label{unified_agn}
\end{figure}

According to the unified model \citep{Antonucci1993}, the structure of an AGN is assumed to be constituted of three essential components, represented in sketch of figure \ref{unified_agn}: 
\begin{itemize}
\item the Central Engine (CE), more likely to be a super massive black hole with a hot accretion disk emitting through unpolarised thermal emission most of the luminosity from ultraviolet to NIR;
\item the optically thick dusty torus, surrounding the CE and blocking the light in the equatorial plane, with its innermost border at sublimation temperature being a strong emitter in the near/mid-infrared;
\item on the polar directions an ionisation cone directly illuminated by the CE, diluted into the interstellar medium to create the narrow line region after few tens of parsecs.
\end{itemize}

NGC 1068 is a typical Seyfert 2 galaxy, which has long been the main target for studies about AGN. It is indeed one of the most luminous and closest active nucleus (about 14.4~Mpc, following recommendation of \citealt{Bland-Hawthorn1997}) and therefore one of the brightest. For this reason we can reach higher resolution and better contrast between the CE and its close environment compared to other active galaxies. For this target, the nucleus can be used as the guide source for the AO system. Such high angular NIR observations have already been published, first obtained with NACO \citep{Rouan2004,Gratadour2006} and more recently using SPHERE with polarimetric information \citep{Gratadour2015}.

\begin{figure}[ht!]
 \centering    
 \includegraphics[width=0.20\textwidth,clip]{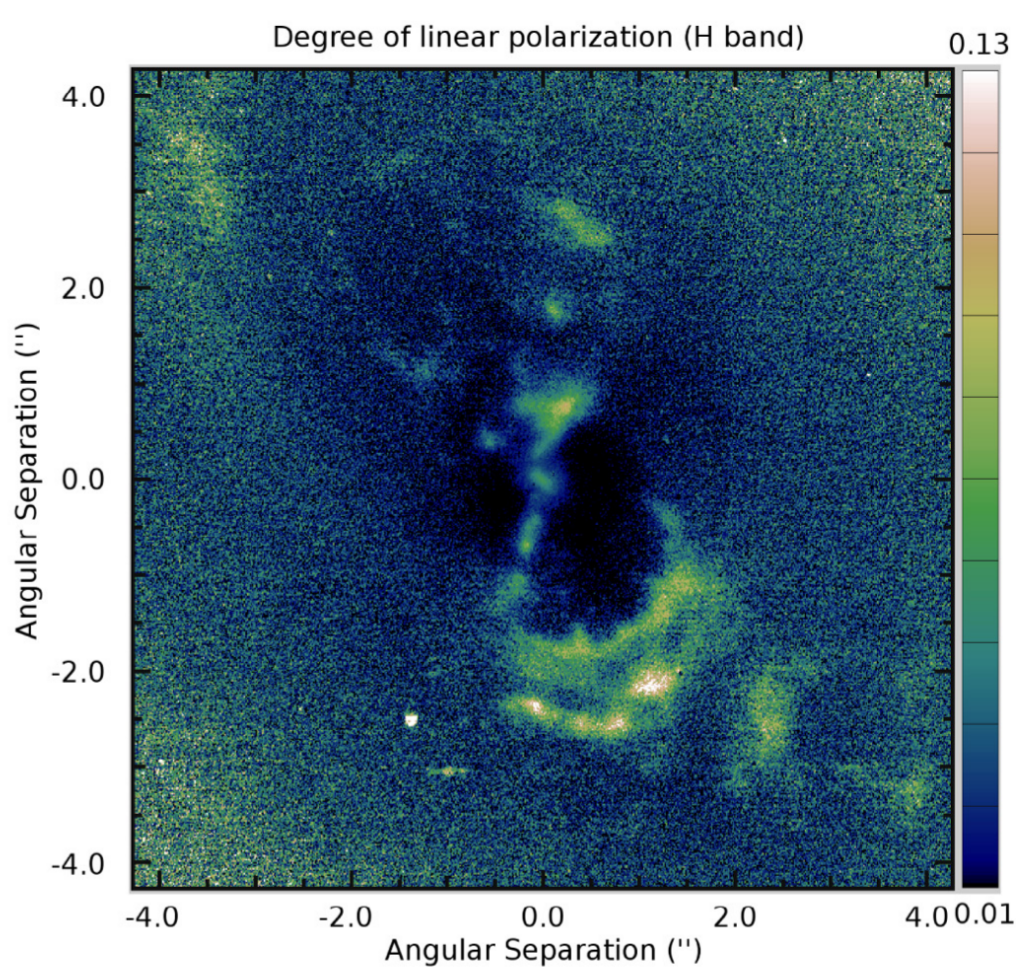}
 \includegraphics[width=0.20\textwidth,clip]{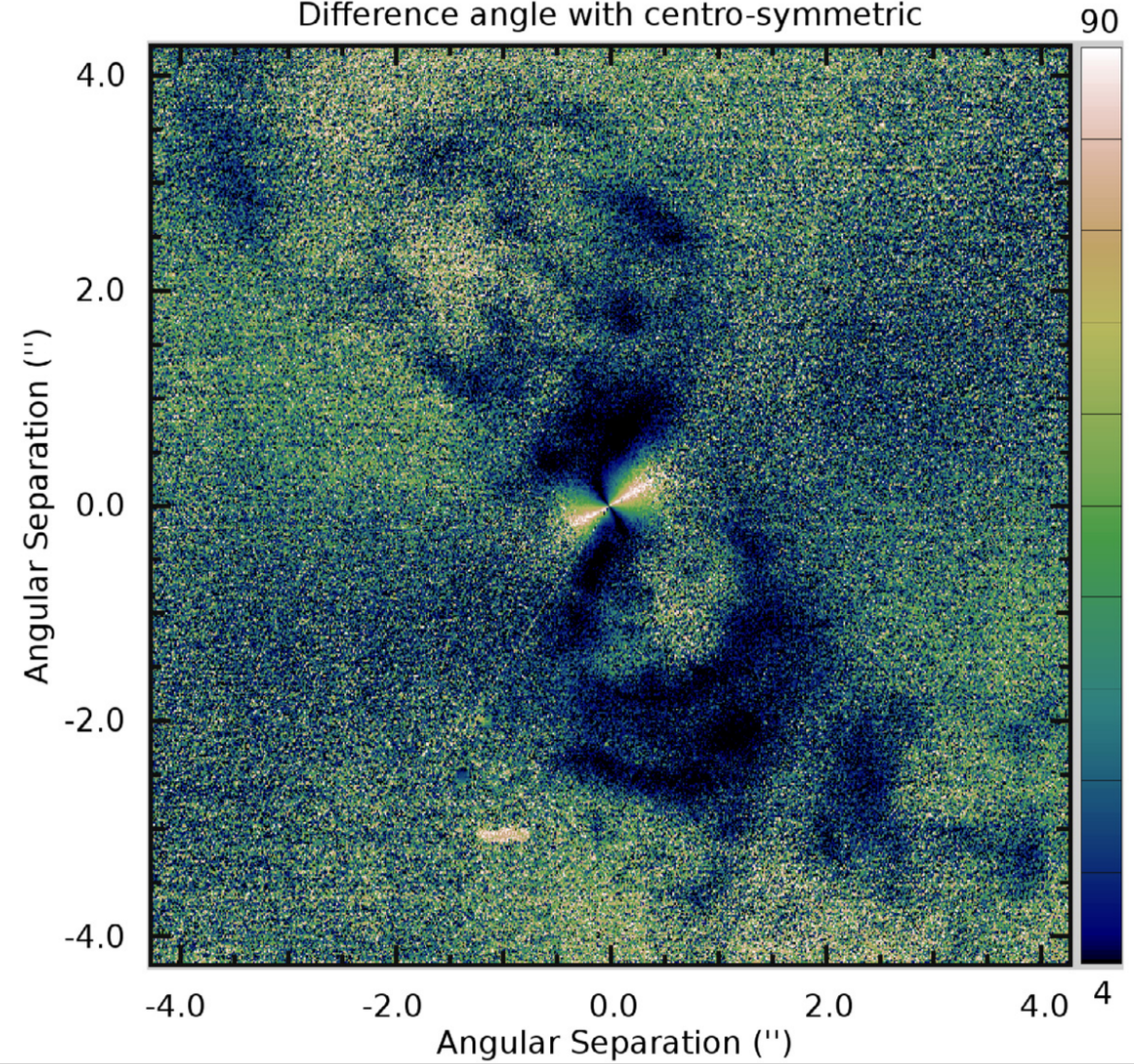}
  \caption{NGC~1068 observed with SPHERE. Left panel: degree of linear polarisation; right panel: result of the difference between the polarisation angle map and a purely centro-symmetric pattern. Images from \cite{Gratadour2015}.}
  \label{NGC1068}
\end{figure}

\begin{figure}[ht!]
 \centering    
 \includegraphics[width=0.20\textwidth,clip]{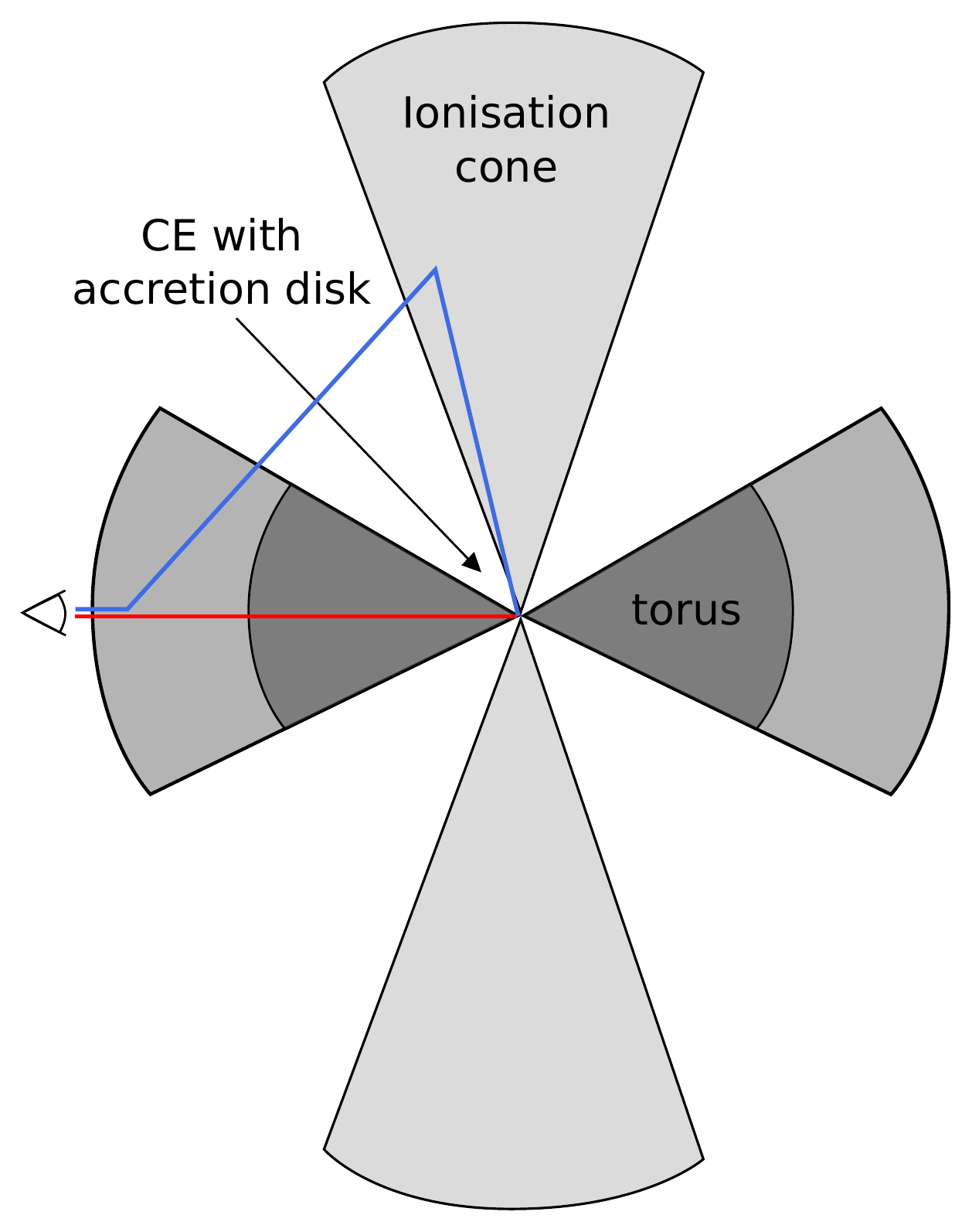}
  \caption{Example of two photon paths on an AGN environment: the blue path will have an integrated optical depth of about 1~-~4 while the red one will be about 50~-~200.}
  \label{path}
\end{figure}

The later observations were conducted during the SPHERE Science Verification (SV) program in December 2014. H and Ks broad band images were obtained, showing a polarisation angle map with a clear centro-symmetric pattern, tracing the ionisation cone. Additionally, a compact area of constant polarisation direction was observed at the centre of this hourglass shape and interpreted as a signature of the torus (see figure \ref{NGC1068}, extracted from \citealt{Gratadour2015}). Our assumption was that the torus, optically thick, would block direct light from the hottest dust and that we would be seeing light scattered twice. The first scattering would take place in the ionisation cone, allowing some photons to be redirected toward the outer part of the equatorial plane and then undergo a second scattering (see figure \ref{path} for illustration). If we detect such photons, the signature would be a polarisation pattern aligned with the equatorial plane. This is an analogous phenomenon to the one observed in the envelope of Young Stellar Objects (YSO) as proposed by \cite{Bastien1990} and \cite{Murakawa2010}. In this configuration, spherical grains can alone produce these signatures in polarisation images. This should constrain the optical depth of the different components as we will investigate hereafter.

The torus is more likely to be clumpy, as the hydrodynamic stability of tens of parsec scale structures is not well explained. Furthermore, fragmentation is usually invoked to explain some of the observations according to recent studies \citep{Mason2009,Muller2009,Nikutta2009,Alonso2011}. However, we will first examine models with uniform structures, each with a constant density, to ease simulations and analysis. We have tested a range of optical depth between $\tau_V$ = 20 and 100 for the torus median plane. We based this choice on the current estimations of about 50 in visible, as derived for instance by \cite{Gratadour2003}. Note that more recently, \cite{Lira2013} or \cite{Audibert2017} found larger optical depth and that both these authors derived them assuming clumpy structures.

\section{MontAGN Simulation code}

Because polarimetry is a complex phenomenon, radiative transfer codes are essential tools for the interpretation of these observations. However as far as we know, no radiative transfer code was able to address of all wavelength ranges. Because of our particular requirements, we decided to develop our own simulation code optimised for AGN observations in the infrared. MontAGN (acronym for “Monte Carlo for Active Galactic Nuclei”) was designed to study whether our assumptions on single and double scattering with a proper torus geometry were valid and if this would allow to reproduce the observed polarisation images of NGC 1068. The MontAGN code is aimed to be available to the public in a near future.

\subsection{Overview}

\begin{figure}[ht!]
 \centering    
 \includegraphics[width=0.40\textwidth,clip]{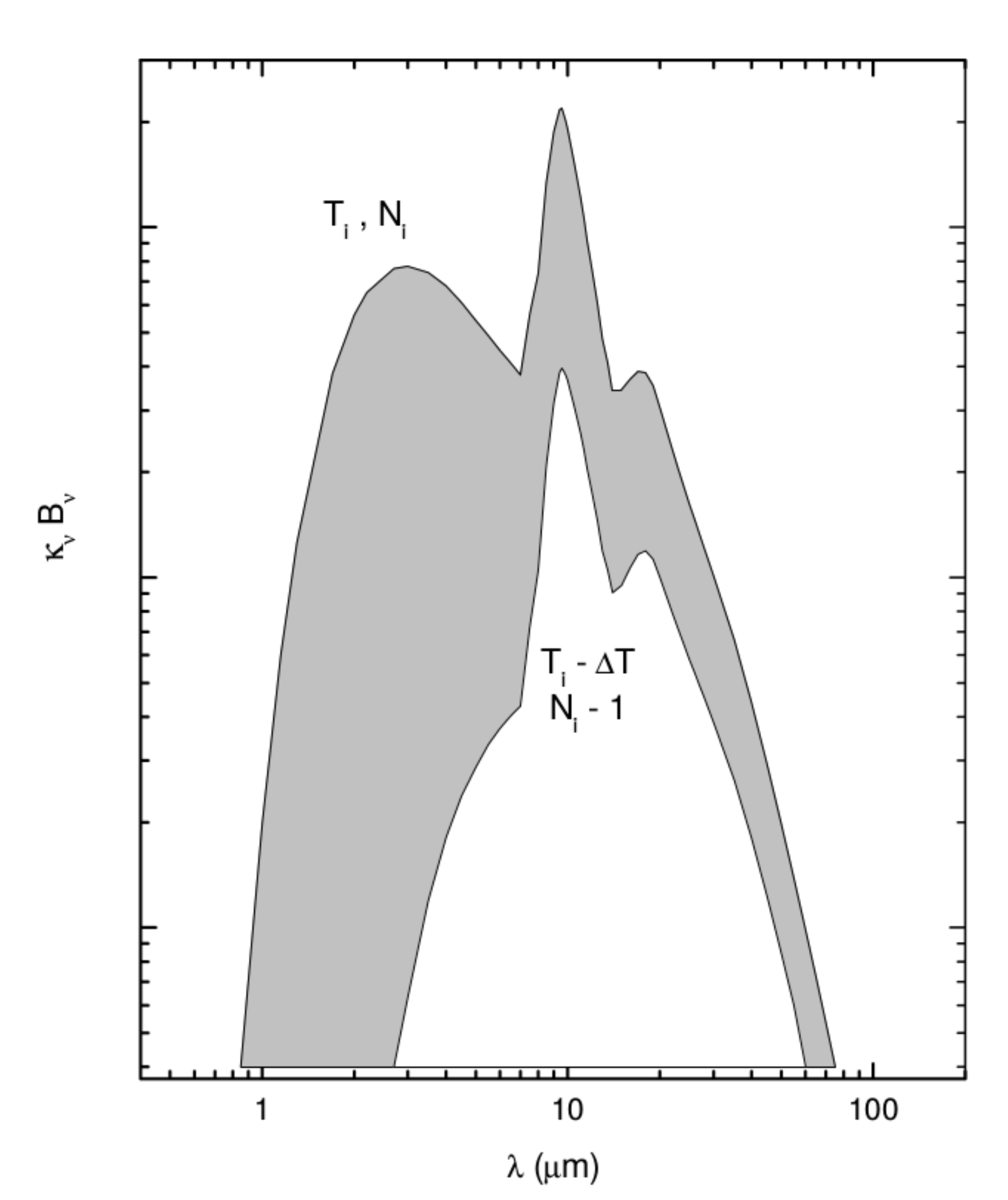} 
  \caption{Temperature correction frequency distribution. Shown are the dust emissivities, j$_\nu$ = $\kappa_\nu$ B$_\nu$(T), prior to and after the absorption of a single photon packet. The spectrum of the previously emitted packets is given by the emissivity at the old cell temperature (bottom curve). To correct the  spectrum from the old temperature to the new temperature (upper curve),  the photon packet should be re-emitted using the difference spectrum  (shaded area), from \cite{Bjorkman2001}.}
  \label{Tadjust}
\end{figure}

MontAGN is written in Python 2.7. It uses photons packets with constant initial energy instead of propagating single photons. This allows to choose between two propagation techniques:
\begin{itemize}
\item If re-emission is disabled, the energy of the packets is modified to take into account, at each event, the photons fraction of the initial packets that continues to propagate in the medium. Therefore at each encounter with a grain, the absorbed fraction of photons is deduced so that only the scattered photons propagate. \cite{Murakawa2010} used the same techniques for the simulations of YSO. 
\item If re-emission is enabled, the code takes into account the absorption and re-emission by dust as well as temperature equilibrium adjustment at each absorption. If absorbed, the photon packet is re-emitted at another wavelength with the same energy but with a different number of photons. This allows to keep the energy constant within the cell. The re-emission wavelength is randomly selected in function of the temperature difference $\Delta T$ in the cell, after the packet's absorption, following the method of \cite{Bjorkman2001}. By doing so, we take into account the previous re-emissions that were not emitted according to the final emission function. We therefore uses the difference of spectrum between the two emissivity for both cell temperatures to emit the photon packets (see figure \ref{Tadjust} extracted from the previously cited paper) and correct the probability distribution function from the previous re-emissions.
\end{itemize}

The user can define whether the re-emission is or not enabled. By disabling it, every packet will propagate until it exits the medium and the number of packets collected at a given wavelength will be larger\footnote{Note that this is only true if the signal in the wavelength range is mainly originating from the centre (hottest dust or CE), it is the opposite when looking at wavelength where emission come mostly from direct dust re-emission (typically in the mid-infrared)}. Disabling this effect allows us to get more statistics at the end of the simulation as every packet is taken into account. But it also requires to have a minimum number of packets in each pixel of the final images as we may obtain in a given pixel only packets with very few photons, because of successive scatterings with a low albedo, a situation that may not be representative of the actual pixel polarisation. This point will be discussed in more details in section \ref{effective_photon}

\subsection{Initialisation}

MontAGN uses a 3D grid of cells, sampling the densities of the different dust species in each cell. Dust grains are considered as dielectric spheres. Before the simulation starts, tables of albedo, phase function, Mueller matrix, absorption and extinction coefficients are generated for a range of grain sizes and wavelengths thanks to Mie theory. For that purpose, we use the bh\_mie module initially written by \cite{Bohren1983}. These tables then just need to be interpolated (linear or logarithmic interpolations) to get the precise value for a given set of grain size and wavelength during the simulation. These interpolations make the execution of the code faster. MontAGN uses the Stokes vectors formalism to represent polarisation of photons packets:

\begin{equation}
S=
\begin{bmatrix}
I \\
Q \\
U \\
V
\end{bmatrix}
\end{equation}

with I the intensity, Q and U the two components of linear polarisation and V the circular polarisation. In MontAGN, we normalise the Stokes vector ($I = 1$) to simplify the Stokes vector propagation. 

\subsection{Photon propagation}

When a packet is emitted, the source is selected according to the respective luminosity of each sources. The packet's wavelength is determined using the source's spectral energy distribution, the photons are initially not polarised (i.e. with Stokes vector of the photons set to [1,0,0,0]). The packet begins to propagate in a randomly chosen direction. At its first encounter with a non empty cell, a value of the optical depth that the photon will be allowed to travel is randomly determined, as in \cite{Robitaille2011}. On its path, the photon's "remaining optical depth" is shrinked as a function of the densities of the dust species in the corresponding cells. When it reaches 0, the packet interact with a grain of the cell: the radius of the grain is randomly chosen using the dust grain size function and from this radius and the wavelength, albedo, phase function, scattering angles and Mueller matrix are determined. In order to make the information on polarisation evolving, we apply to the Stokes vector the Mueller matrix and the rotation matrix: 

\begin{equation}
S_{final} = M\times R\times S_{init}
\end{equation}

with M the Mueller matrix constructed from Mie theory and R the rotation matrix, both depending on the scattering angles. More information about Matrices and Stokes formalism can be found in appendix \ref{Mueller}. A new optical depth is randomly determined after each scattering.

When the photon packet exits the simulation box, it is recorded with its four Stokes parameter, the number of interaction, its last interaction position and its final direction of propagation, registered in the form of two angles, inclination $\theta$ (ranging from 0 to 180 $^\circ$) and azimuth $\phi$ (ranging from 0 to 360 $^\circ$). 
From these data, Q and U maps are created, summing photons packets (with a possible selection on altitude and azimuthal angles) and combined in maps of polarisation angle and degree using:

\begin{equation}
P_{lin} = \frac{\sqrt{U^2+Q^2}}{I}
\end{equation}

\begin{equation}
P_{circ} = \frac{V}{I}
\end{equation}

\begin{equation}
\theta_{pol} = \frac{1}{2}  ~\mathrm{atan2}(U,Q)
\end{equation}

If the model is axi-symmetric, MontAGN allows to increase the signal obtained from simulations following two assumptions. First, in the case of cylindrical symmetries, all the photons exiting the simulation with the same inclination angle $\theta$ but different azimuthal angle $\phi$ are equivalent. We can therefore use all photons with a $\theta$ in the proper range by a rotation around the vertical axis of the photon's final position.

If the model is also symmetric according to the equatorial plane, photons recorded above the equatorial plane are equivalent to those below the plane. At the end we can add photons of the four quadrants (up-left, up-right, bottom-left and bottom-right) as long as we properly change their polarisation properties according to the symmetry. It is however mandatory to observe at an inclination angle of 90$^\circ$ to use this last method.

Note that it is possible to define an upper bound for the number of absorptions a photons packet undergo in order to keep only packets transporting significant energy and have faster simulations. As this information is kept, it is also possible to get access to maps of averaged number of interactions undergone in a particular direction.

\subsection{Maps}

As we record packets with different numbers of photons, we need to take these differences into account when creating the observed maps. All maps (including those referred to as "averaged") are generated summing photons instead of packets. The parameters\footnote{mainly the intensities I, Q, U and V, but also for example the number of scatterings} of all packets are summed with a factor 
proportional to the fraction of remaining photons in each packet $i$ (in each resolution element). As the number of photons is proportional to the energy, the energy $E_i$ of the packets is used to weight the packets. The averaged number of scatterings $n$ in a pixel of the map will therefore be computed from the number of scatterings of each packet $n_i$ as:

\begin{equation}
n = \frac{\Sigma^i E_i \times n_i}{\Sigma^i E_i}
\end{equation}

Packets are selected for computing the maps on a certain range of inclinations. In the following, we used the range $\pm5^\circ$ and the images computed at 90$^\circ$ are therefore including photons exiting with an inclination in the range [85$^\circ$,95$^\circ$].

\section{Models}
\subsection{First tests and toy models}

We first conducted validation tests of the MontAGN code on YSO, using the geometry described in \cite{Murakawa2010}. Our results were compatible with those shown in their paper in terms of polarisation degree and angles, albeit a higher Poissonian noise due to the very high optical depth involved in this model (about $6\times10^4$ in K band). The clear polarisation angle pattern seen in \cite{Murakawa2010}, see their figure 6, needs more photons to be achieved. For a proper validation on AGNs, we defined a set of three models to compare the results of simulations between MontAGN and STOKES codes \citep{Goosmann2007,Marin2012,Marin2015} in a simple case. Detailed informations about the comparison of both codes can be found in \cite{Grosset2016sf2a}. First analysis and conclusions are also available in \cite{Marin2016sf2a}. As the two codes are not aimed to work in the same spectral bands, namely the infrared for MontAGN and the NIR, visible, UV and X-rays for STOKES, we selected intermediate and overlapping wavelengths: 800, 900 and 1000 nm. Our baseline models were filled with only silicates grains in the dusty structures and electrons in the ionisation cone. All grains data, including graphites used in other models, come from \cite{Draine1985}. All dust distributions were set to dielectric spheres with a radius ranging from 0.005 to 0.250 $\mu$m with a power law of -3.5, following MRN distribution as defined in \cite{Mathis1977}.

The three models are based on simple geometrical features with constant silicates or electrons densities: 
\begin{itemize}
\item A flared dusty torus, ranging from 0.05 pc to 10 pc, a value consistent with what is currently estimated \citep{Antonucci1993,Kishimoto1999,Garcia-Burillo2016}. It has a half opening angle of 30$^\circ$ from the equatorial plane and an optical depth of 50 at 500 nm in the equatorial plane.
\item An ionised outflow, composed of electrons in a bi-cone of half opening angle of 25$^\circ$ from the symmetry axis of the model, from 0.05 pc to 25 pc at an optical depth of 0.1.
\item An outer dusty shell ranging from 10 pc to 25 pc, outside of the ionisation cone, with a radial optical depth of 0.5 at 500 nm.
\end{itemize}

\begin{figure}[ht!]
 \centering    
 \includegraphics[width=0.20\textwidth,clip]{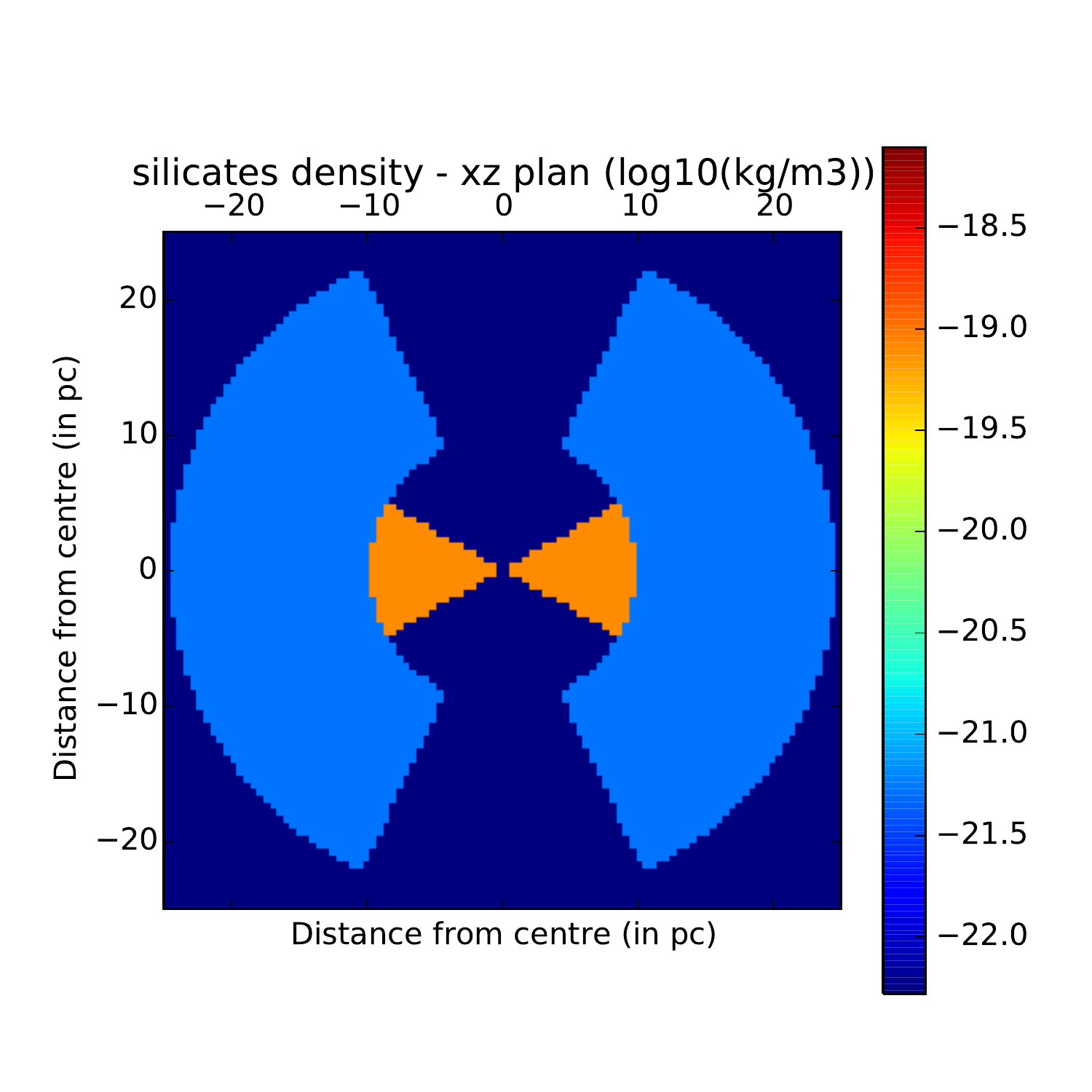} 
 \includegraphics[width=0.20\textwidth,clip]{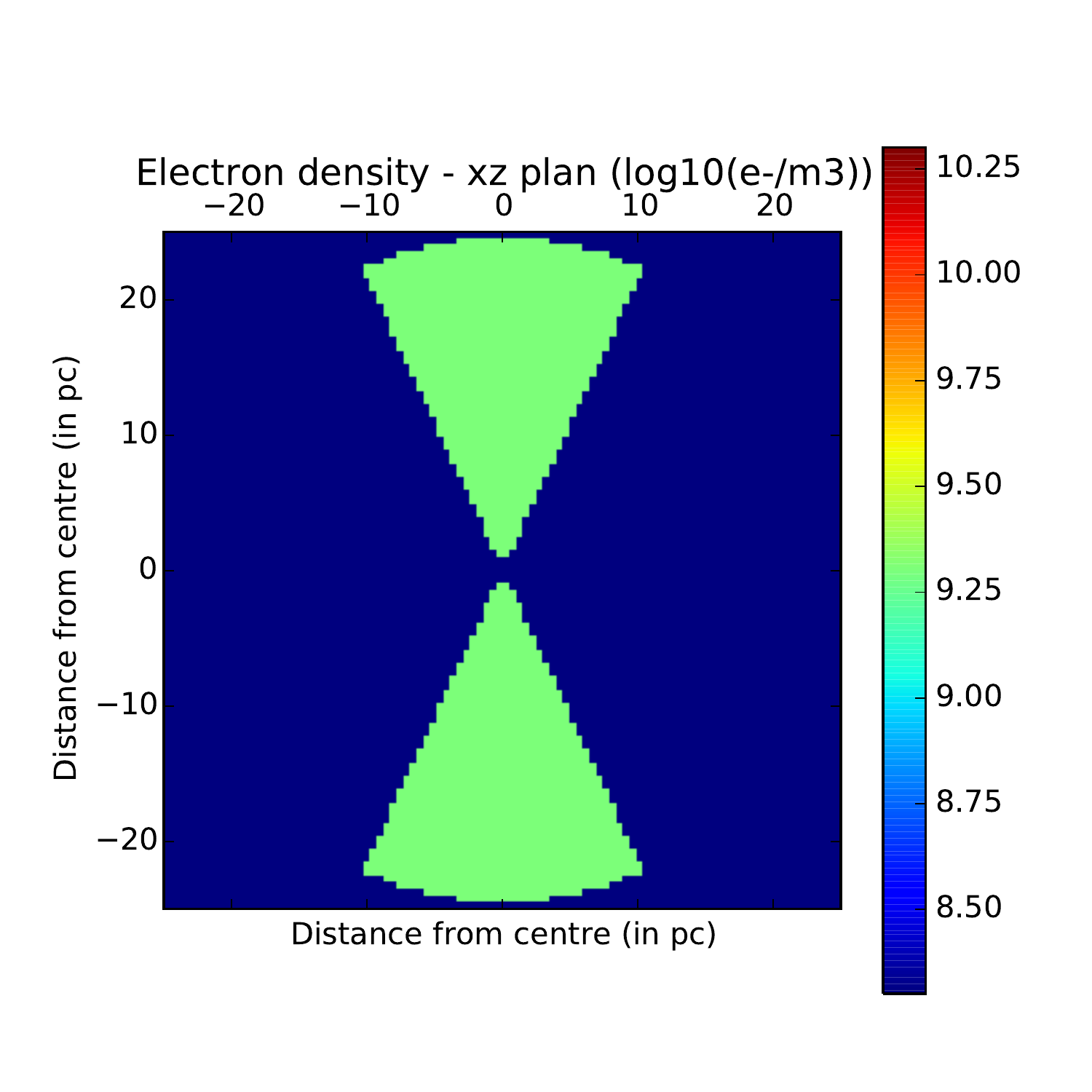} 
 
 \includegraphics[width=0.20\textwidth,clip]{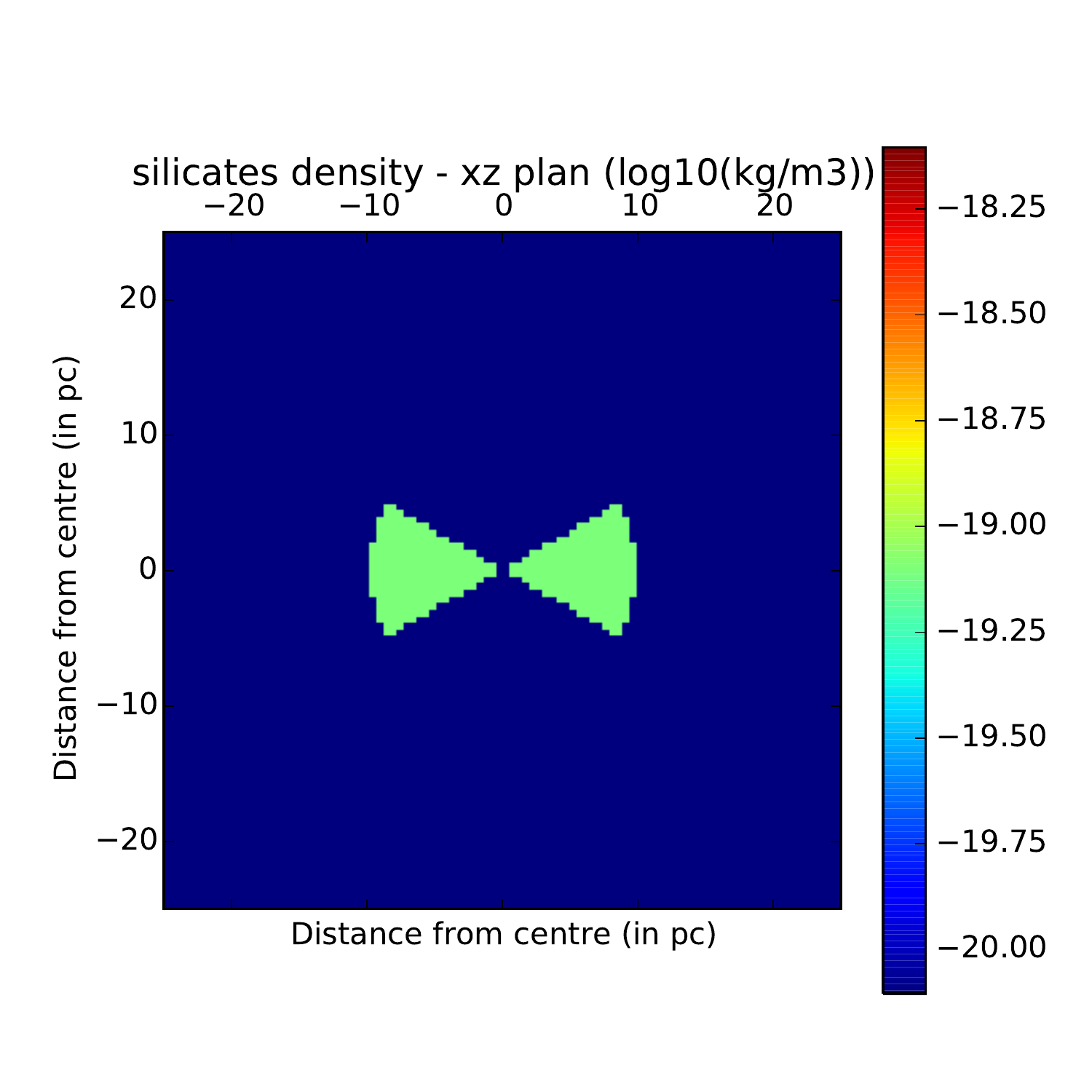} 
 \includegraphics[width=0.20\textwidth,clip]{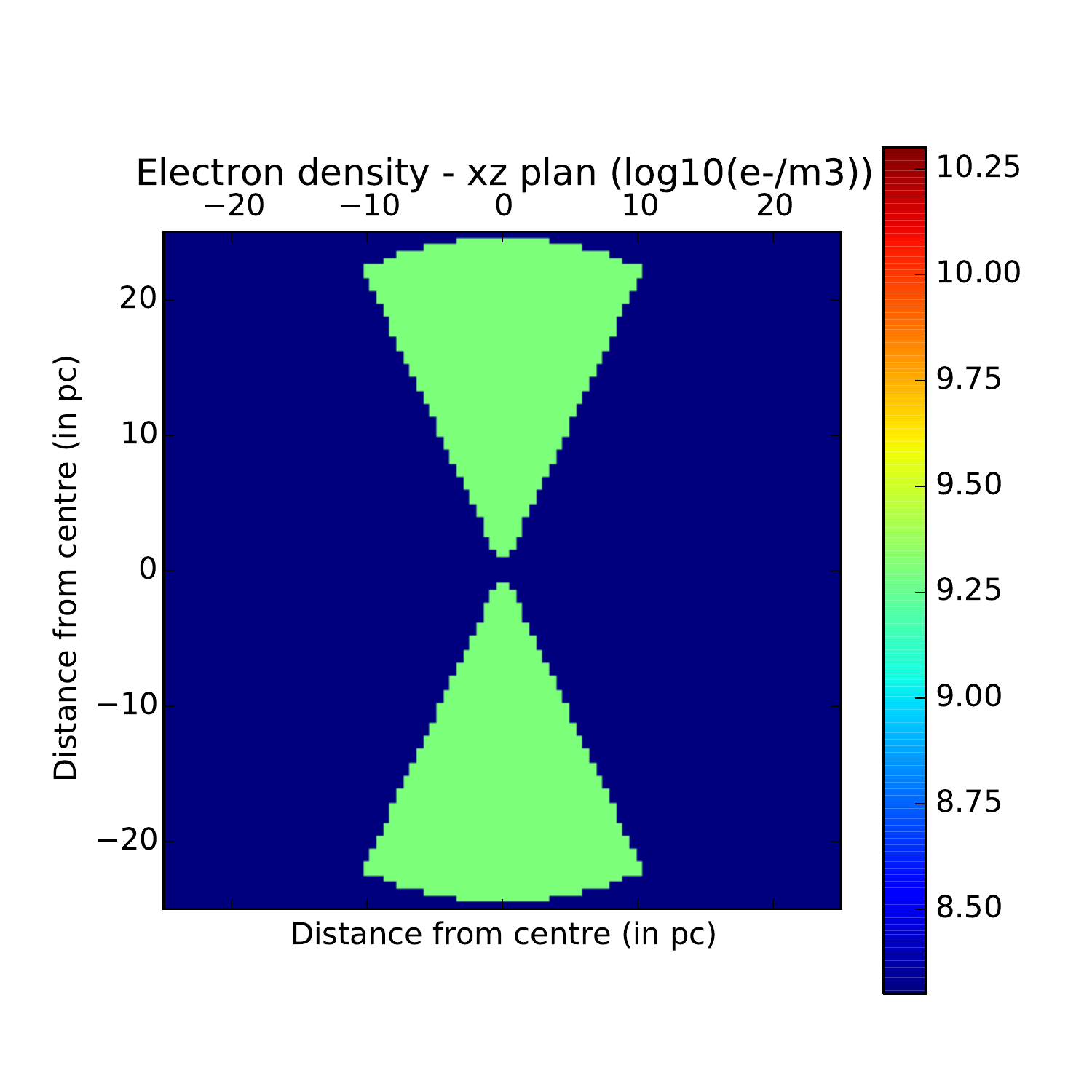}
  
 \includegraphics[width=0.20\textwidth,clip]{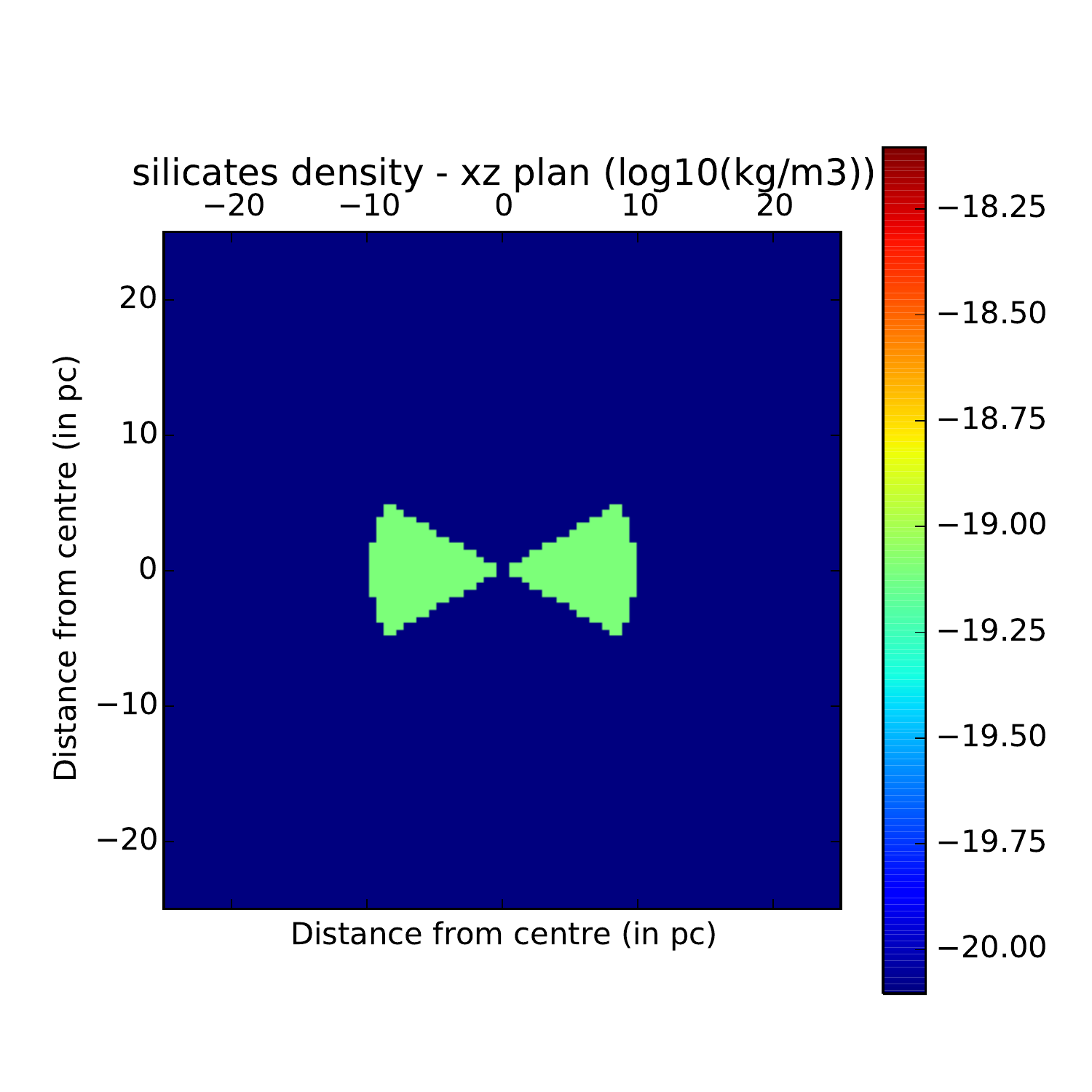} 
 \includegraphics[width=0.20\textwidth,clip]{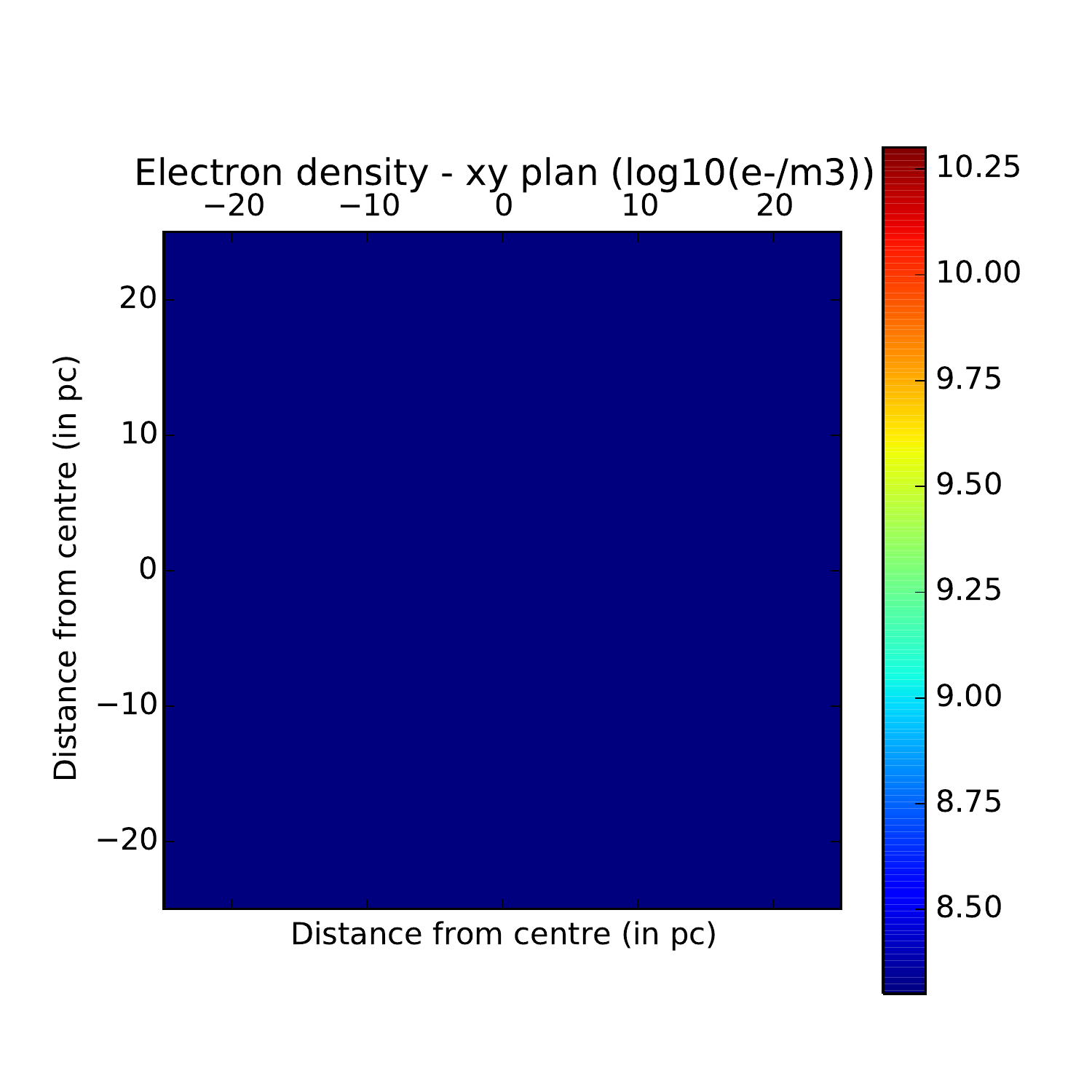} 
  \caption{Vertical slices of grain density of silicates (in log$_{10}$(kg/m$^3$), first column) and density of electrons (in log$_{10}$(m$^{-3}$), second column) set for model 1 (first row), model 2 (second row) and model 3 (third row).}
  \label{model1}
\end{figure}

The first model is built with these three components (model 1), a second one with only the torus and the outflow (model 2) and the third is reduced to the torus only (model 3). See figure \ref{model1} for density maps of each model.

\begin{figure}[ht!]
 \centering    
 \includegraphics[width=0.19\textwidth,clip]{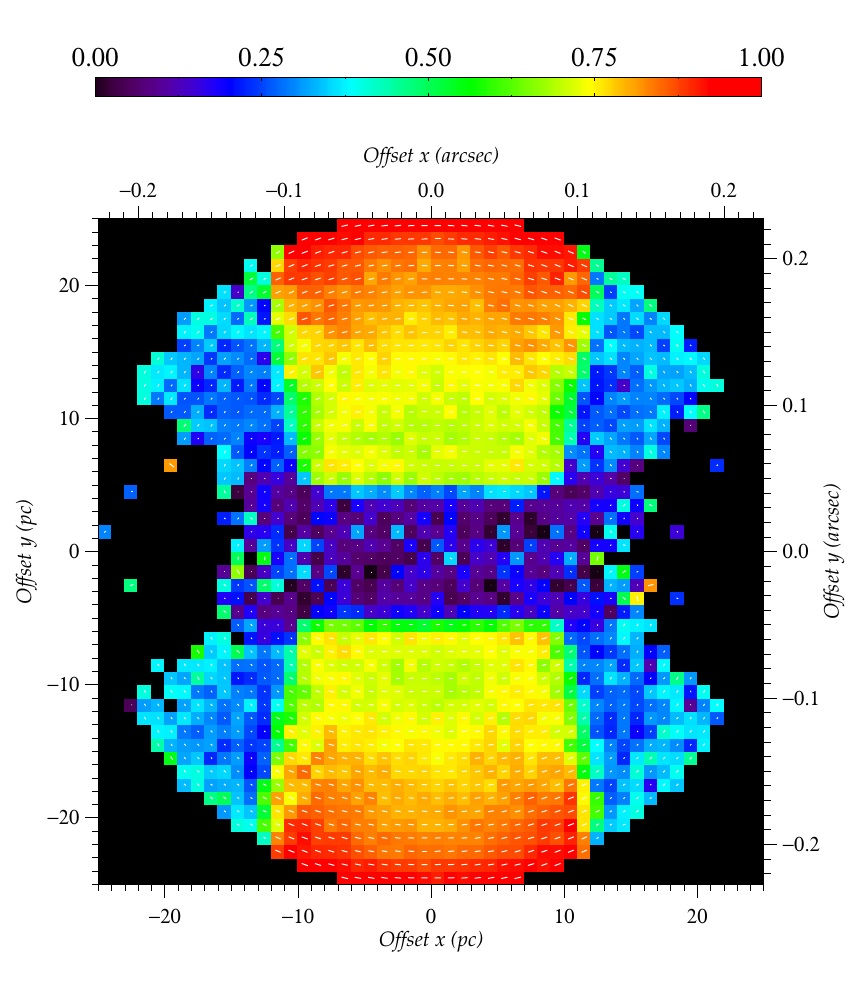}
 \includegraphics[width=0.25\textwidth,clip]{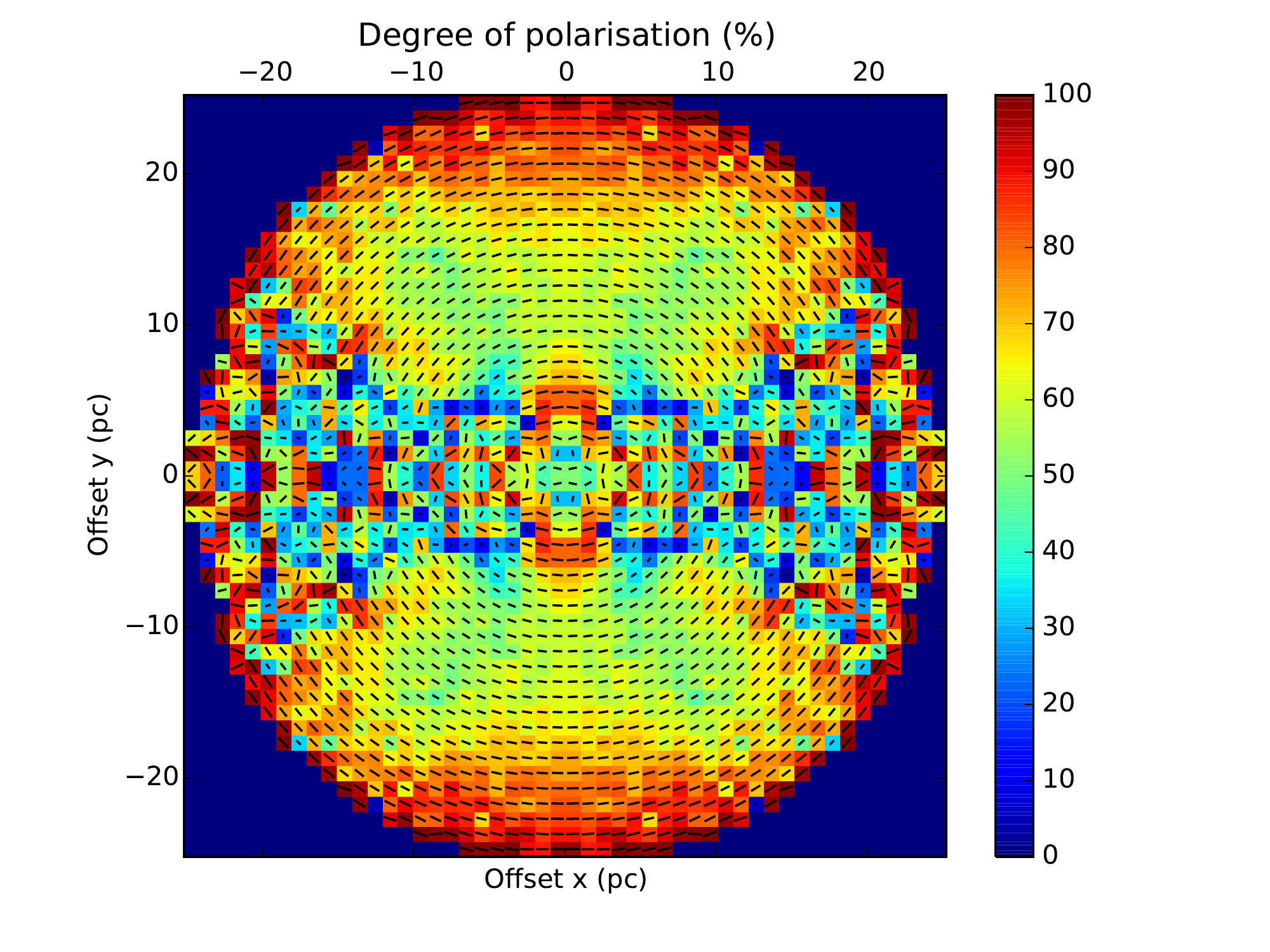}
 
 \includegraphics[width=0.19\textwidth,clip]{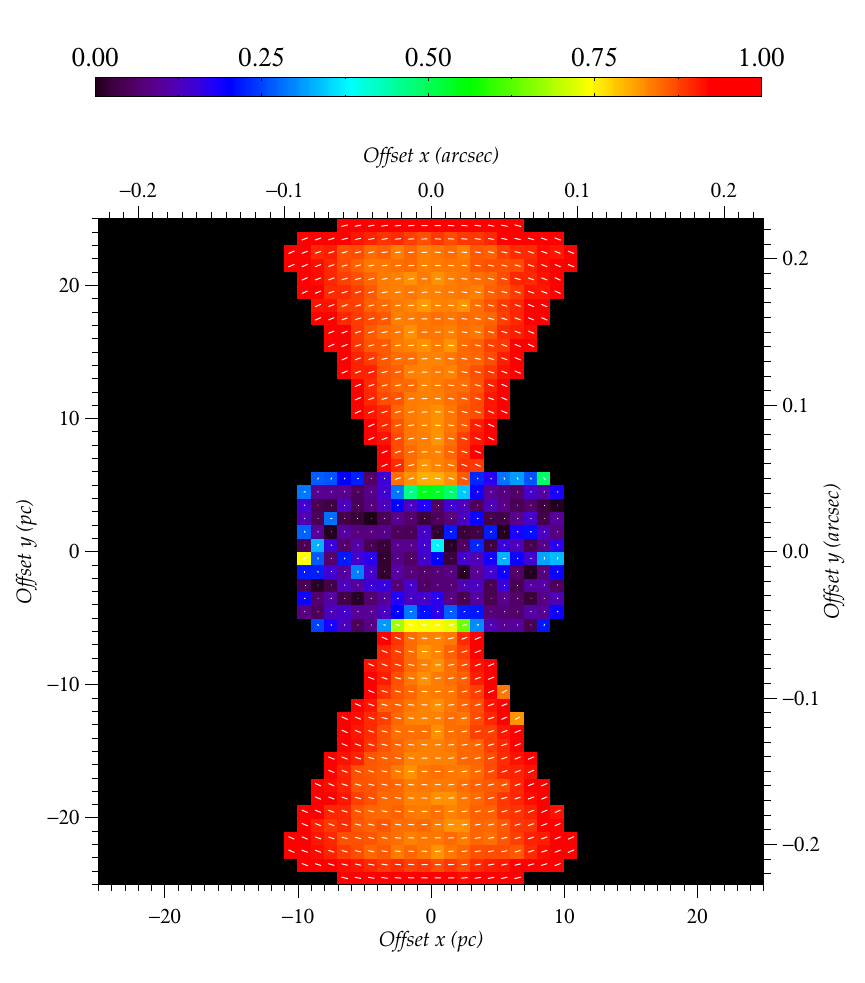}
 \includegraphics[width=0.25\textwidth,clip]{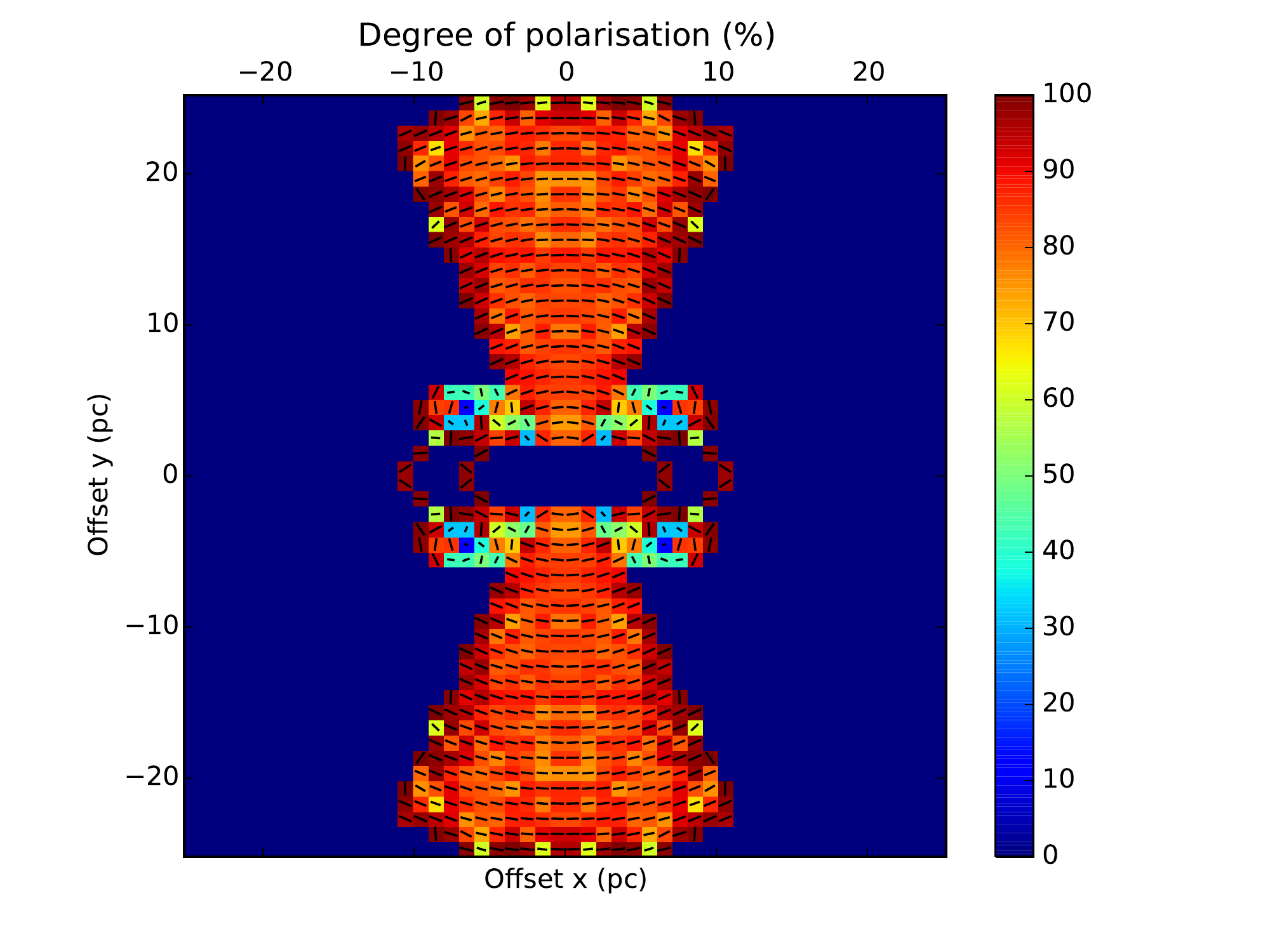}
 
 \includegraphics[width=0.19\textwidth,clip]{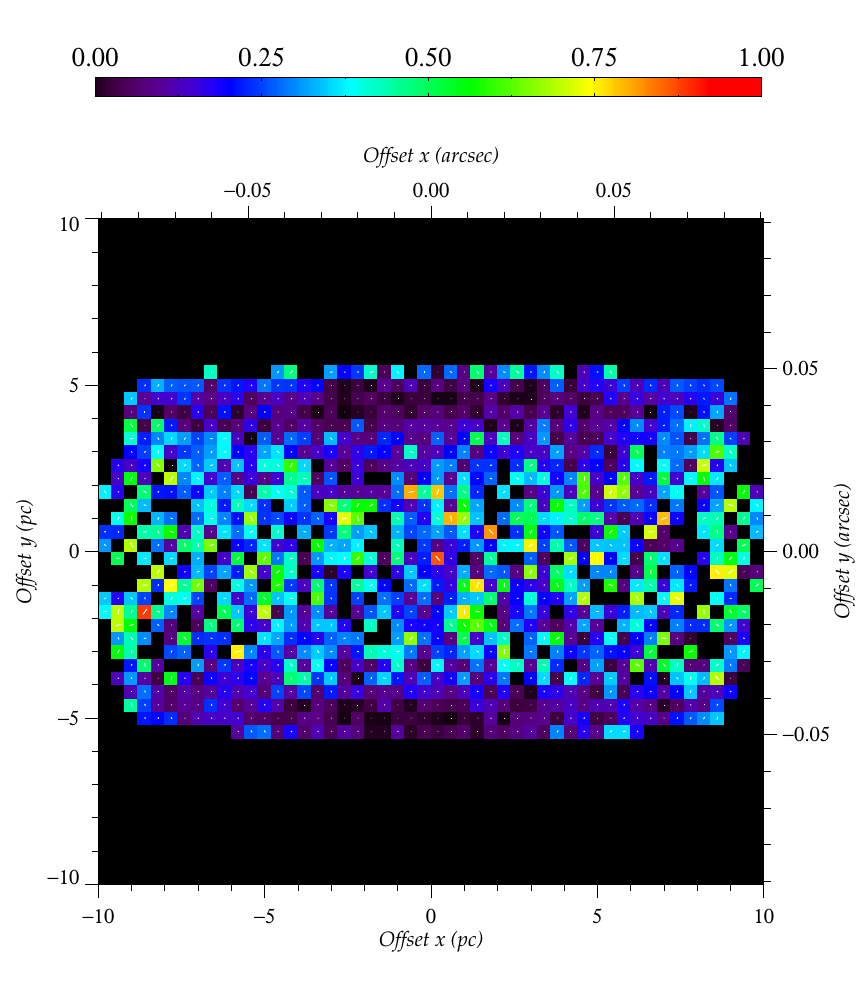}
 \includegraphics[width=0.25\textwidth,clip]{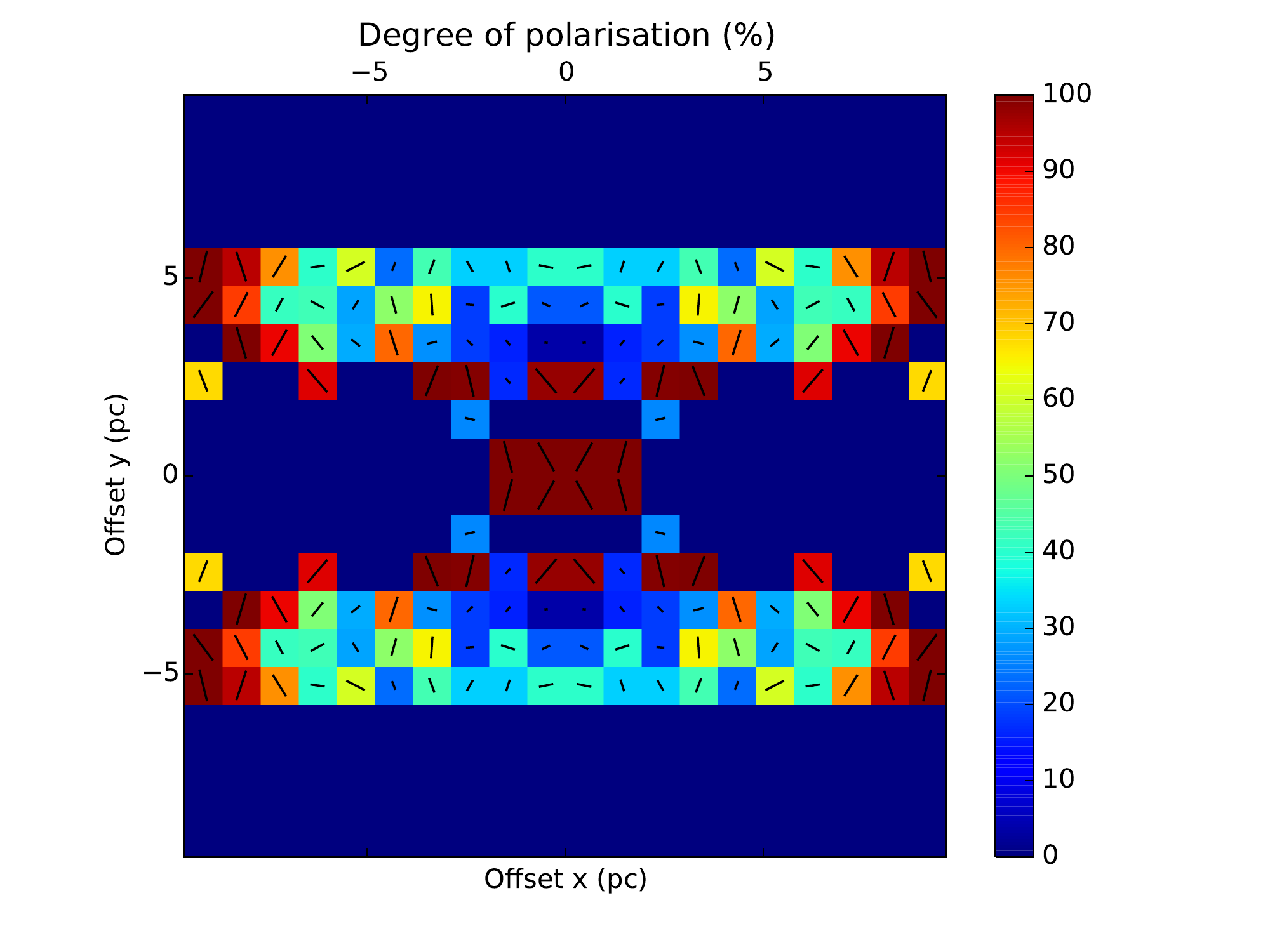}
  \caption{Maps of polarisation degree with an inclination angle of 90$^\circ$ at 800 nm for model 1, 2 and 3. First column shows maps from STOKES, second column corresponds to MontAGN maps (in \%). First row is for model 1, second row for model 2 and third row shows model 3 results. Polarisation vectors are shown, their length being proportional to the polarisation degree and their position angle representing the polarisation angle}
  \label{STOKES800}
  \end{figure}

\begin{figure}[ht!]
 \centering    
 \includegraphics[width=0.30\textwidth,clip]{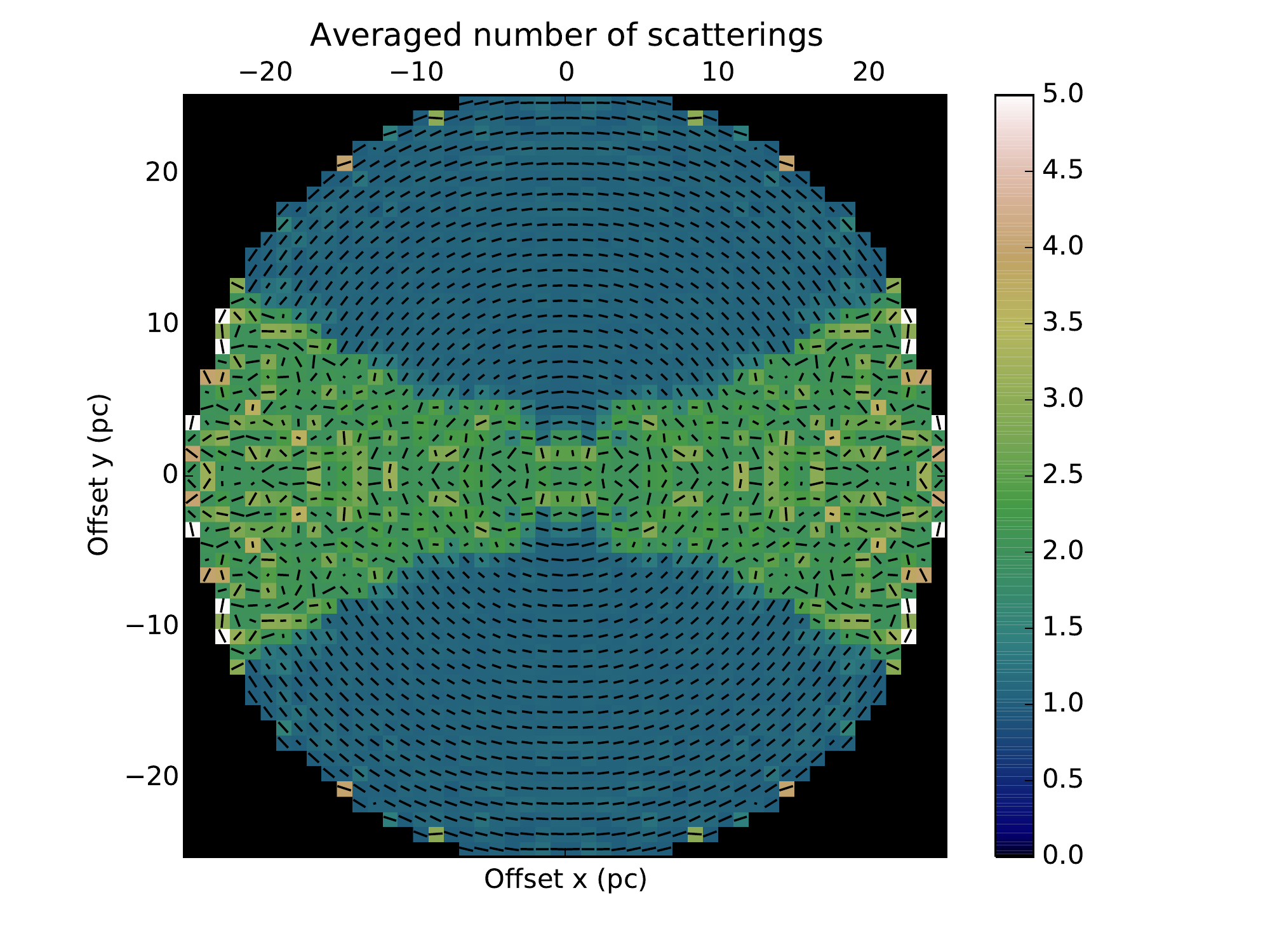}
 \includegraphics[width=0.30\textwidth,clip]{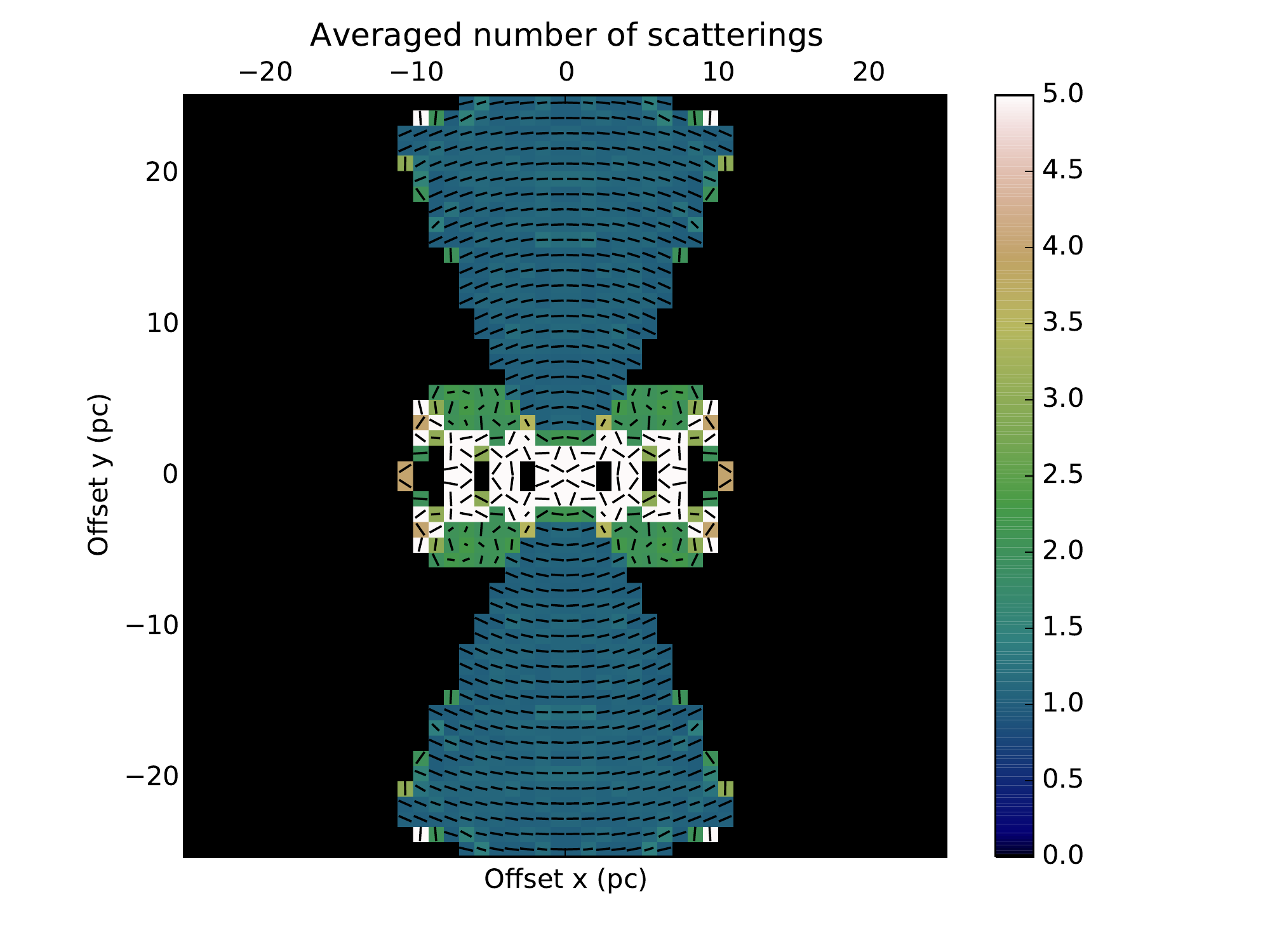}
 \includegraphics[width=0.30\textwidth,clip]{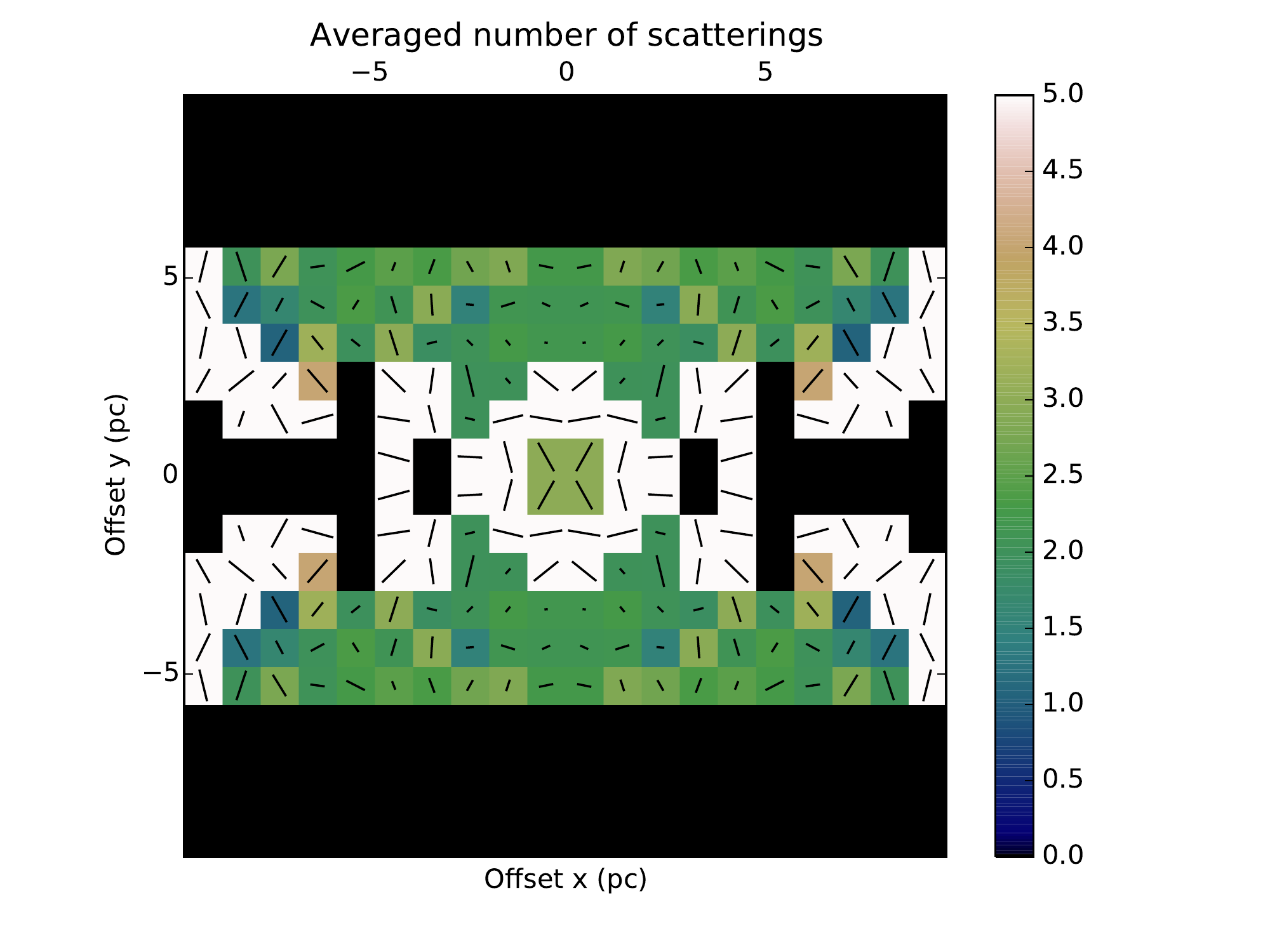}
  \caption{Maps of observed averaged number of scatterings with an inclination angle of 90$^\circ$ at 800 nm for model 1, 2 and 3 with MontAGN. Polarisation vectors are shown, their length being proportional to the polarisation degree and their position angle representing the polarisation angle ($p=1$ is represented by a length of a pixel size).}
  \label{toy800}
  \end{figure}

We launched for each of these models 10$^7$ packets, disabling re-emission and considering only the aforementioned wavelengths. All inclination angles were recorded. We only show in figure \ref{STOKES800} and \ref{toy800} the maps obtained at 800 nm at an observing angle of 90$^\circ$ (edge on).
  
As showed in figure \ref{STOKES800}, results between the two codes are in fairly good agreement, as expected from \cite{Grosset2016sf2a} and \cite{Marin2016sf2a}. The few observed differences arise likely from the divergence in the simulation strategy. Because MontAGN (in the configuration used here) propagates photons packets that may have suffered strong absorption, the pixels with a low number of photons may not contain reliable information in some cases (see section \ref{effective_photon}). However, as the packets number increases, the maps converge toward a more realistic result, close to those produced with STOKES. In the central regions, we only have few photons recorded. This is expected because escaping photons here have been scattered several times, their number decreasing strongly at each interaction because of the low albedo. For this reason, the central regions differ slightly between STOKES and MontAGN results, especially in their outer parts.

In model 1, we get two different regions. First a central region mainly constituted of photons that have been scattered twice. The second region is separated into two polar areas of single scattering. These appear in blue in the maps of figure \ref{toy800}. As it is expected in this configuration (discussed in \citealt{Fischer1996} and \citealt{Whitney1993} for example), we observe at north and south two well define regions of centro-symmetric polarisation position angle. In these two zones, photons are scattered with an angle close to 90$^\circ$ which ensure a high degree of polarisation and a polarisation position angle orthogonal to the scattering plane, leading to this characteristic pattern.

The central belt is mostly dominated by the photons scattered twice. Therefore it traces the regions that photons can hardly exit without interaction, because of the optical depth of the medium: these are the areas obscured by the torus. However there is no clear privileged direction of polarisation angle in this central band. This is likely to be due to the surrounding medium, allowing the first scattering to take place in any direction around the torus, a conclusion that leads us to the definition of model 5 (described in next section).

Model 2 only differs by the lack of the dust shell surrounding the torus. The signal clearly shows the ionisation cone as a region of single scattering, and the torus region. In this second area, the central signal is not tracing photons scattered twice, but photons that undergo multiple scattering (5 or more). The role of the shell is therefore critical for the central signal we observe with model 1. It provides an optically thin region where photons can be scattered, without travelling through the optically thick torus (see sketch on figure \ref{path}).

This is confirmed by model 3, which reproduces quite well the results of model 2 in the torus field of view despite having no ionisation cone where the photons can undergo their first scattering.

\subsection{Double scattering Models}

\begin{figure}[ht!]
 \centering    
 \includegraphics[width=0.20\textwidth,clip]{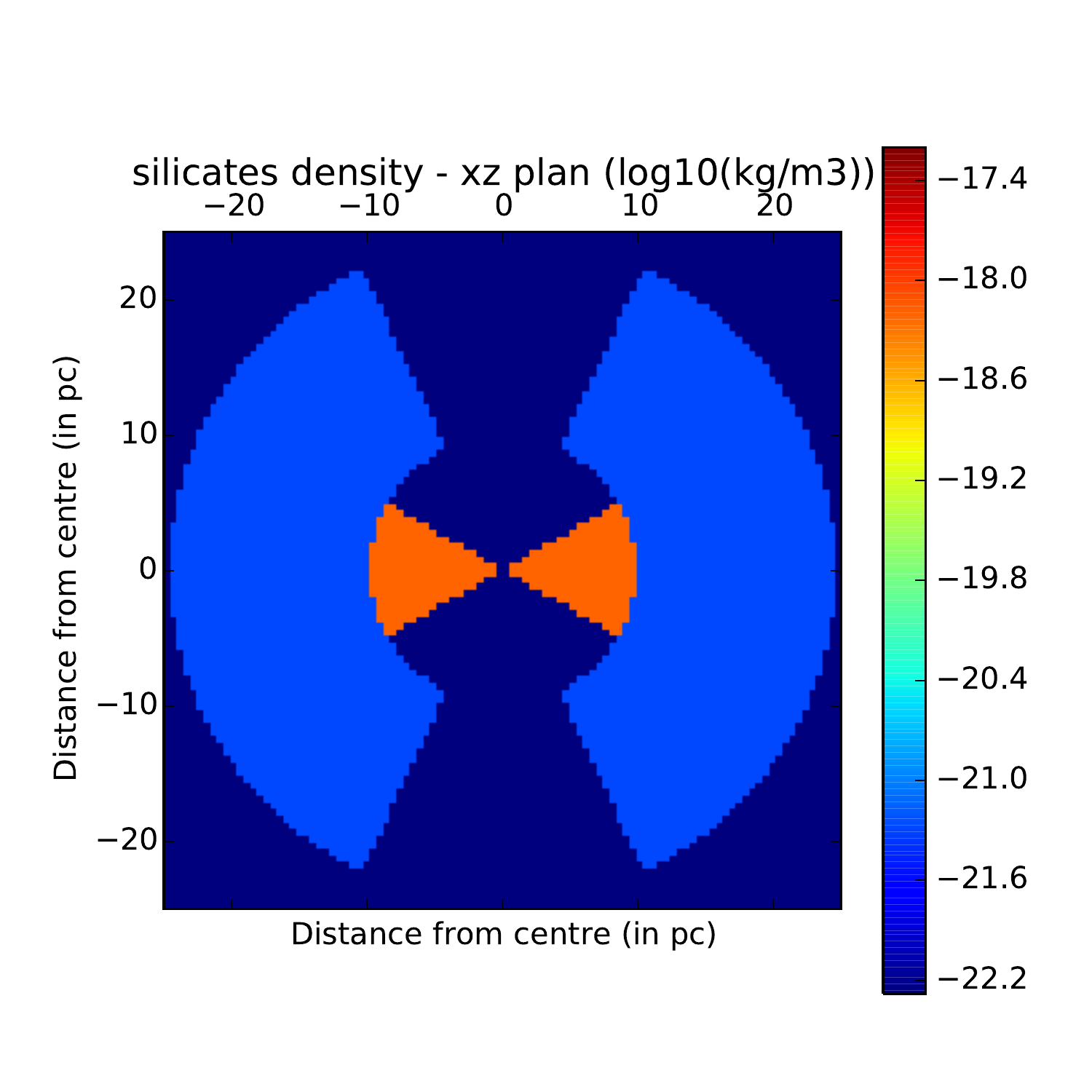} 
 \includegraphics[width=0.20\textwidth,clip]{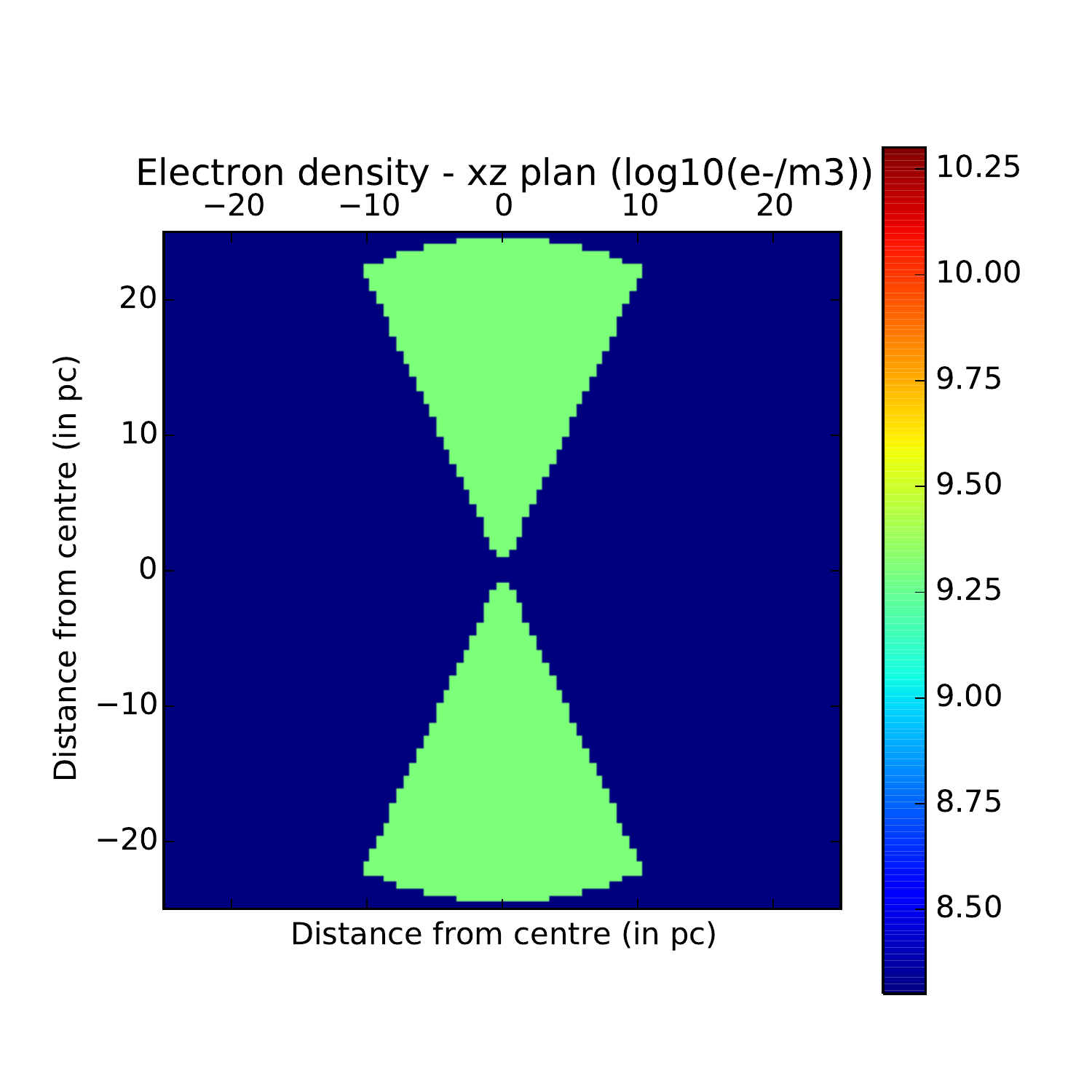} 
 
 \includegraphics[width=0.20\textwidth,clip]{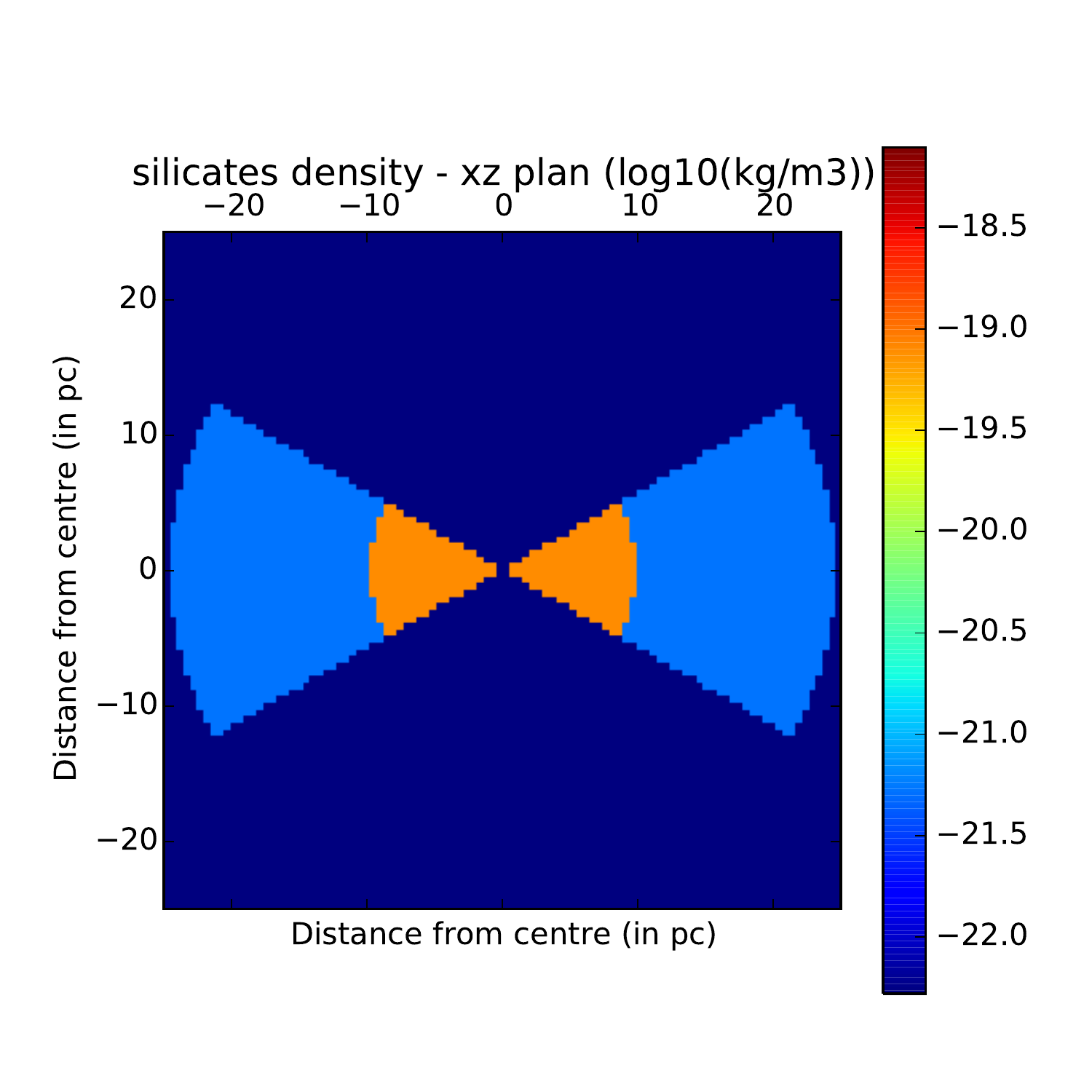} 
 \includegraphics[width=0.20\textwidth,clip]{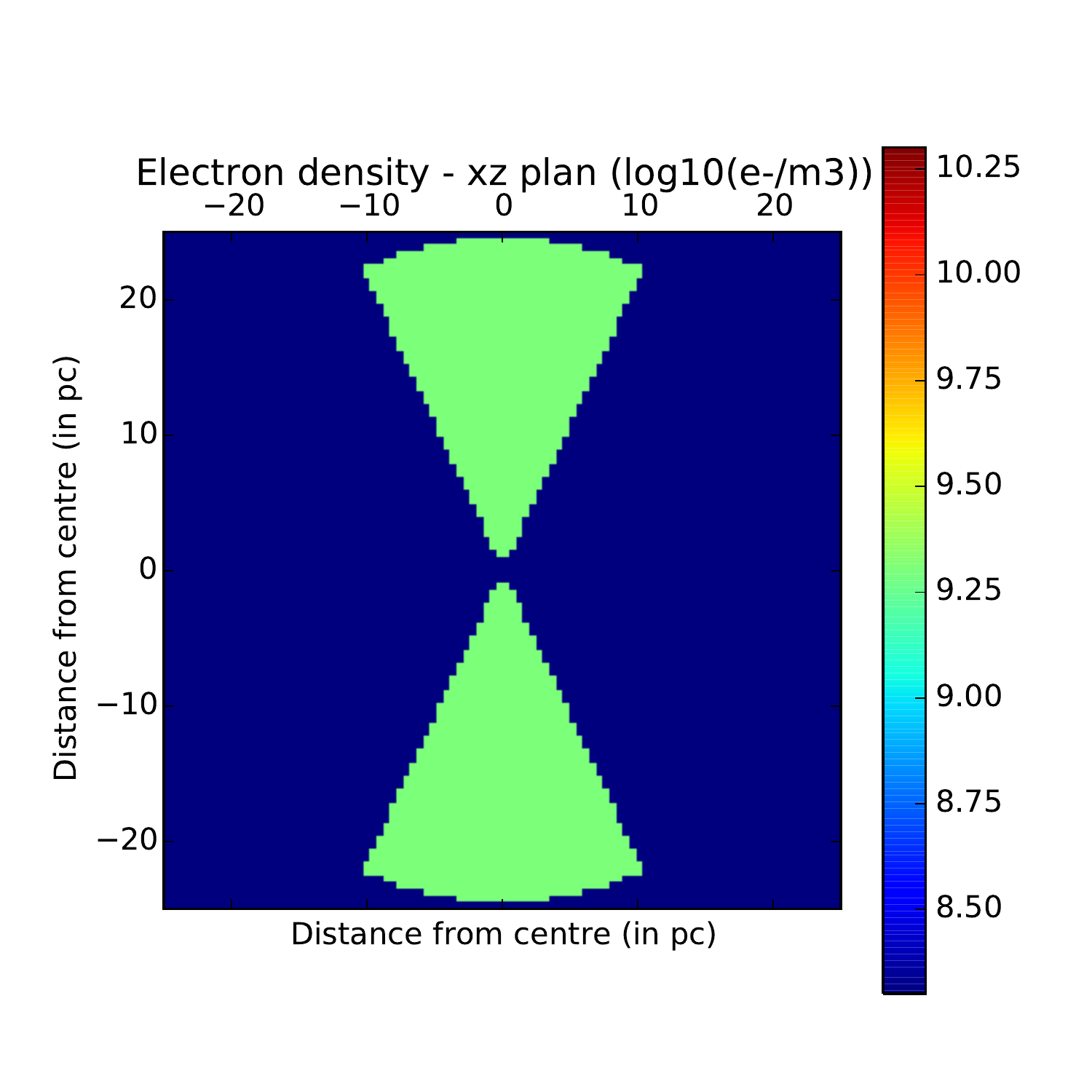} 
  \caption{Grain density of silicates (in log$_{10}$(kg/m$^3$), first column) and electron density (in log$_{10}$(m$^{-3}$), second column) set for model 4 (first row) and model 5 (second row) (shown here for the silicates-electron composition).}
  \label{model2}
\end{figure}

As these simple models were not able to reproduce well enough the observed horizontal polarisation in the central inner parsecs of \cite{Gratadour2015}, we then added some features. Namely we increased the density of the torus to an optical depth of nearly 150 at 800 nm (model 4) and replaced in a second model the outer shell by an extension of the torus with lower density (model 5). See figure \ref{model2} for the density maps.

We used for these models two different dust compositions, only silicates on one hand and a mixture of graphites (parallel and orthogonal) and silicates grains in the other hand. The ionisation cone contains only electrons in both cases. In case of silicates and graphites, we used the following number ratio: 37.5\% silicates, 41.7\% orthogonal graphites (electric field oscillates perpendicular to the graphite plane) and 20.8\% parallel graphites (electric field oscillates parallel to the graphite plane), based on \cite{Goosmann2007}. Other ratio could be considered, for example by using the work of \cite{Jones2013}. Silicate and mixture models share the same $\tau_R$.

\begin{figure}[ht!]
 \centering    
 \includegraphics[width=0.30\textwidth,clip]{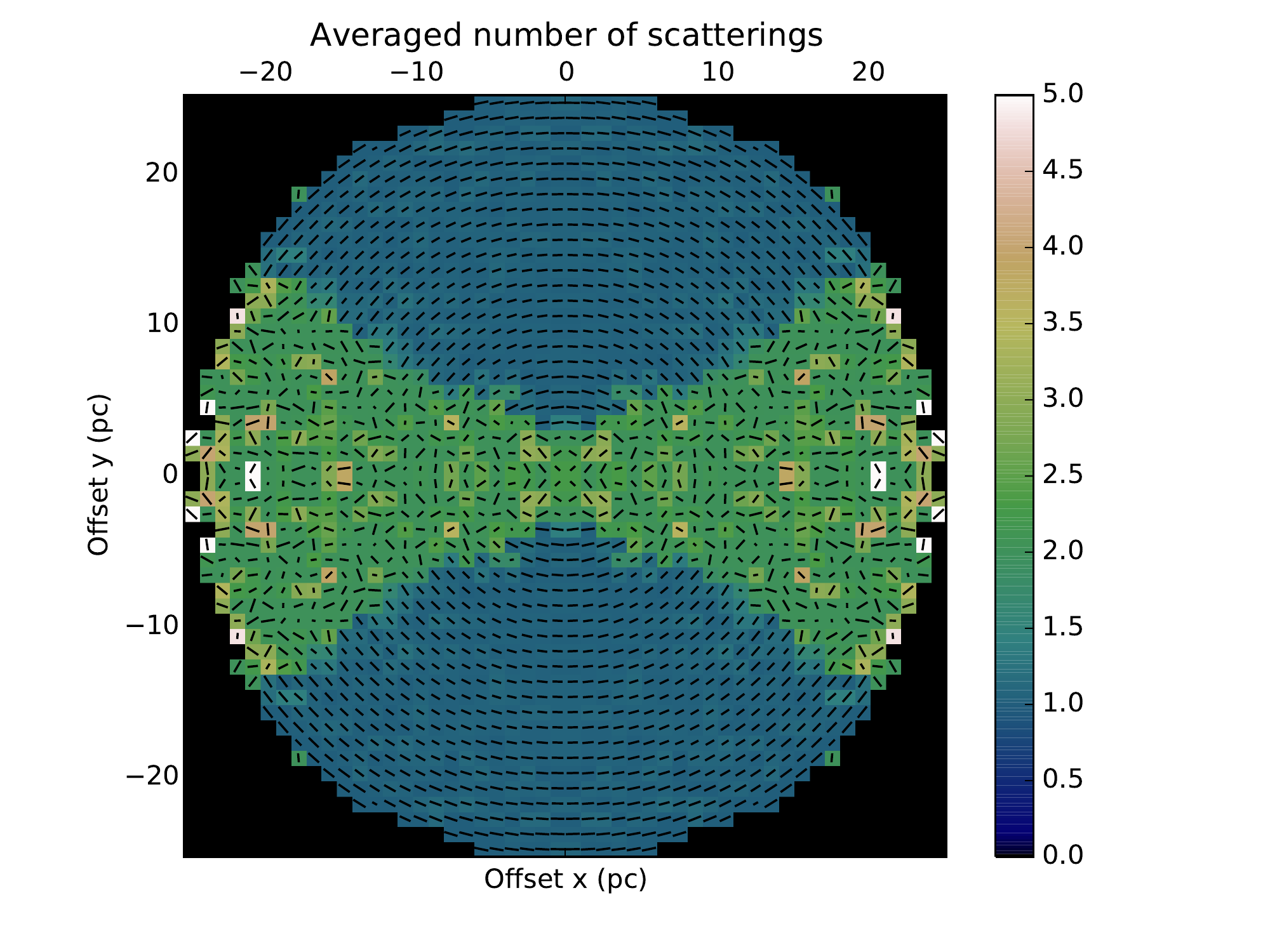}
 \includegraphics[width=0.30\textwidth,clip]{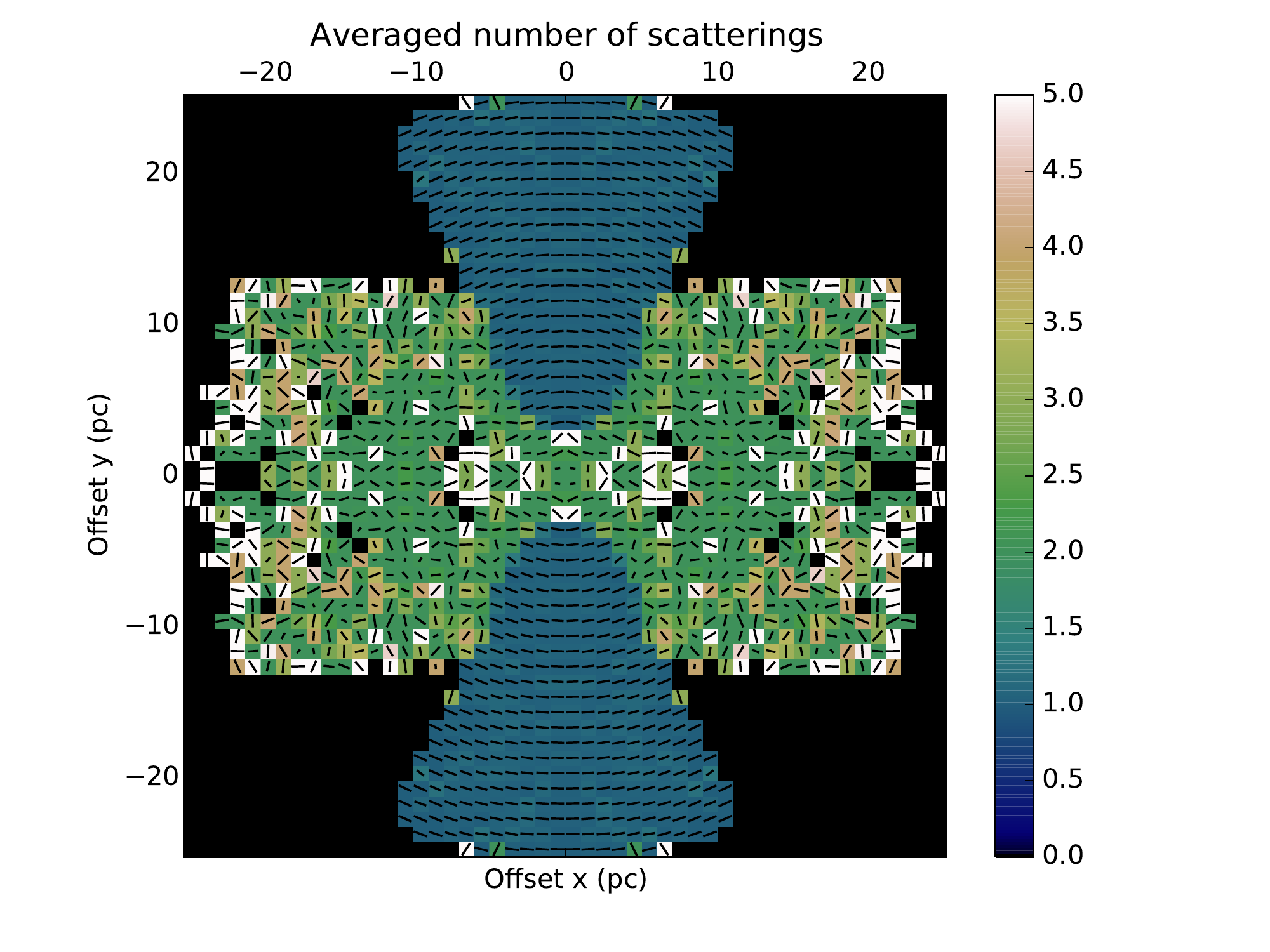}
  
 \includegraphics[width=0.30\textwidth,clip]{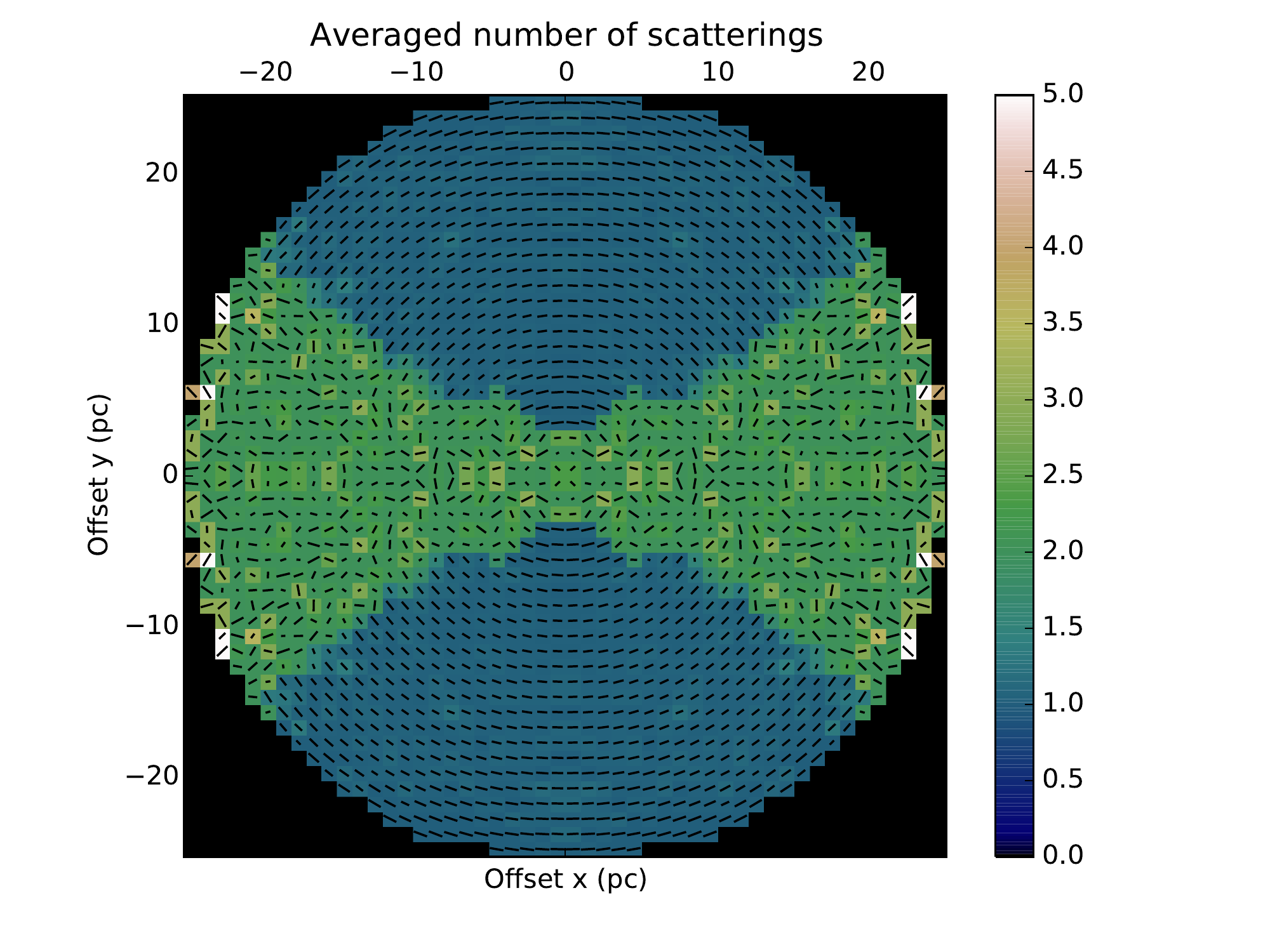}
 \includegraphics[width=0.30\textwidth,clip]{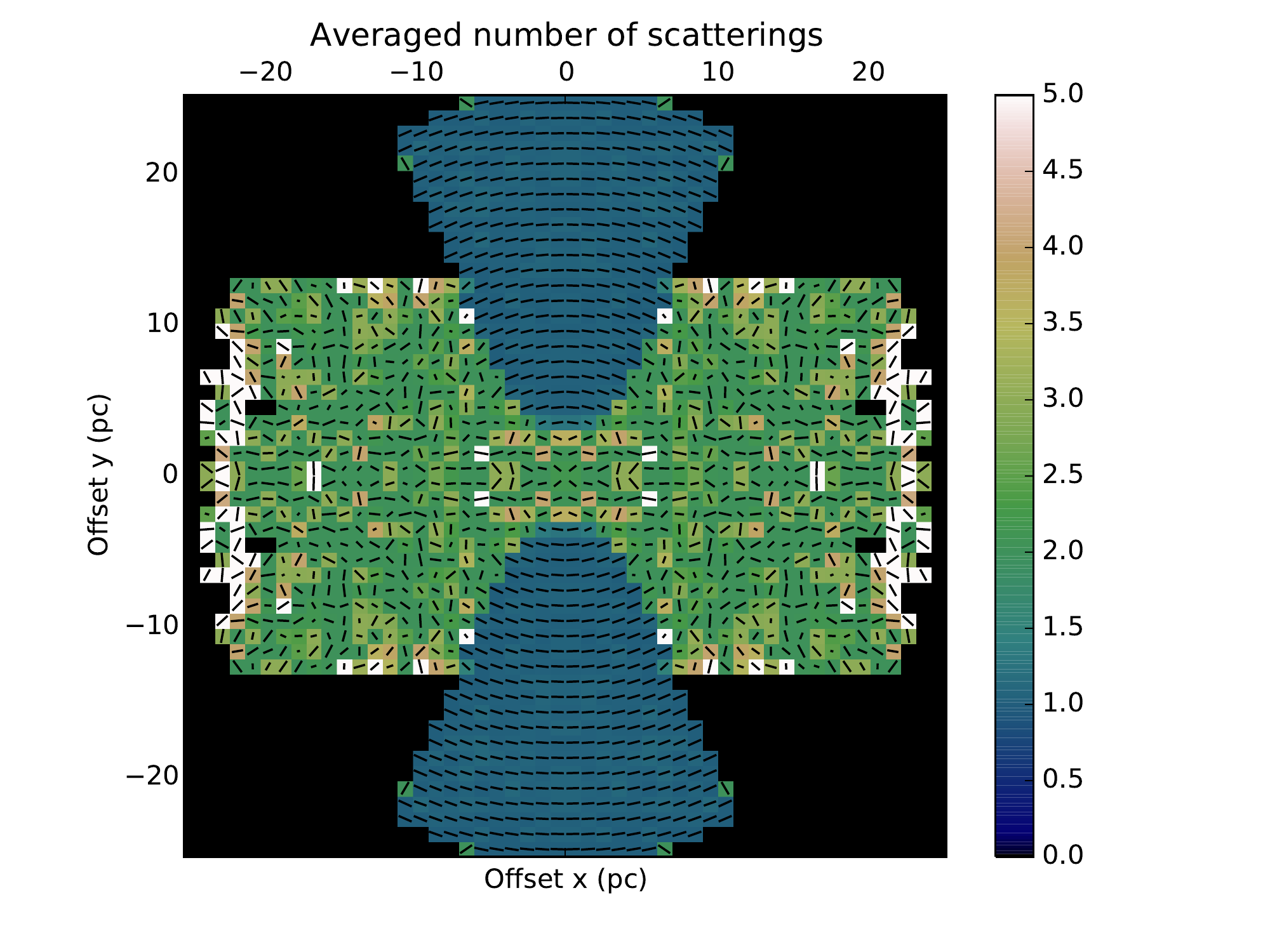}
  \caption{Maps of observed averaged number of scattering with an inclination angle of 90$^\circ$ at 800 nm with MontAGN. Polarisation vectors are shown, their length being proportional to the polarisation degree and their position angle representing the polarisation angle. First two images are for models with silicate and electrons, last two images for models with silicates graphites and electrons. First and third images correspond to model 4 and the second and fourth ones to model 5.}
  \label{res800}
  \end{figure}

\begin{figure}[ht!]
 \centering   
  \includegraphics[width=0.30\textwidth,clip]{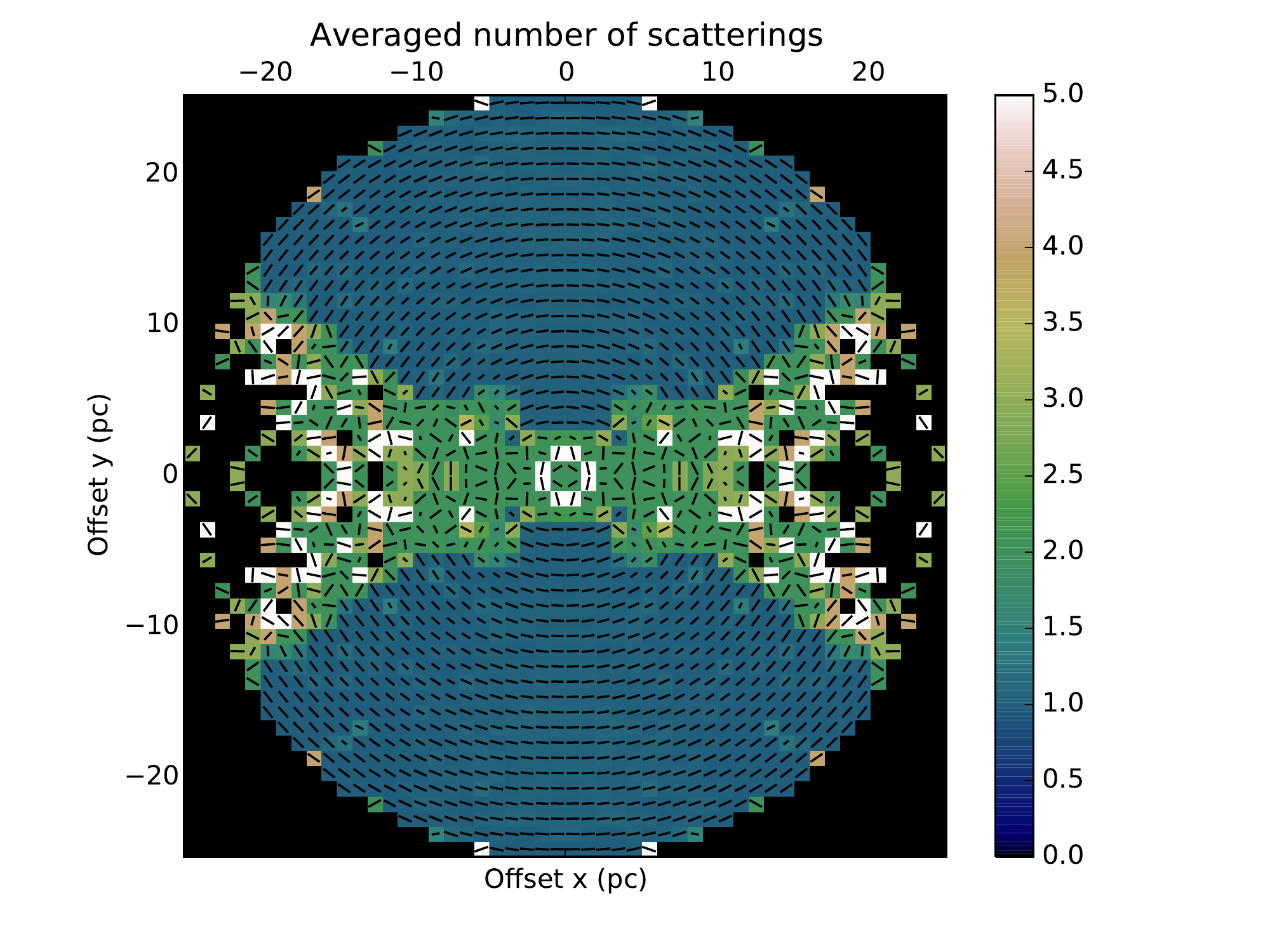}
  \includegraphics[width=0.30\textwidth,clip]{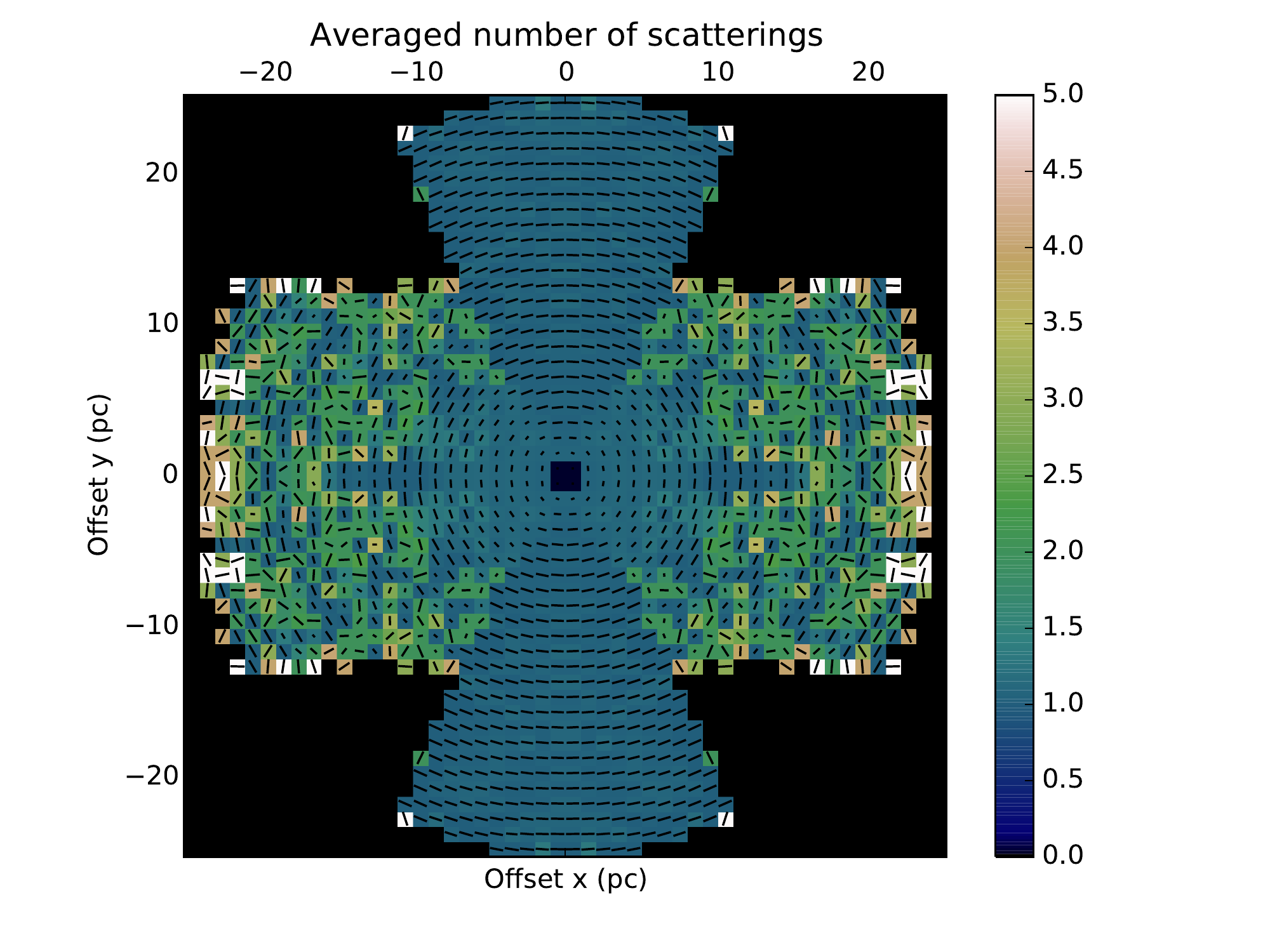}
  
  \includegraphics[width=0.30\textwidth,clip]{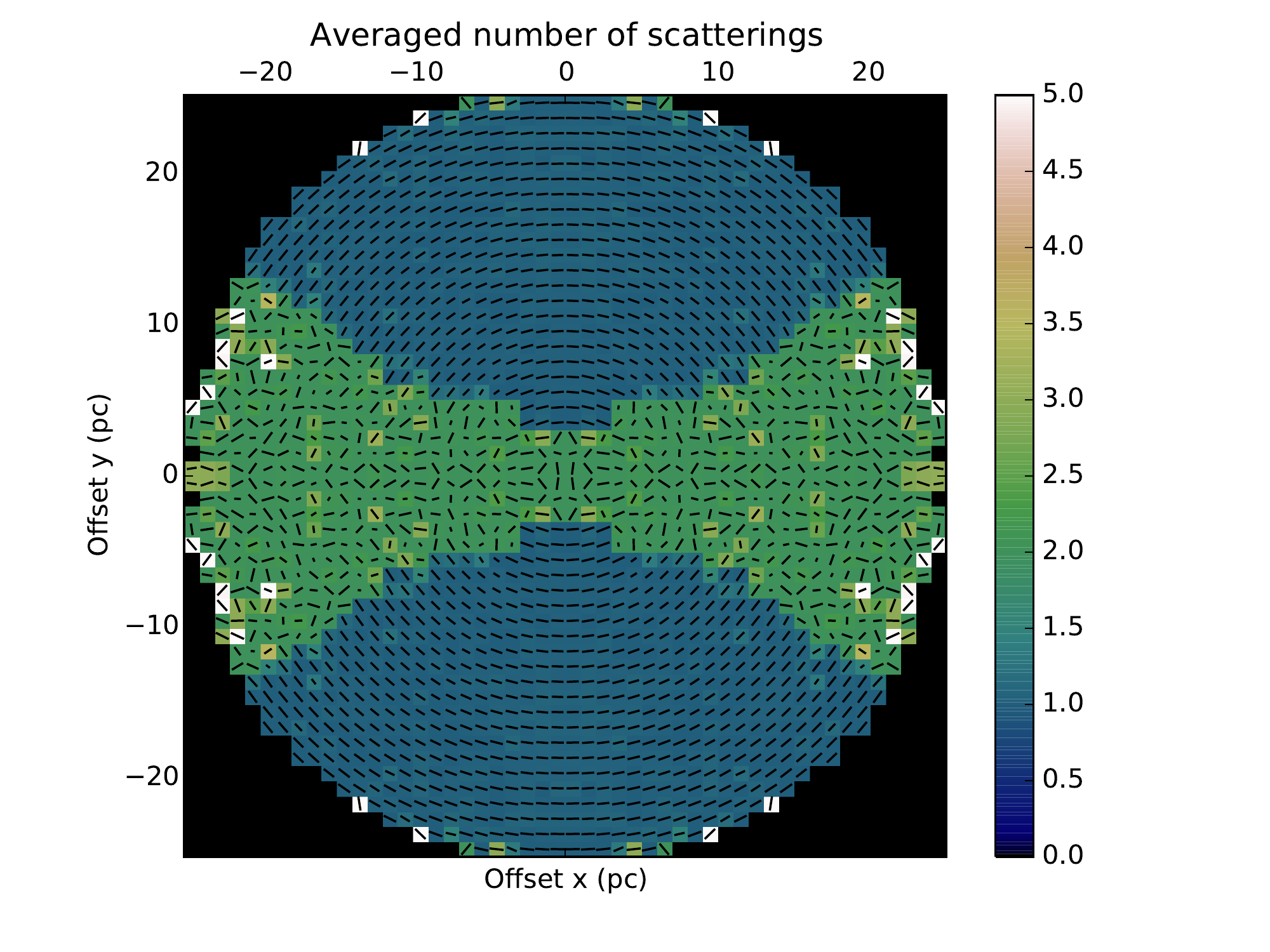}
  \includegraphics[width=0.30\textwidth,clip]{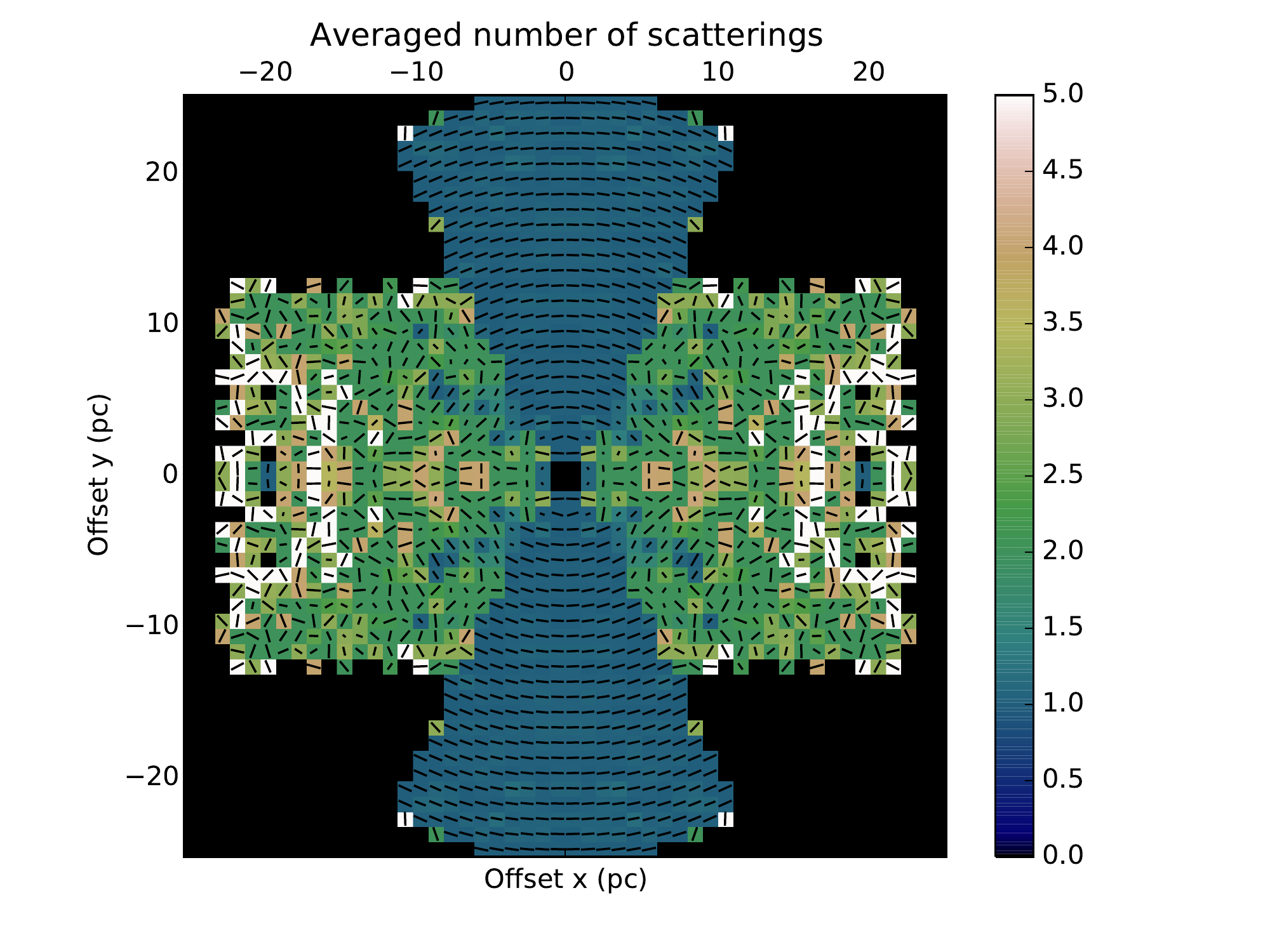}
   \caption{Same as figure \ref{res800} but at 1600 nm.}
   \label{res1600}
\end{figure}

We ran each of these models with 10$^7$ photons packets launched, without re-emission and considering only the wavelengths 800 nm and 1.6 $\mu$m. All inclination angles were recorded. We show in figures \ref{res800} and \ref{res1600} maps derived from the MontAGN simulations. These correspond to maps of averaged number of scatterings with polarisation vectors ($p=1$ is represented by a length of a pixel size), for models with the two dust compositions.

One should first notice that the dust composition slightly affects the results, mostly on maps of figure \ref{res1600}. This is likely due to the difference in optical depth introduced by the difference in composition\footnote{The optical depth is the same in R but evolves differently depending on the dust composition} as it will be discussed later in section \ref{composition}.

The centro-symmetric regions of model 4 are similar for both dust composition at 800 nm and 1.6 $\mu$m and are similar to results of model 1. As the optical depth of the torus should not affect this part of the maps, this is consistent. However, the results found with model 4 are very close to those observed with model 1, even in the central belt. In this region, most of the photons undergo two scatterings at 800 nm for model 4, the only difference arises from the model with silicates where we see that a slightly larger fraction of photons have undergone more than two scattering events. At 1.6 $\mu$m, there are almost no photon coming from the regions shadowed by the torus on the outer central belt, with the pure silicates model. Again this is an effect of optical depth (see section \ref{composition})

Model 5 differs (second and fourth images of figure \ref{res800} and \ref{res1600}) more from previous models. By adding a region, spatially limited, where photons can be scattered a second time, hidden by the torus from the emission of the source, we see a large region of double scattered light. Polarisation in this area in map at 800 nm shows a clear constant horizontal polarisation. This is not visible on the 1.6 $\mu$m map, where the optical depth of the torus is not high enough to block photons coming directly from the AGN centre.

\subsection{Optical Depth}
\label{optical_depth}

\begin{figure}[ht!]
 \centering   
  \includegraphics[width=0.30\textwidth,clip]{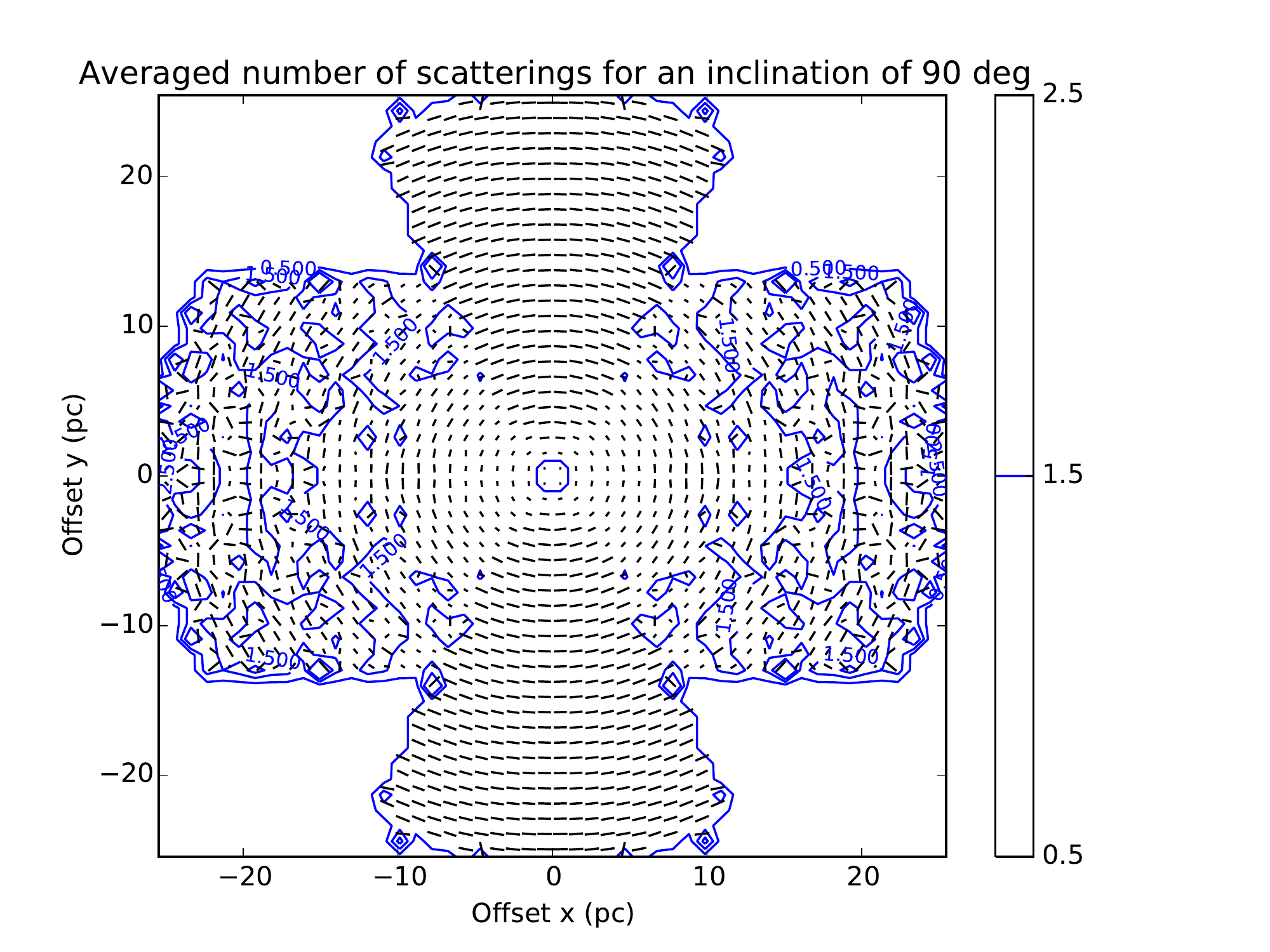}
  \includegraphics[width=0.30\textwidth,clip]{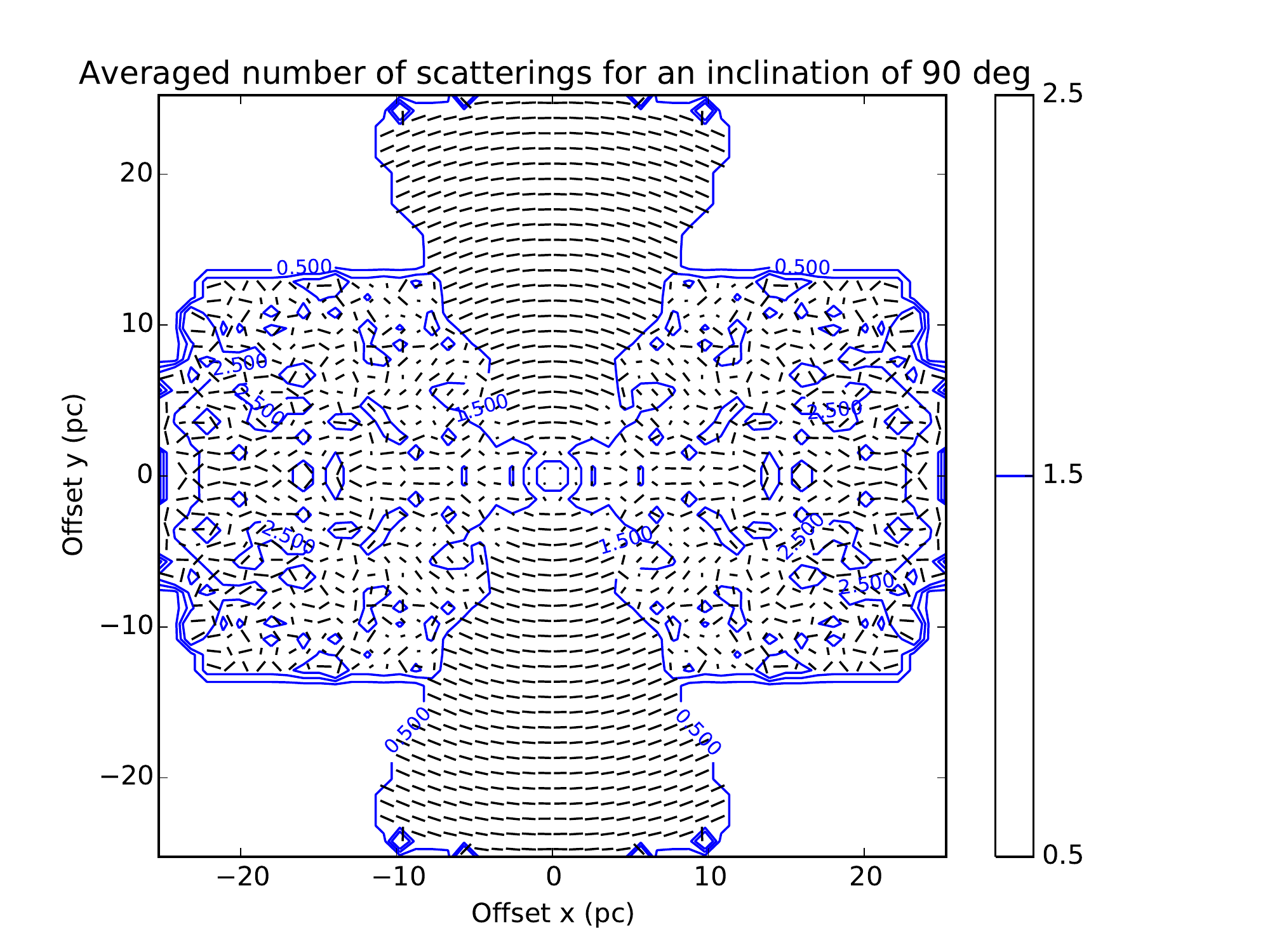}
  \includegraphics[width=0.30\textwidth,clip]{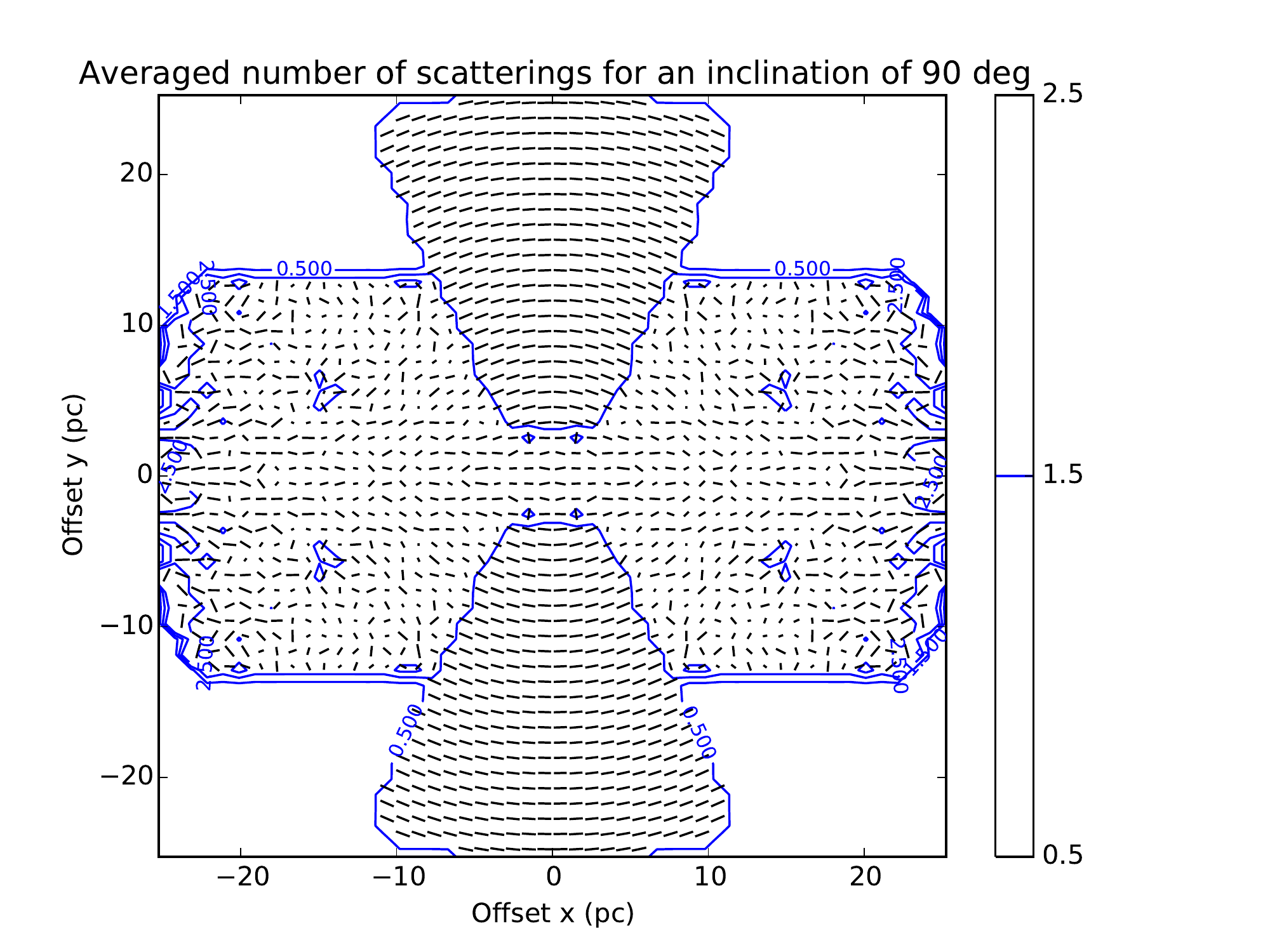}
  \includegraphics[width=0.30\textwidth,clip]{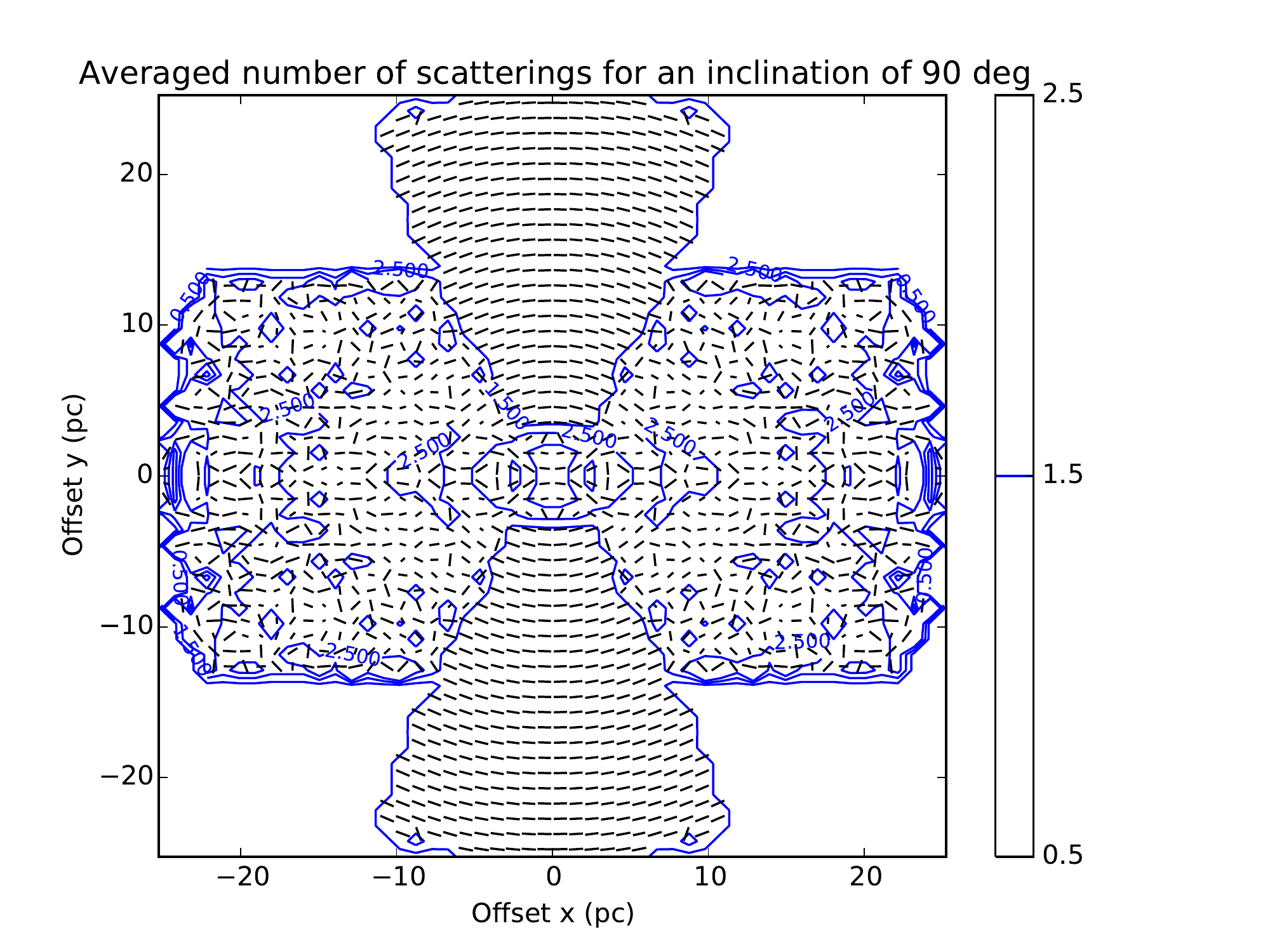}
  \includegraphics[width=0.30\textwidth,clip]{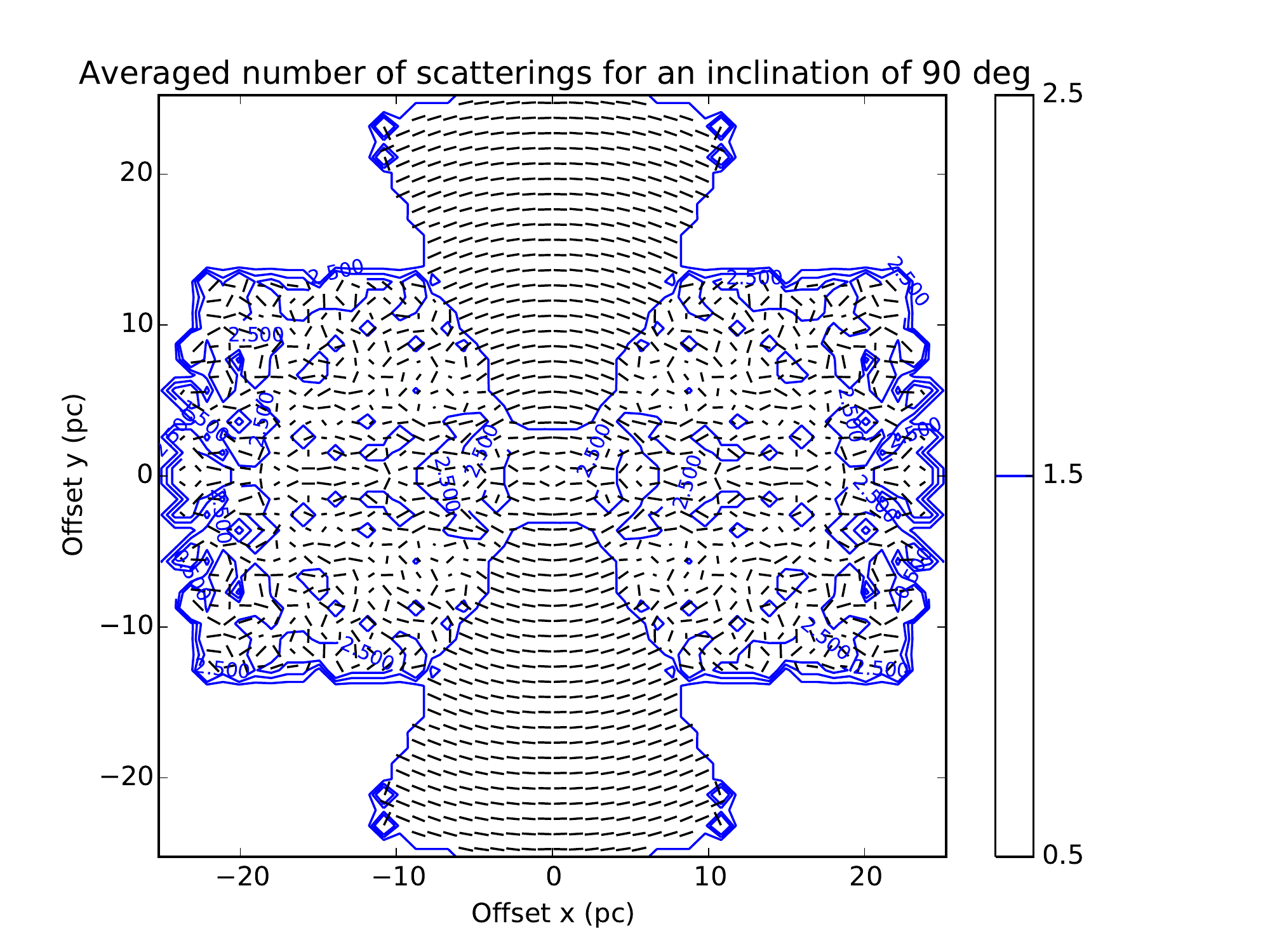}
   \caption{Maps of observed averaged number of scattering with an inclination angle of 90$^\circ$ at 1.6 $\mu$m with MontAGN. Optical depth of the cone is fixed to 0.1, those of the extended torus to 0.8 and the torus one is respectively 5, 10, 20, 50 and 100 (in H band)}
   \label{tau_torus}
\end{figure}

\begin{figure}[ht!]
 \centering   
  \includegraphics[width=0.30\textwidth,clip]{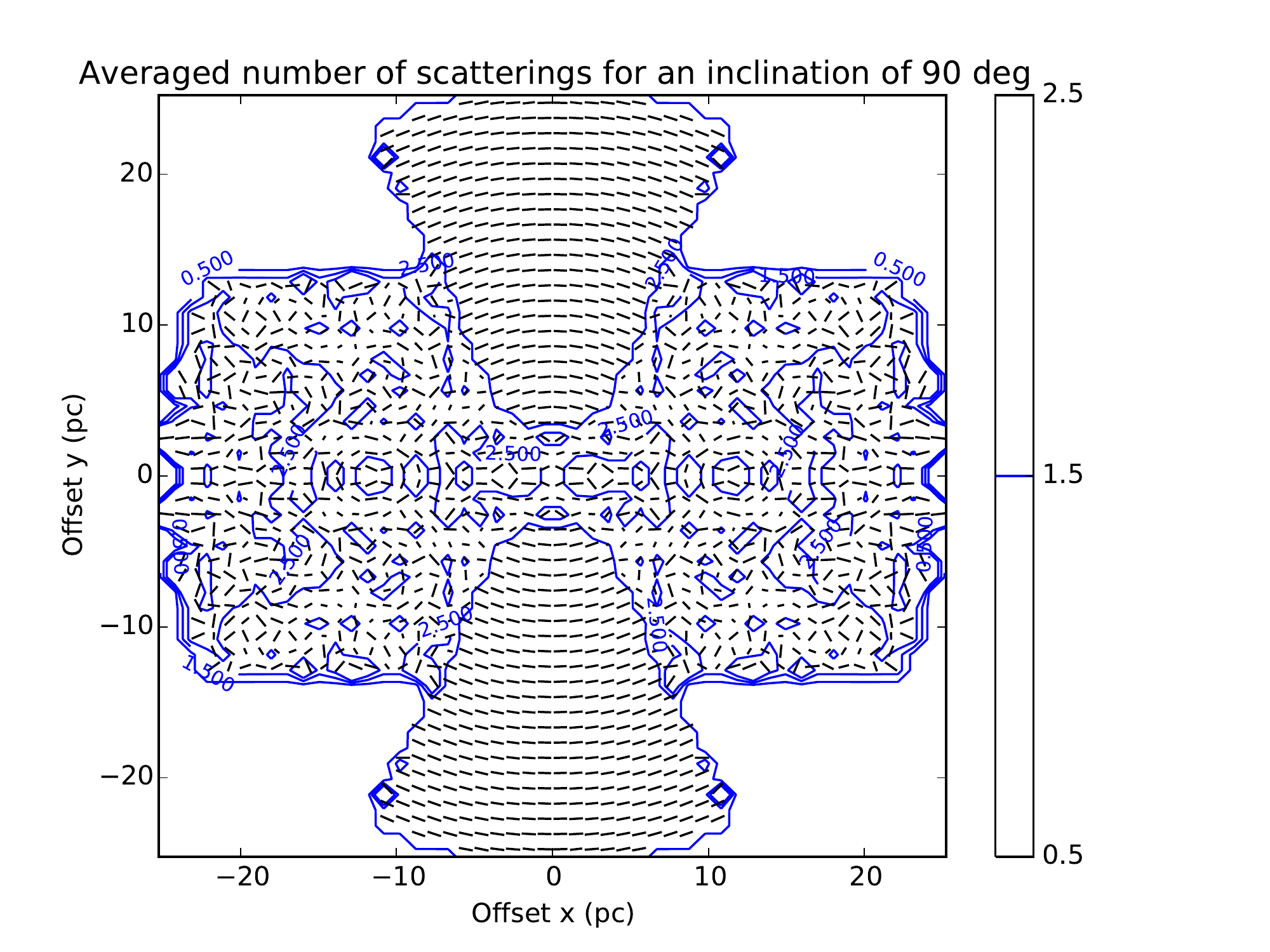}
  \includegraphics[width=0.30\textwidth,clip]{mod23-322_1600nm_090_cndiff_vect.pdf}
  \includegraphics[width=0.30\textwidth,clip]{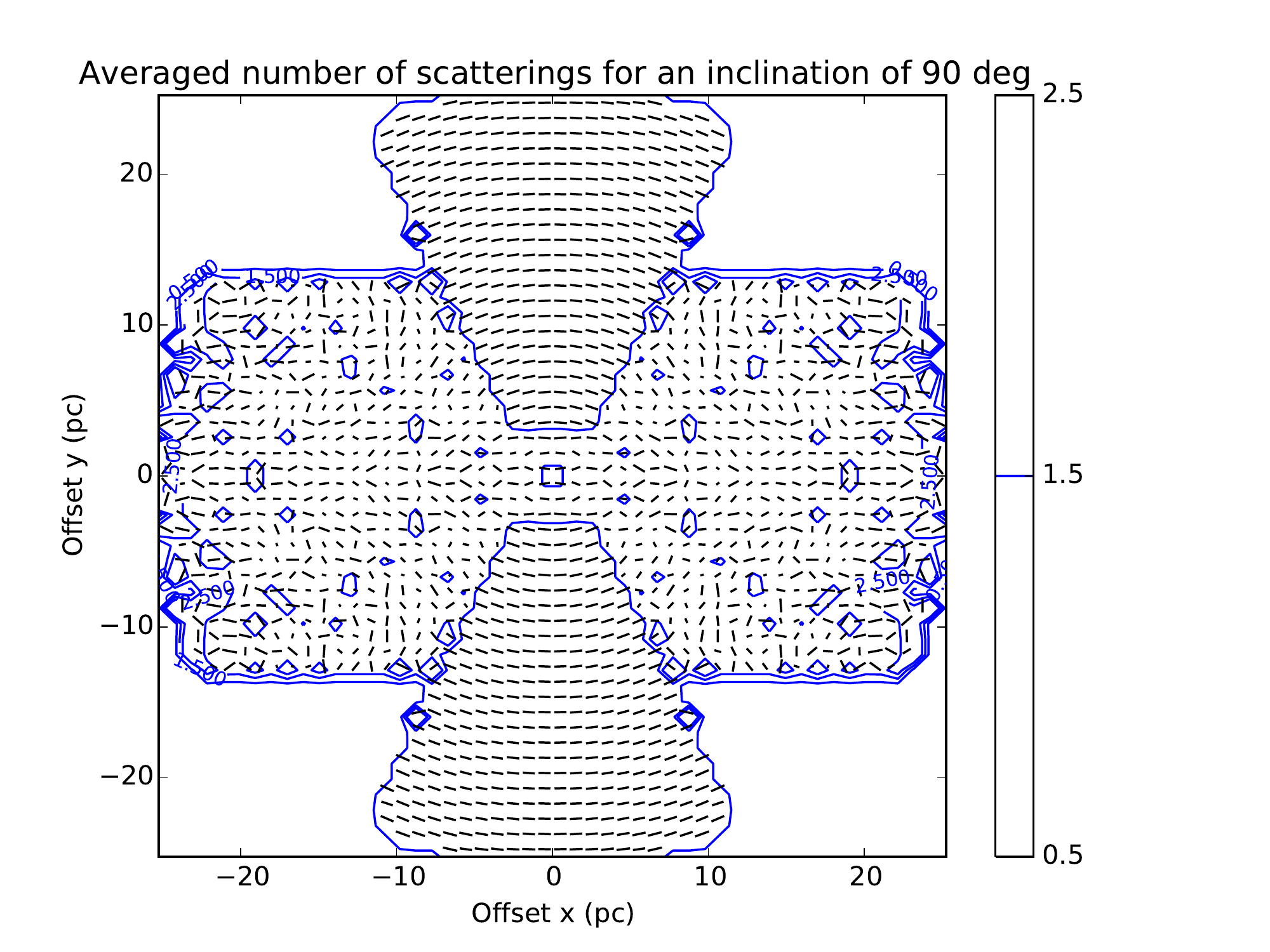}
  \includegraphics[width=0.30\textwidth,clip]{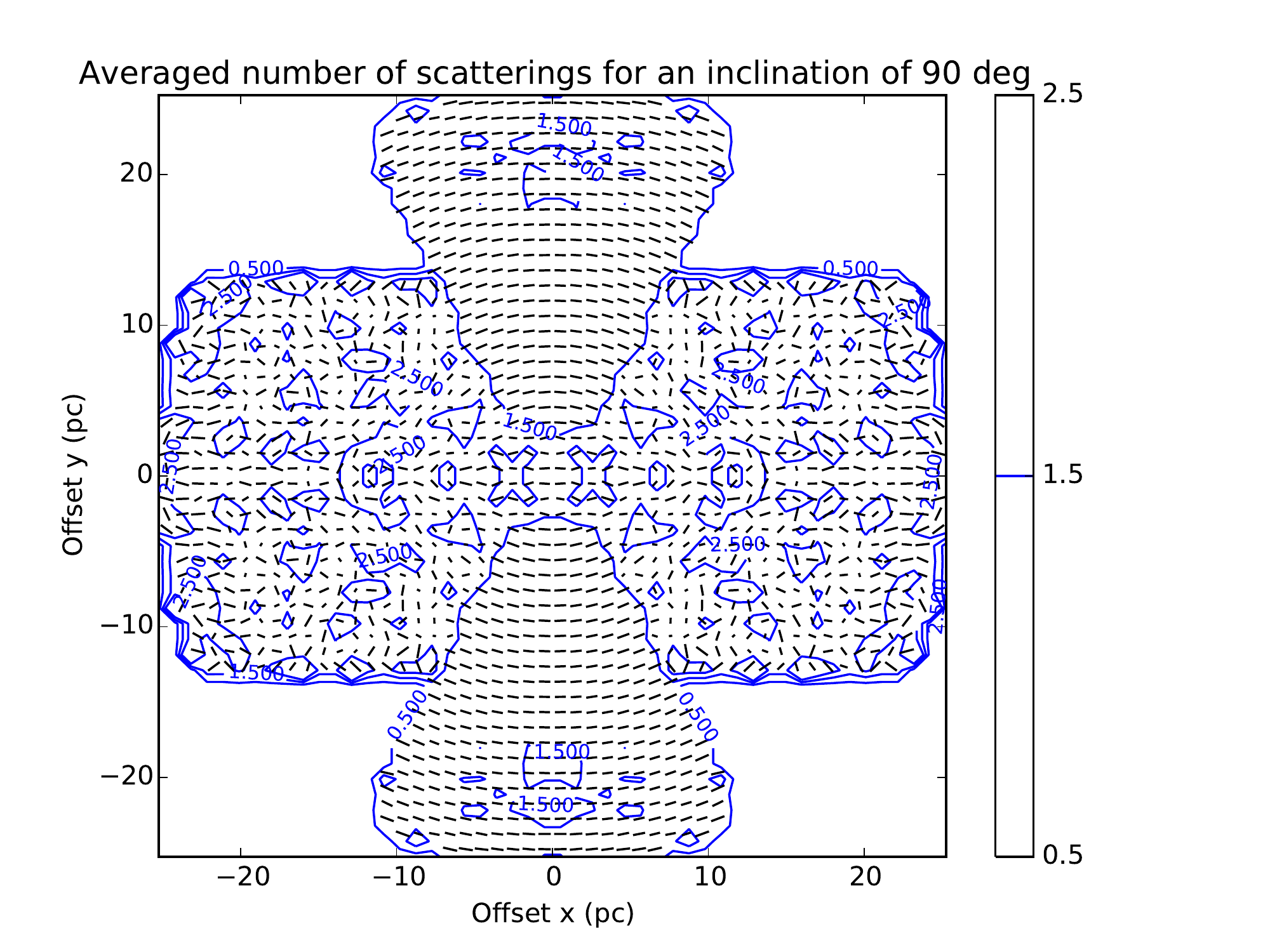}
  \includegraphics[width=0.30\textwidth,clip]{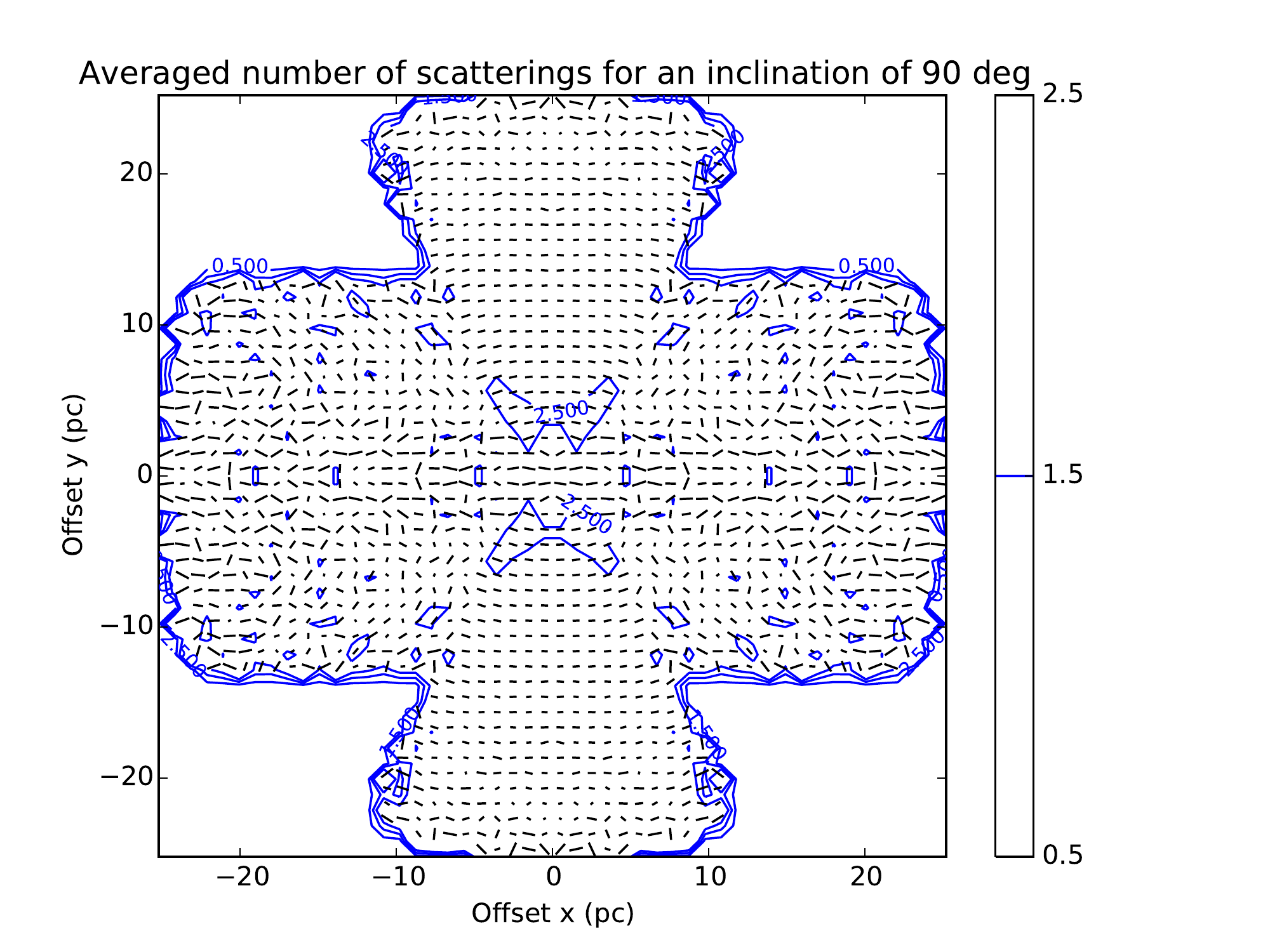}
   \caption{Same as figure \ref{tau_torus}. Optical depth of the cone is respectively 0.05, 0.1, 0.5, 1.0 and 10.0, those of the torus and its extended part are fixed to 20 and 0.8 (in H band)}
   \label{tau_cone}
\end{figure}

\begin{figure}[ht!]
 \centering   
  \includegraphics[width=0.30\textwidth,clip]{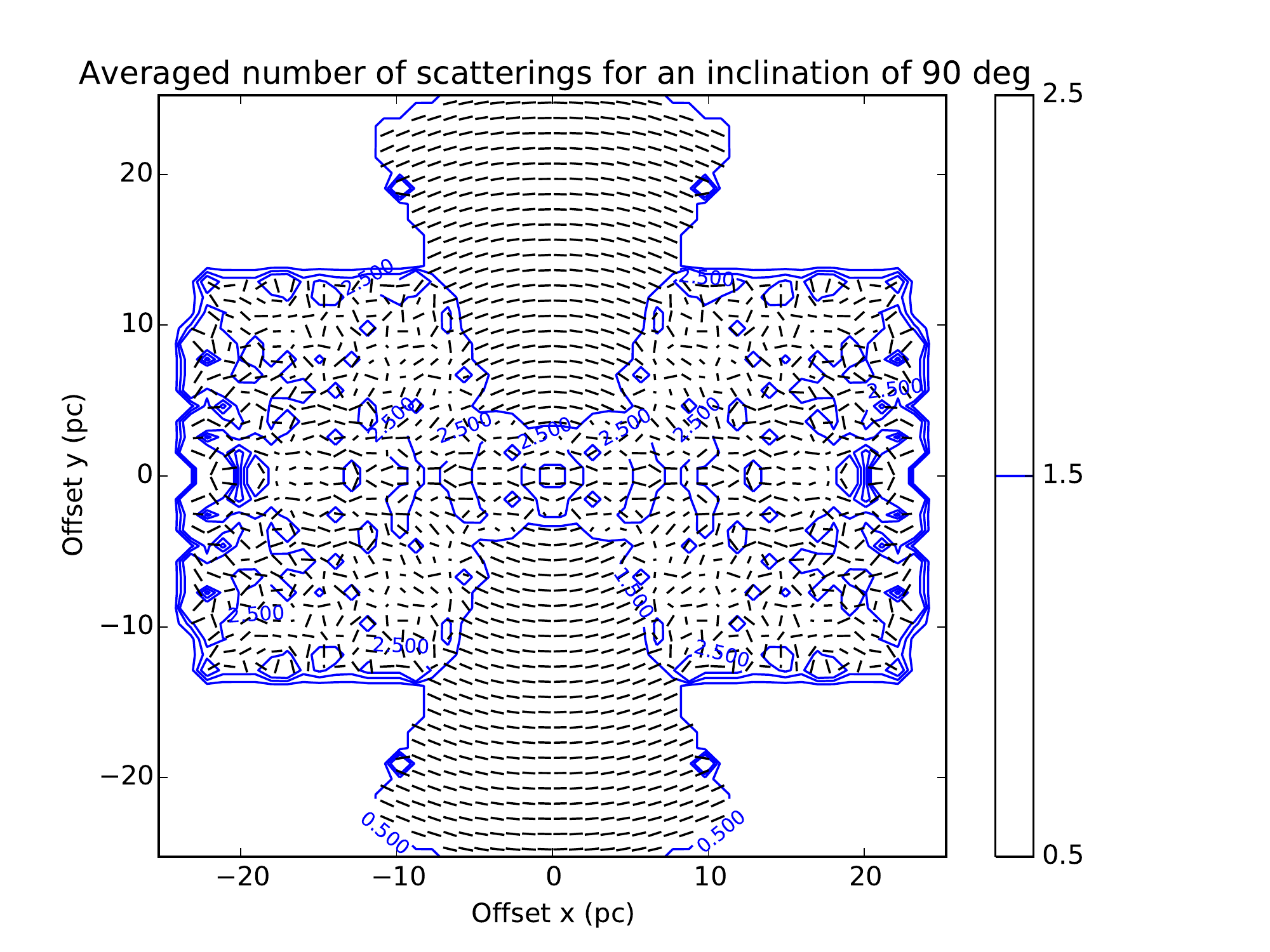}
  \includegraphics[width=0.30\textwidth,clip]{mod23-322_1600nm_090_cndiff_vect.pdf}
  \includegraphics[width=0.30\textwidth,clip]{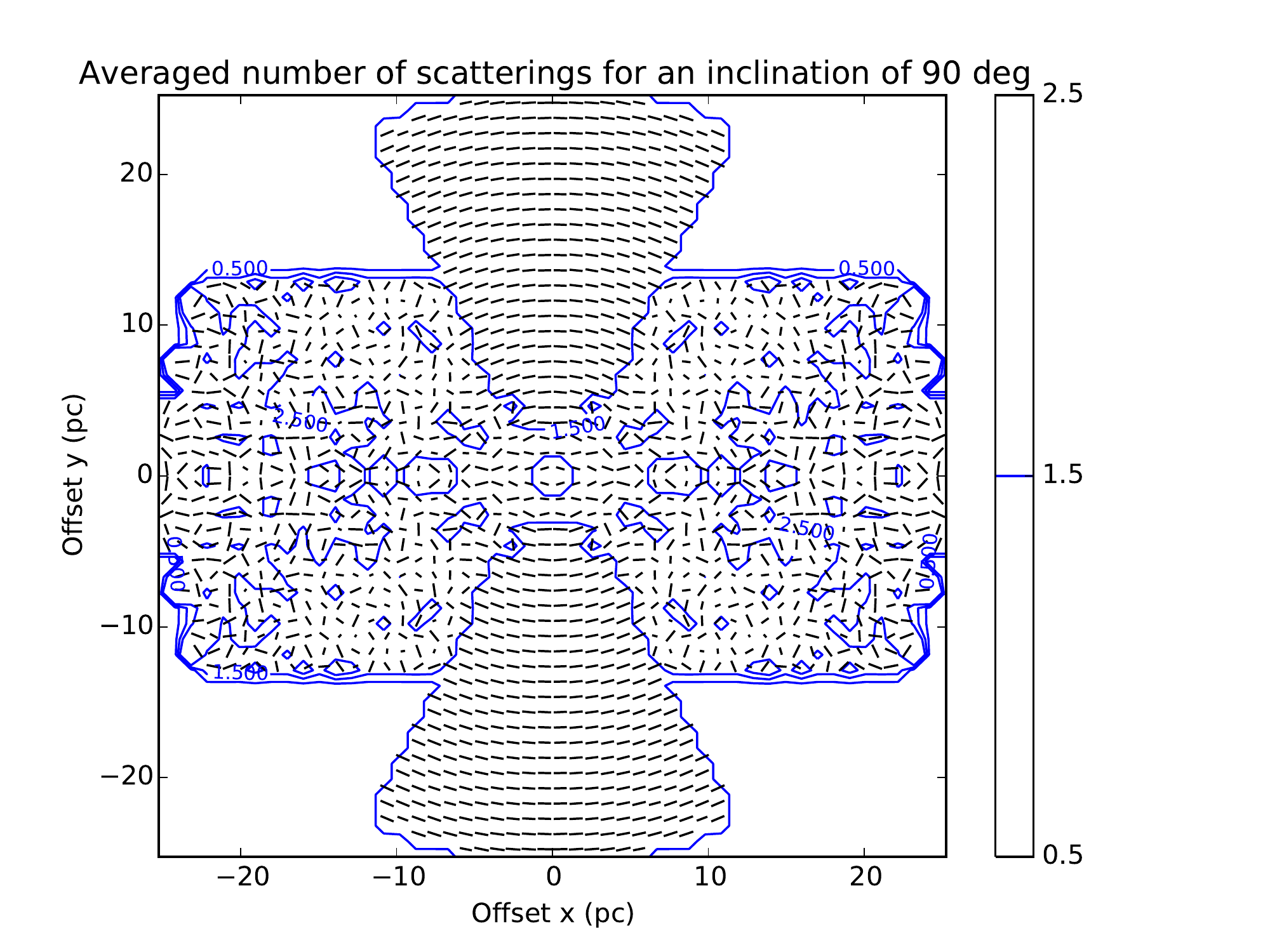}
  \includegraphics[width=0.30\textwidth,clip]{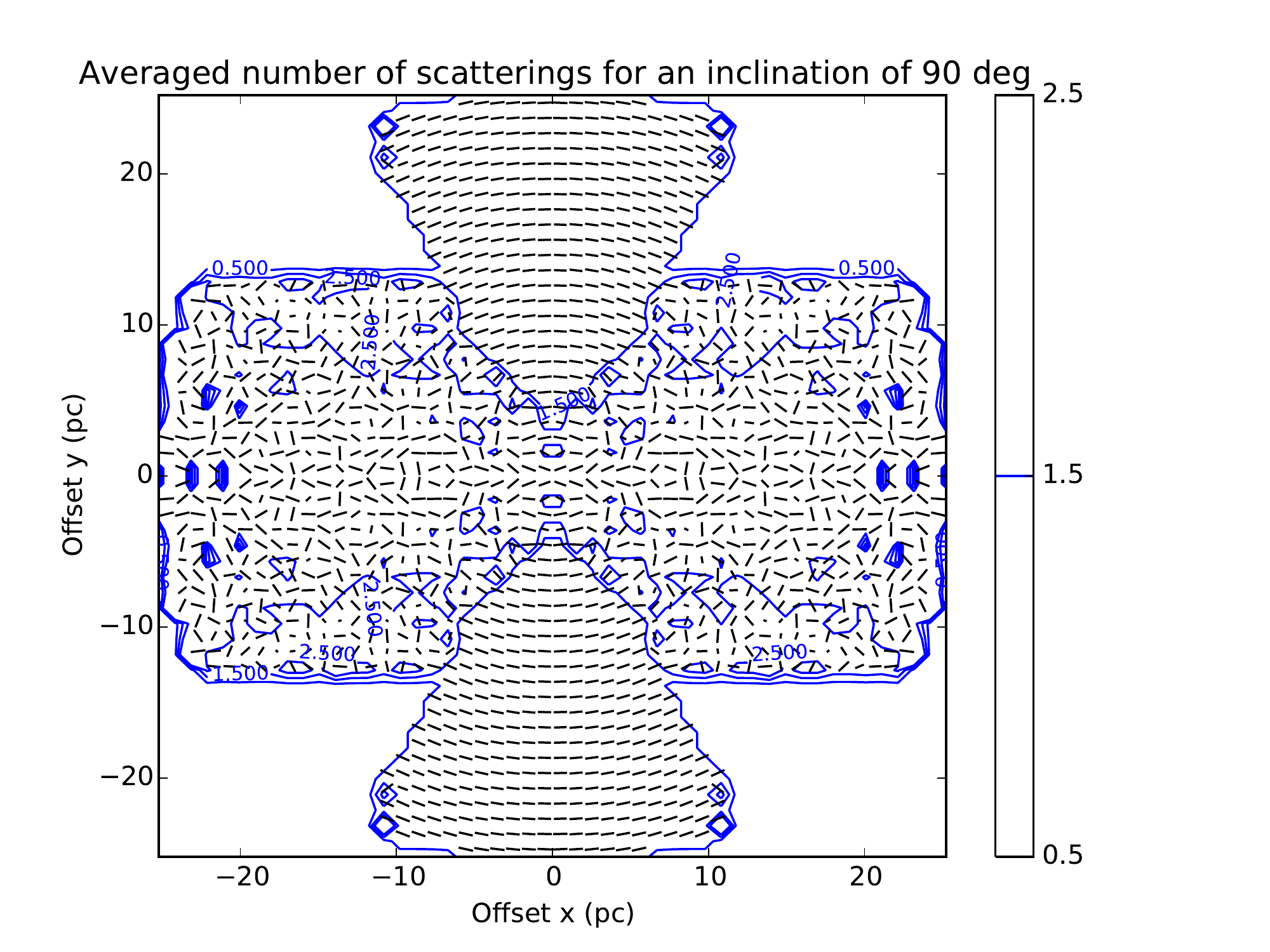}
  \includegraphics[width=0.30\textwidth,clip]{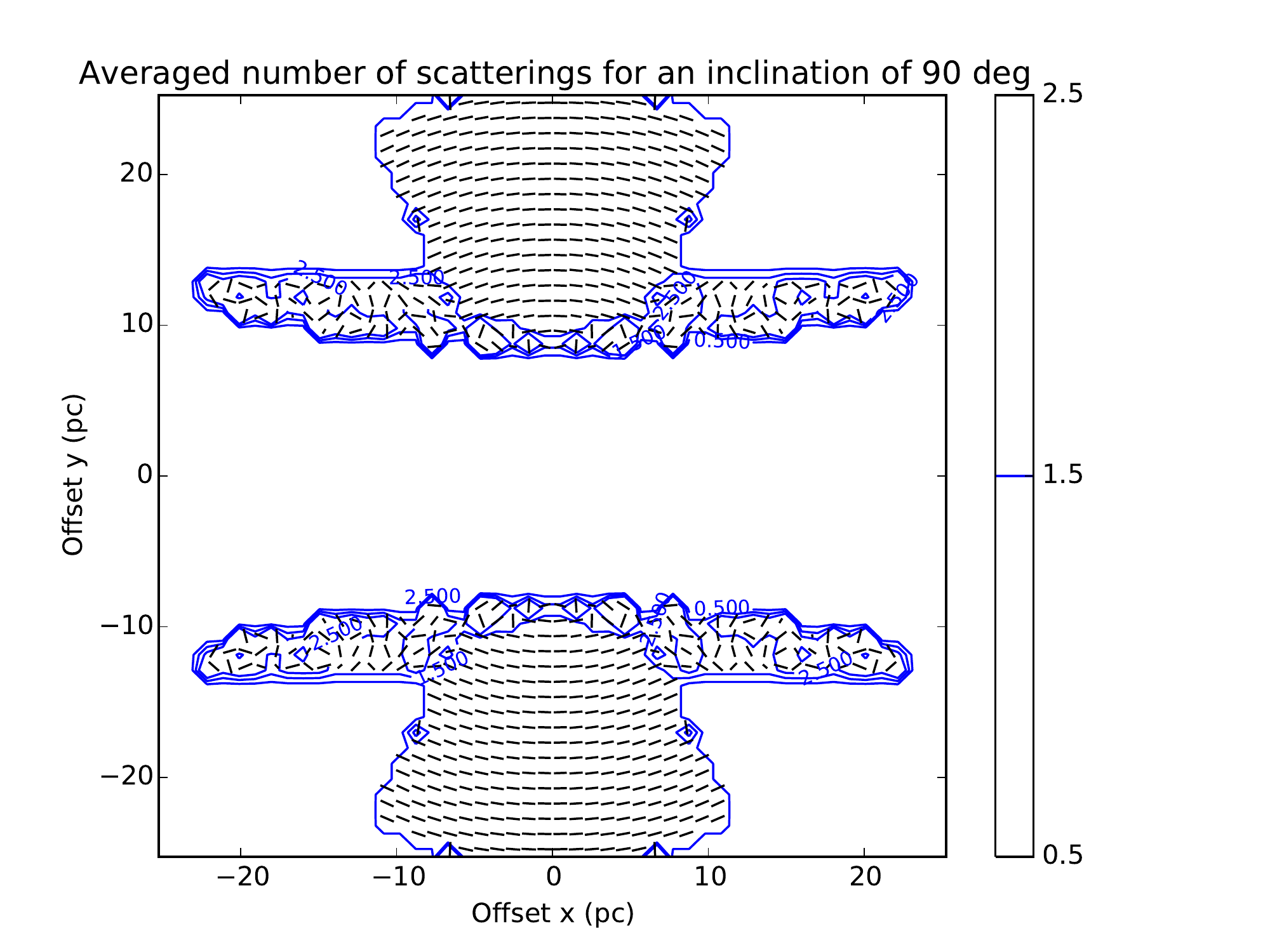}
   \caption{Same as figure \ref{tau_torus} and \ref{tau_cone}. Optical depth of the cone is 0.1, those of the extended torus respectively 0.4, 0.8, 4.0, 8.0 and 80.0. The torus one is fixed to 20 (in H band)}
   \label{tau_torext}
\end{figure}

In order to measure the impact and the importance of the optical depth, we ran a series of models based on model 5, differing only by the optical depth of their components. Among the three main parameters, the ionisation cone, the torus and its extended region, we kept fixed the density of two of them and changed the last one's. First we took models with varying the optical depth of the torus. In a second time, we changed the optical depth of the cone, and third we study a variation of those of the extended torus. The results are shown in figures \ref{tau_torus}, \ref{tau_cone} and \ref{tau_torext}, respectively. Polarisation vectors are shown, their length being proportional to the polarisation degree and their position angle representing the polarisation angle. All these models contains the same mixtures of silicates and graphites as in model 4 and 5. $2 \times 10^7$ packets were launched per model, all at 1.6 $\mu$m where the optical depth is set. Some additional maps are available in appendix \ref{add_map}.

With the first series of simulation, (figure \ref{tau_torus}), we obtain similar results as on the previous cases. For a low optical depth of the torus, photons can travel without interaction through the torus and produce a centro-symmetric pattern, even in the central region. By increasing the density, the importance of these photons decreases until the double scattered light become predominant (around $\tau \approx$ 20). At this step, increasing the optical depth does not seem to change significantly the observed features.

With the second and third series, the density of the two other structures must be in a certain range to obtain the constant horizontal polarisation pattern we are looking for. If these structures have an optical depth too low or too high, the central belt shows multiple scatterings as seen in both figures \ref{tau_cone} and \ref{tau_torext}. In extreme cases with very high optical depth in the outer torus, we have no photons in this region, a case close to simulations of model 2 (see figure \ref{toy800}). These two conditions seem to be important to produce the features we are expecting.

A last remark is that the photons detected from the ionisation cone are not any more just scattered once at high optical depth in the cone (figure \ref{tau_cone}). However the centro-symmetric pattern seems to be approximatively conserved for double scattering in the cone.

\subsection{Reproducing the NGC 1068 Observations}

Based on previous simulations, we tried to reproduce the SPHERE observations of NGC 1068. As we observed constant polarisation orientation in the core of NGC 1068 both in H and Ks bands, we need in our simulation an optical depth $\tau \geqslant 20$ at 2.2 $\mu$m to be able to generate such polarisation features. We changed our previous model 5 to adapt it to these larger wavelength. Here, the torus has an optical depth $\tau_{Ks}$=19, which gives $\tau_H$=35 and $\tau_V$=169. Note that this is in the range of the current estimation of the density of dust in the torus as it will be discussed in section \ref{concl_obs}. We used for this simulation a mixture of silicates and graphites to rely on a realistic dust composition for an AGN, inspired from \cite{Wolf1999}. We adapted their mass ratio to number ratio according to \cite{Guillet2008The} leading to 57\% silicates, 28.67\% parallel graphites, 14.33\% orthogonal graphites. 

\begin{figure}[ht!]
 \centering   
  \includegraphics[width=0.40\textwidth,clip]{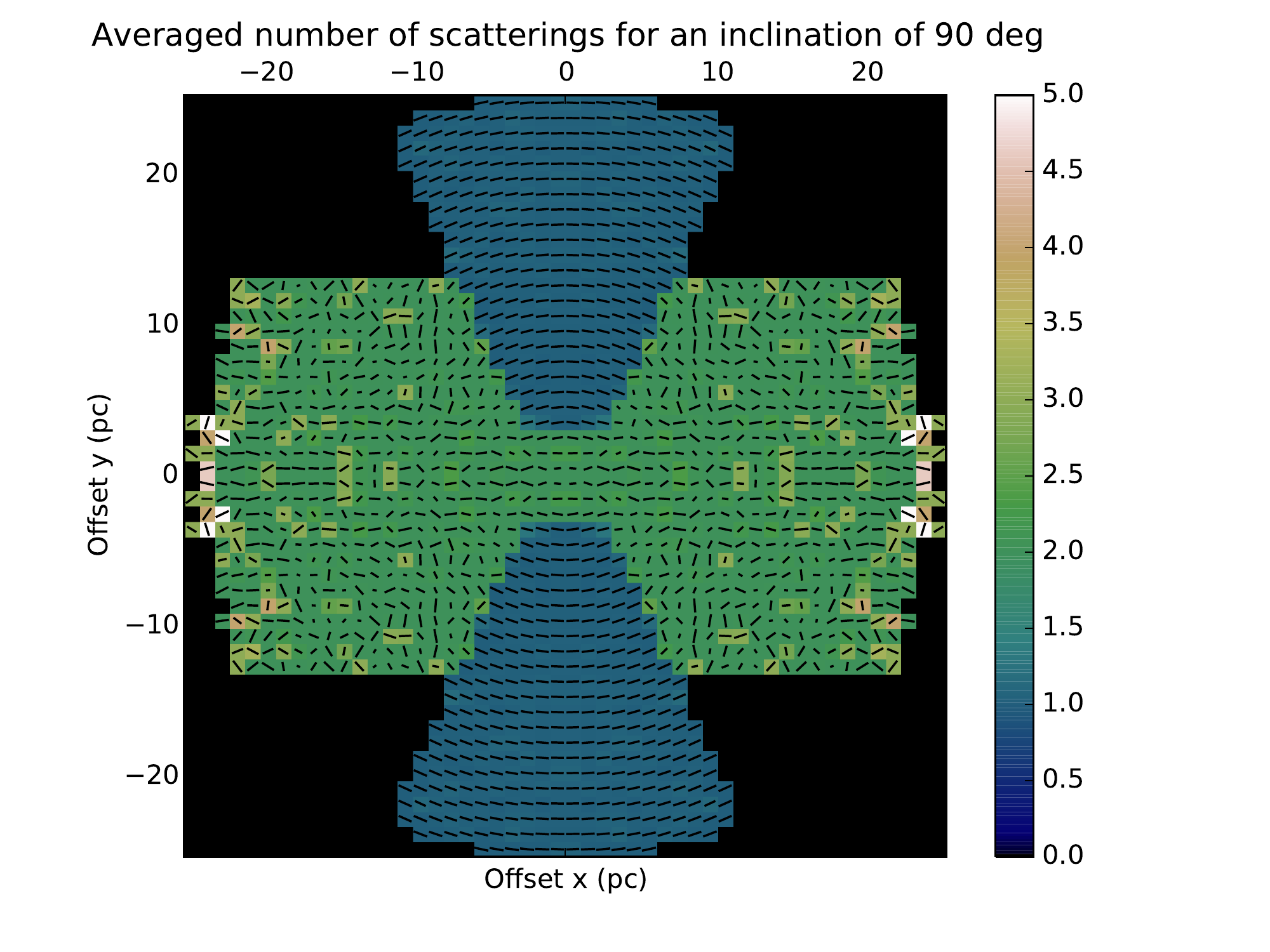}
  \includegraphics[width=0.40\textwidth,clip]{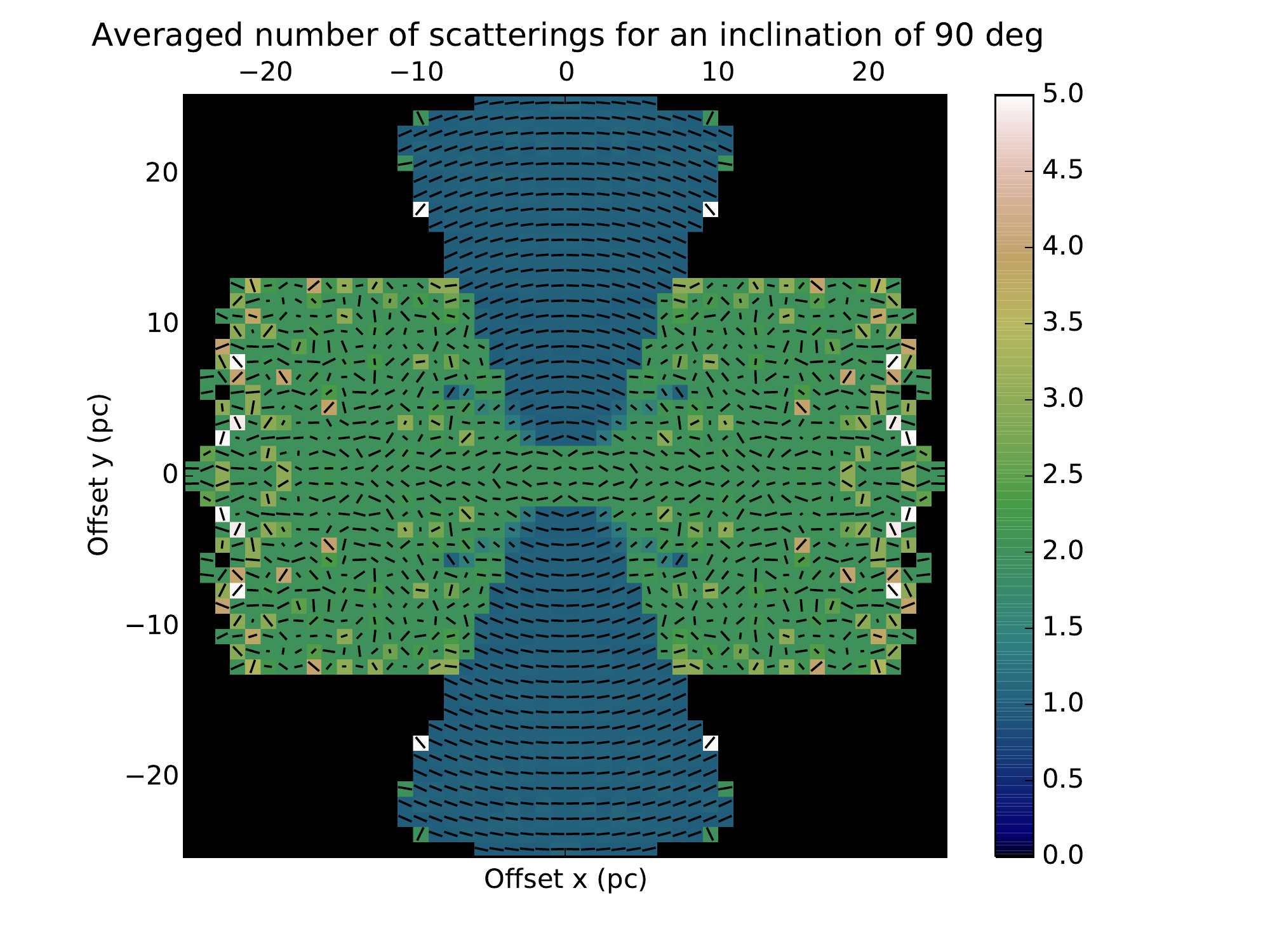}
   \caption{Maps of observed averaged number of scattering with an inclination angle of 90$^\circ$ at 1.6 (top) and 2.2 $\mu$m (bottom). Polarisation vectors are shown, their length being proportional to the polarisation degree and their position angle representing the polarisation angle.}
   \label{simu_1068}
\end{figure}

With these parameters, maps of figure \ref{simu_1068} are now able to reproduce some of the observed features. Namely, the nearly constant horizontal polarisation in the central part is similar to the one observed on NGC 1068 at 1.6 and 2.2 $\mu$m. We are however not able yet to reproduce the ridge that appears on the very centre of the SPHERE observations.

\section{Discussion}

\subsection{Photons significance}
\label{effective_photon}

One major concern is the simulation strategy of conserving the packets until they exit the simulation box whatever their number of interactions. We must verify to what extend the detected photons are relevant. The energy of each packet is decreased according to the albedo at each interaction, as is the number of photons in the packet. Therefore it is common at the end to record photons packets with energies $10^4$ times lower than the initial energy. If a high energy packet reaches a pixel previously populated only with low energy ones, it will totally dominate the pixel so that Q and U values and the polarisation parameters will be essentially linked to this packet. Indeed, most of the photons in the pixel will be from this packet.

As opposed to simulations including absorption, where all the recorded photons have the same probability, it is not any more the case here. A pixel of an output image could have stacked many low energy packets before a high energy one, more representative of the actual polarisation, reach it. But it is also possible that no high power photon reaches this cell, in the course of the simulation leading to a wrong estimate.

To analyse our results, we need first to be sure to disentangle between photons representative of the actual polarisation, and those which are not. Limiting the number of scattering is a good way to ensure that under a given limit of number of photons, packets will stop to be recorded. We also analysed maps of "effective number" of photons packet per pixel, as shown in appendix \ref{add_map}. These maps are generated by dividing the energy received in each pixel by the highest energy of packets recorded in the pixel. If a single packet dominates the pixel's information, the effective number of packets in the pixel will be close to 1, indicating a non reliable pixel.

As the energy of packets is decreasing with the number of scatterings, regions of low number of scatterings are more likely to be reliable. All regions of single scattered photons, with the exception of the central pixels, are reliable as long as photons can not reach the pixel in this region directly without being scattered. For the same reasons, all the regions with photons scattered twice on average are very likely to be representative, because only single scattered packets have higher energies.

\subsection{Effect of Geometry}

In the ionisation cone, we are able to reproduce the observed centro-symmetric pattern at large distance from the centre. This is well expected from photons scattered only once (see e.g. \citealt{Marin2012}). The maps of the average number of scatterings confirm that in all these regions, we see mainly photons only scattered once.

More interesting is the central region. We are able to reproduce with model 5 (figure \ref{res800}) a horizontal polarisation, at least at 800 nm. This is comparable to what was obtained in the case of YSO by \cite{Murakawa2010} (see for example their figure 6). The case of YSO is however very different from AGN in terms of optical depth, which is about 6.0x10$^5$ in K band in the equatorial plane. However the geometry of the dust distribution has some similarities, with two jets in the polar directions and a thick dusty environment surrounding the central source. We based our interpretation of the pattern observed in NGC 1068 on the effect called "roundabout effect" by \cite{Bastien1990} and used by \cite{Murakawa2010} to describe the photons path in YSO. In AGN environments, despite having an optically thick torus, we do not expect such high optical depth. Studies tend to argue for optical depth lower than 300 in visible as discussed in section \ref{concl_obs}.

As said before, to be able to see horizontal polarisation with only spherical grains, it is necessary to have a region, beyond the torus, with lower optical depth at the considered wavelength (typically under 10), for the photons to be scattered a second time toward the observer. This is realistic because one can expect the external part of the torus to be diluted into the interstellar medium, with a smooth transition. The theory of disk wind being at the origin of torus, described for example in \cite{Elitzur2009}, would support such dilution.

The ionisation cone is an important piece as it is required for the first scattering. However, while comparing results of model 1 and 5, it seems important to have a collimated region for this first scattering to happen because all photons will have almost the same scattering plane which therefore leads to a narrow range of polarisation position angle. This is not the case in model 1 where photons could interact not only in the ionisation cone but in all the outer shell. 

Note that all the tested geometries are rather simple in terms of structures and their edges are sharp. One of the advantages of using a 3D grid for sampling dust density is the capability to reproduce more complex structures with filaments or clumps. These are however more complex in terms of polarisation signatures (see \citealt{Marin2015,Marin2017}) and we therefore limited ourselves to simple models in this first paper. For further studies, we aim at developing more realistic models. In particular, MontAGN will be able to incorporate density structures from AGN hydrodynamical simulations (\citealt{Rybicki1990,Pereyra1997,Proga2000,Hopkins2012}), which could help for example to constrain the wind geometry and launching mechanism (see \citealt{Marin2013})

\subsection{Dust Composition}
\label{composition}

\begin{figure}[ht!]
 \centering   
  \includegraphics[width=0.50\textwidth,clip]{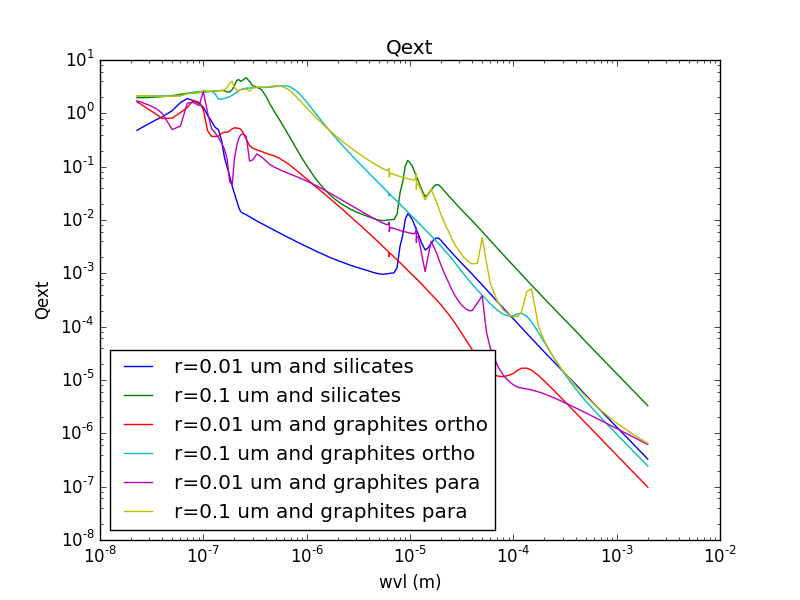}
   \caption{Extinction coefficient $Q_{ext}$ in function of wavelength, grain radius and grain type (data from \citealt{Draine1985}).}
   \label{Qext}
\end{figure}

Dust composition also has an effect on the polarisation maps. As the graphites and silicates do not have the same absorption coefficient dependence on wavelength, for a given optical depth in the V band, the optical depth will be different in H or K band. For instance with $\tau_V$ $\approx$ 50, we have $\tau_H$ $\approx$ 2.7 in the silicates case and $\tau_H$ $\approx$ 11.6 with the mixture of graphites and silicates (because of difference in extinction coefficient as seen in figure \ref{Qext}). 

This fact also explains why there are almost no photons in the equatorial belt of upper panel of figure \ref{res1600} (only silicates), as opposed to the case with silicates and graphites. The extended torus region indeed has in the pure silicate model a low optical depth at 1.6 $\mu$m and photons have a low probability to interact in this area, making the signal in the equatorial belt hard to detect. This will be discussed and used later to study the role of the optical depth in the outer torus.

We also observed that in the cone, the flux due to scattering on electrons is stronger than the one coming from scattering on silicate grains. This difference stems directly from the absorption properties. Silicates through Rayleigh scattering and electrons with Thomson scattering have the same scattering phase functions. However electrons have an absorption coefficient of 0 while silicate's is close to 1 and depends on the wavelength. 

\subsection{Thick Torus}

In order for the photons scattered twice to be dominant in the central region, it is necessary to have an optical depth high enough in the equatorial plane. If not, photons coming directly from the source, with a single or no scattering will contaminate the observed polarisation and even become dominant, as said in section \ref{effective_photon}. The switch from one regime to the other is when the probability for a photon to emerge from the dust having suffered at most one scattering is of the same order as the probability of a photon being scattered twice. Note that because electrons do not absorb photons, they are much more efficient than dust grains for a same optical depth. Even with electrons in the cone, photons have to be redirected in the correct solid angle toward the equatorial plane and to be scattered again without being absorbed in the observer's direction. The favourable cases to observe this signal seems to be for optical depth higher than 20 in the equatorial plane, as determined from figure \ref{tau_torus}.

This conclusion is in agreement with the observation that no broad line emission is detected in total flux when the AGN is viewed edge-on. With lower optical depth torus, we would have been able to detect such broad lines directly through the torus. This is not the case as demonstrated for example by \cite{Antonucci1985} on NGC 1068 and \cite{Ramos-Almeida2016} on a larger sample, who revealed these hidden lines only thanks to polarimetry.

\subsection{Optical Depth of Scattering Regions}

Another important parameter is the density of the matter in the cone and in the extended part of the torus. If both areas are at low density, photons have a lower probability to follow the roundabout path and are less likely to produce the horizontal pattern, as seen in figure \ref{tau_cone}. But if the optical depth is too high in the cone and/or in the outskirts of the torus, typically of the order of 1~-~2, photons have a low probability to escape from the central region and will never be observed. We can see on the last two panels of figure \ref{tau_cone} that the averaged number of scatterings in the cone is larger than 1, and that the signal of the photons scattered twice becomes weaker in the central belt. The range of values which favour the constant polarisation signal in the centre is around 0.1~-~1.0 in the cone and 0.8~-~4.0 in the extended region of the torus. This corresponds for the outer torus to a density about 20 times lower than in the torus.

We can therefore underline the importance of these two parameters and especially the extension of the torus. When comparing these results to the previous models, we noticed that on the first panel of figure \ref{res800} and \ref{res1600}, we observe few photons in the centre of the 800 nm image and almost none on the 1.6 $\mu$m one. The optical depth of the torus is lower in the second case and this region is therefore more easily reached by photons in this configuration. The only factor that can explain this lack of photon at 1.6 $\mu$m is the optical depth difference in the outer shell, which is of the order of 0.5 at 800 nm and much lower in the NIR, to be compared to the 0.8 lower limit determined previously. The structure, composition and density of the torus (including its outer part) are therefore all intervening to determine the polarisation pattern in the median plane. To be observable, the horizontal polarisation requires a proper combination of the parameters within a rather narrow range.

\subsection{Consequences for observations}
\label{concl_obs}

On figure \ref{simu_1068}, one can see that there are very few differences between H band and Ks band images. The optical depth of 169 in V band required for these images is acceptable, being in the range of present estimations. \cite{Marin2015} use values of optical depth in the range 150~-~750 in visible, while \cite{Gratadour2003} derived $\tau_V=40$. Assuming clumpy structures, \cite{Alonso2011} obtained optical depth of about 50 in V per cloud fitting mid-IR spectral energy distributions of NGC 1068. \cite{Lira2013,Audibert2017} more recently found integrated values of optical depth of about 250, based on fits of the NIR and mid-IR spectral energy distributions of samples of Seyfert 1 and 2 galaxies. We have to keep in mind that all these optical depth are derived assuming clumpy structure so that we should be careful when comparing them to optical depths of continuous dust distributions, the masses of dust being very different. It is in our plans to explore the effect of a clumpy structure, a capability already implemented in MontAGN.

A part of the polarisation may arise from elongated aligned grains as studied by  \cite{Efstathiou1997}. If so, the required optical depth could change. This might explain the observed ridge on the polarimetric images of NGC 1068 and aligned elongated grains are therefore something we should investigate further. However, the exact mechanism of this alignment in this configuration on a rather large scale (60 pc) is still to be established. For instance if due to magnetic field, it would require a toroidal component because we expect the polarisation pattern to be parallel with the magnetic field in case of dichroic absorption as discussed in \cite{Lopez2015}.

Our interpretation does not require special properties of grains (elongated and/or aligned) and indeed, spherical dust grains in the torus coupled to electrons in the ionisation cone are able to reproduce the polarisation orientation in the central belt. We constrain in this case the optical depth of the structures to a rather narrow range of values. In the cone, the range of optical depth is in between 0.1 to 1.0. A value of 0.1 gives an electrons density of $2.0 \times 10^9$ m$^{-3}$ which is consistent with the estimated range in AGN ($10^8$ - $10^{11}$ m$^{-3}$ for NGC 1068 according to \cite{Axon1998,Lutz2000}).

The only feature that our simulations do not seem to reproduce is the ridge at the very centre of NGC 1068 \citep{Gratadour2015}. Investigating a model with non spherical grains could potentially solve this problem through dichroism.

\section{Conclusion and Prospectives}

In this paper, we described and used the radiative transfer code with polarimetric capabilities MontAGN to study the inner dust structures of AGN, especially NGC 1068, in which the observed polarisation pattern is compatible with the idea of photons scattered twice proposed by \cite{Bastien1990} and later validated by \cite{Murakawa2010} on YSO. This was the basis of our analysis of our observations of NGC 1068 \citep{Gratadour2015}. The code allows us to simulate NIR polarimetric observations of an AGN featuring an ionisation cone, a torus and an extended envelope.

Despite limiting ourselves to a simple case where the various structures have a constant density of dust or electrons, we are able with spherical grains only, to constrain the optical depth of the different dust structures so that obtaining a similar polarisation pattern as the one observed on NGC 1068.

We highlight the important role of both the ionisation cone and the dusty torus. The cone allows photons to be scattered toward the equatorial plane. In order for this redirection to be efficient, we found that electrons are much more satisfactory than dust grains, because they are non absorbing, a hint that these cones are more likely populated by electrons, as suspected by \cite{Antonucci1985}. We estimate the optical depth in the ionisation cone, measured vertically, to be in the range 0.1 - 1.0 in the first 25 parsecs from the AGN.

We found that for the light directly coming from the central region of the AGN to be blocked, the torus should have an in-plane integrated optical depth greater than 20 in the considered band, so $\tau_{Ks} \geqslant 20$ in the case of NGC 1068. Furthermore, the torus should not only be constituted of an optically thick dense part, but also of an almost optically thin extended part. We find satisfactory results with an outer part with optical depth $0.8<\tau_H<4.0$ and an extension ranging from 10 to 25 pc from the centre. We considered structures with constant density, something unrealistic, but we expect similar results for a more continuous torus with a density decreasing with distance to the centre. This is a direction in which we aim at carrying our study in the near future. 

There are other improvements we will be able to conduct. One major amelioration will be to include elongated aligned grains. As already mentioned, another improvement is the introduction of a clumpy torus instead of constant density structures. We also aim to include the ``peeling-off'' technique (\citealt{Yusef-Zadeh1984}) in MontAGN in order to obtain larger statistics for a given inclination when simulating the polarisation signal of a peculiar object. A net gain of computation time is expected, at the expense of all other inclinations.

\begin{acknowledgements}
The authors would like to acknowledge financial support from the Programme National Hautes Energies (PNHE) and from "Programme National de Cosmologie and
Galaxies" (PNCG) funded by CNRS/INSU-IN2P3-INP, CEA and CNES, France.
\end{acknowledgements}

%
%

\bibliographystyle{aa} 
\bibliography{grosset} 

\begin{appendix} 
\section{Stokes formalism and Mueller matrices}
\label{Mueller}

In order to compute the evolution of polarisation through scattering, MontAGN is using Stokes formalism. It requires two matrices to compute the new Stokes vector after the scattering event. The Mie theory gives the values of $S_1$ and $S_2$ which are determined for each wavelength, grain type, grain radius and scattering angle. These two quantities are linked to the electrical properties of the material and are used to compute the elements of the Mueller matrix:

\begin{equation}
S_{11}= \frac{1}{2}(|S_{2}|^2+|S_{1}|^2)
\end{equation}
\begin{equation}
S_{12}= \frac{1}{2}(|S_{2}|^2-|S_{1}|^2)
\end{equation}
\begin{equation}
S_{33}= Re(S_2 \overline{S_1})
\end{equation}
\begin{equation}
S_{34}= Im(S_2 \overline{S_1})
\end{equation}

with the Mueller matrix being define as:

\begin{equation}
M=
\begin{pmatrix}
S_{11}& S_{12}& 0& 0\\
S_{12}& S_{11}& 0& 0\\
0& 0& S_{33}& S_{34}\\
0& 0& -S_{34}& S_{33}\\
\end{pmatrix}
\end{equation}

By applying the Mueller matrix to the Stokes vector, we take into account the change in the polarisation introduced by the main scattering angle ($\alpha$) between the incoming ray and the new direction of propagation.

\begin{equation}
S_{final}=M\times S_{inter}\\
\end{equation}

which gives:

\begin{equation}
\begin{bmatrix}
I_f\\
Q_f\\
U_f\\
V_f\\
\end{bmatrix}
= 
\begin{pmatrix}
S_{11}& S_{12}& 0& 0\\
S_{12}& S_{11}& 0& 0\\
0& 0& S_{33}& S_{34}\\
0& 0& -S_{34}& S_{33}\\
\end{pmatrix}
\begin{bmatrix}
I_{inter} \\
Q_{inter} \\
U_{inter} \\
V_{inter}
\end{bmatrix}
\end{equation}

However, as the scattering is not necessary in the previous polarisation plane of the photon, we need to translate the old photon's polarisation into the new frame used for the scattering. We use for this purpose a rotation matrix defined as follow:

\begin{equation}
R(\beta)=
\begin{pmatrix}
1& 0& 0& 0\\
0& \mathrm{cos}(2\beta)& \mathrm{sin}(2\beta)& 0\\
0& -\mathrm{sin}(2\beta)& \mathrm{cos}(2\beta)& 0\\
0& 0& 0& 1\\
\end{pmatrix}
\end{equation}

The rotation matrix is mandatory to modify the polarisation plane according to the scattering geometry and depend on $\beta$, the azimuthal scattering angle. We then switch to $S_{final}$ from $S_{initial}$ by applying both matrices:

\begin{equation}
S_{final}=M\times R(\beta)\times S_{initial}\\
\end{equation}

and in detailed form:

\begin{equation}
\begin{bmatrix}
I_f\\
Q_f\\
U_f\\
V_f\\
\end{bmatrix}
= 
\begin{pmatrix}
S_{11}& S_{12}& 0& 0\\
S_{12}& S_{11}& 0& 0\\
0& 0& S_{33}& S_{34}\\
0& 0& -S_{34}& S_{33}\\
\end{pmatrix}
\begin{pmatrix}
1& 0& 0& 0\\
0& \mathrm{cos}(2\beta)& \mathrm{sin}(2\beta)& 0\\
0& -\mathrm{sin}(2\beta)& \mathrm{cos}(2\beta)& 0\\
0& 0& 0& 1\\
\end{pmatrix}
\begin{bmatrix}
I_{i}\\
Q_{i}\\
U_{i}\\
V_{i}\\
\end{bmatrix}
\end{equation}

\section{Additional maps}
\label{add_map}

Here are shown some of the additional maps computed from the model 5 of section \ref{optical_depth} at 1.6 $\mu$m in which optical depth of the cone is fixed to 0.1, those of the extended torus to 0.8 and the torus one is 20 (in H band).

\begin{figure}[ht!]
 \centering
  \includegraphics[width=0.30\textwidth,clip]{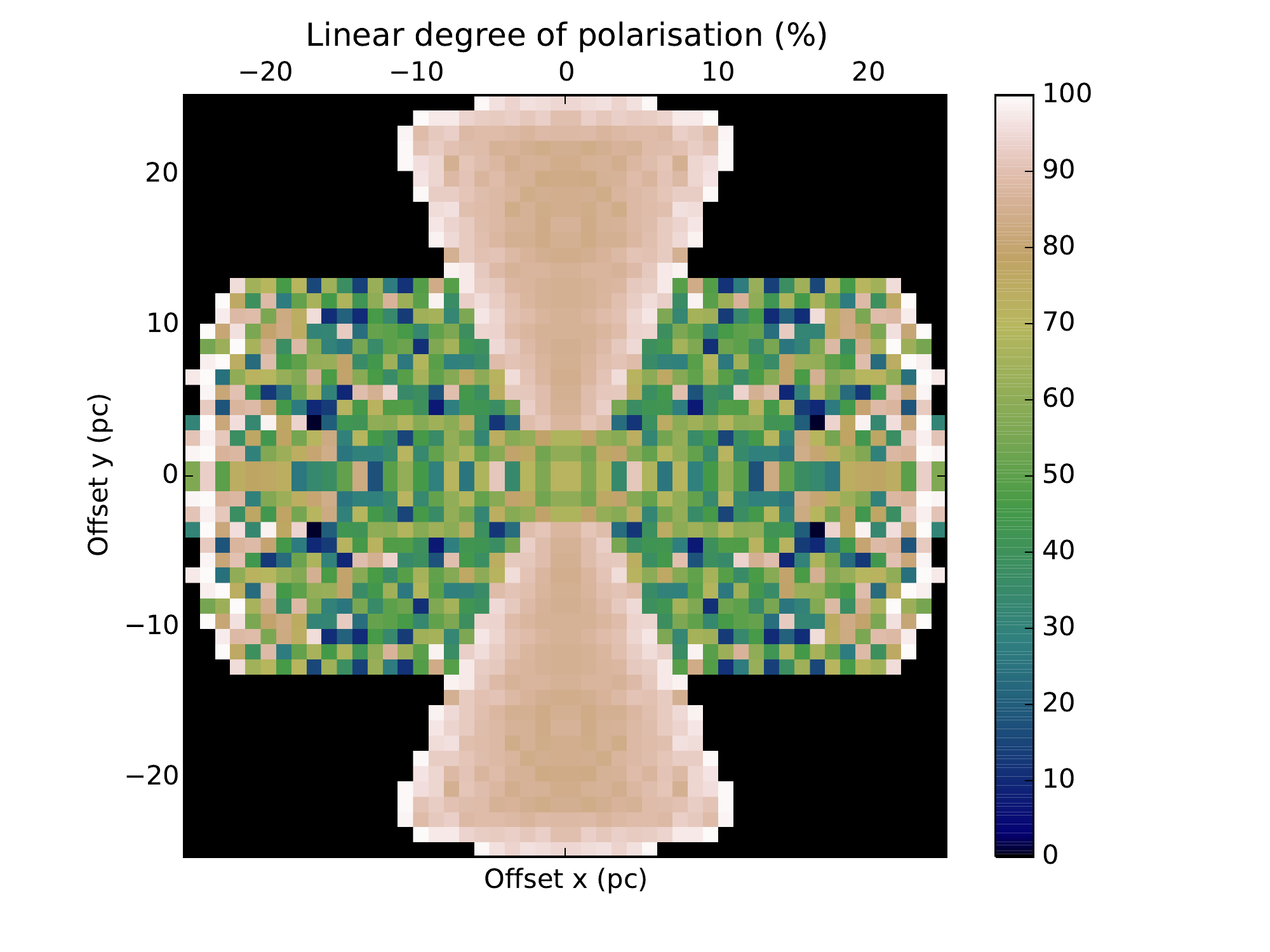}
  \includegraphics[width=0.30\textwidth,clip]{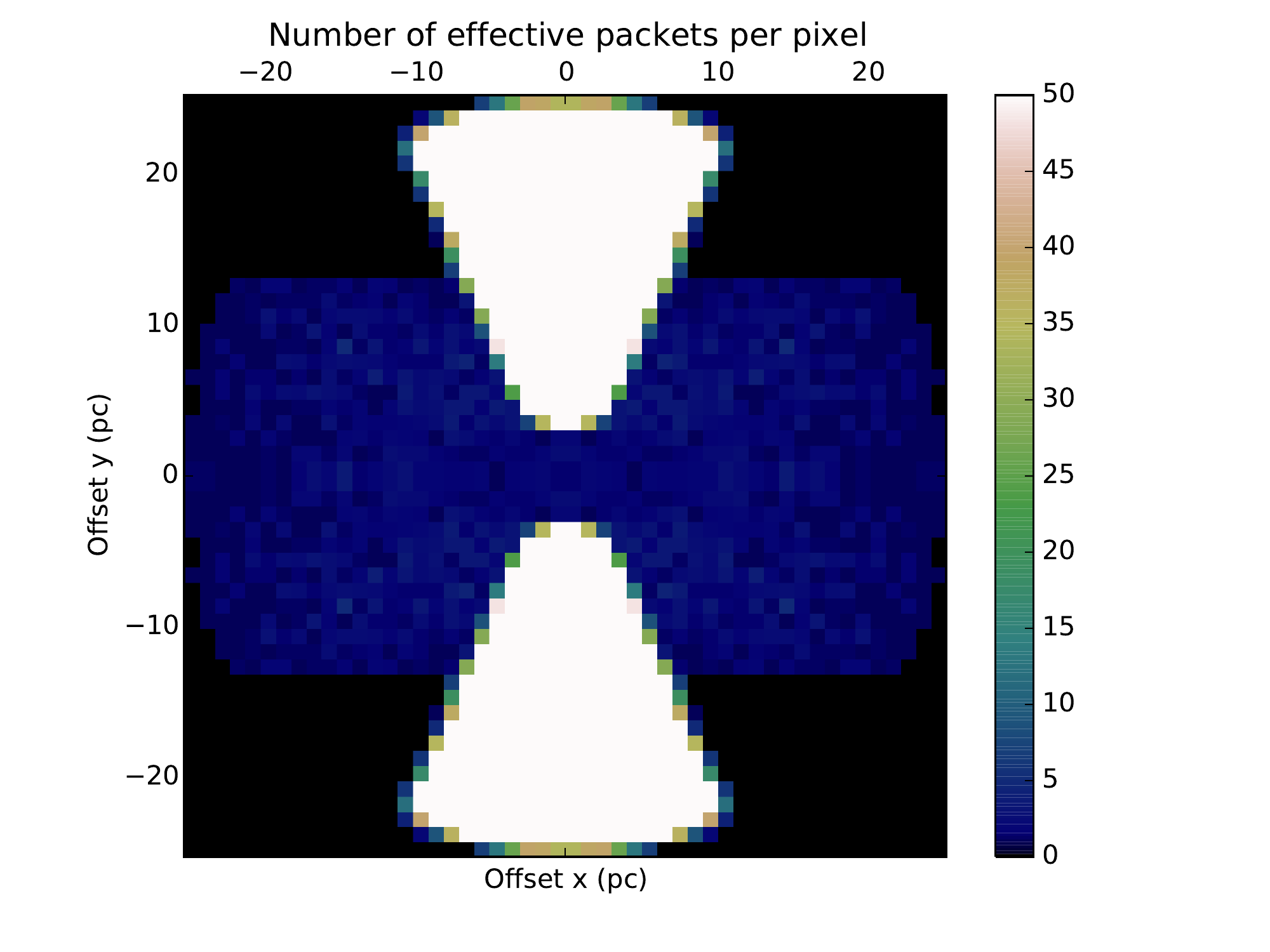}
   \caption{Maps of observed linear polarisation degree (first image) and of effective number of packets (second one, effective number of packets is defined in section \ref{effective_photon}) with an inclination angle of 90$^\circ$ at 1.6 $\mu$m.}
   \label{simu_1068_other_maps}
\end{figure}
\vspace{0.5cm}

\begin{figure}[ht!]
 \centering
  \includegraphics[width=0.30\textwidth,clip]{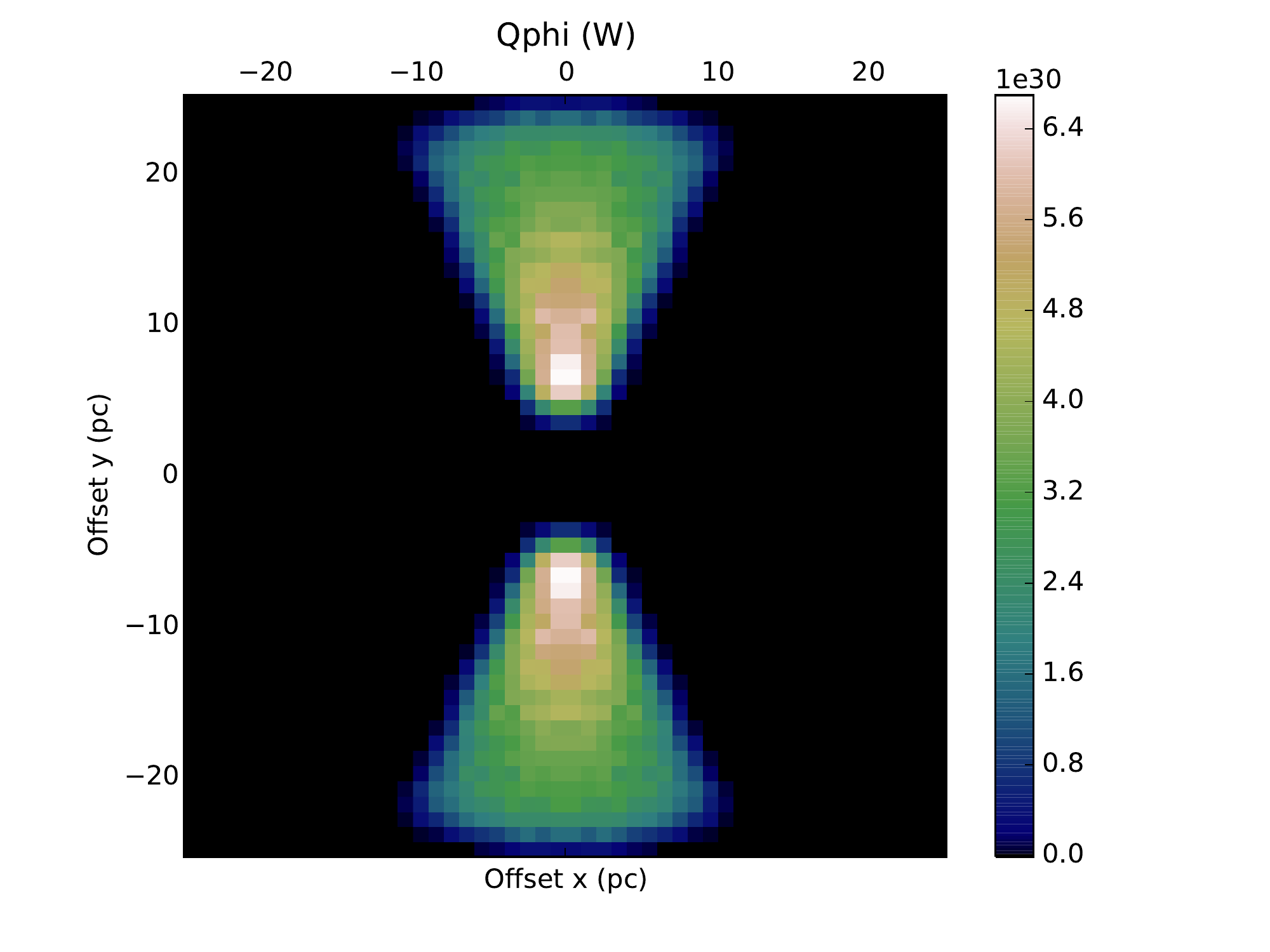}
  \includegraphics[width=0.30\textwidth,clip]{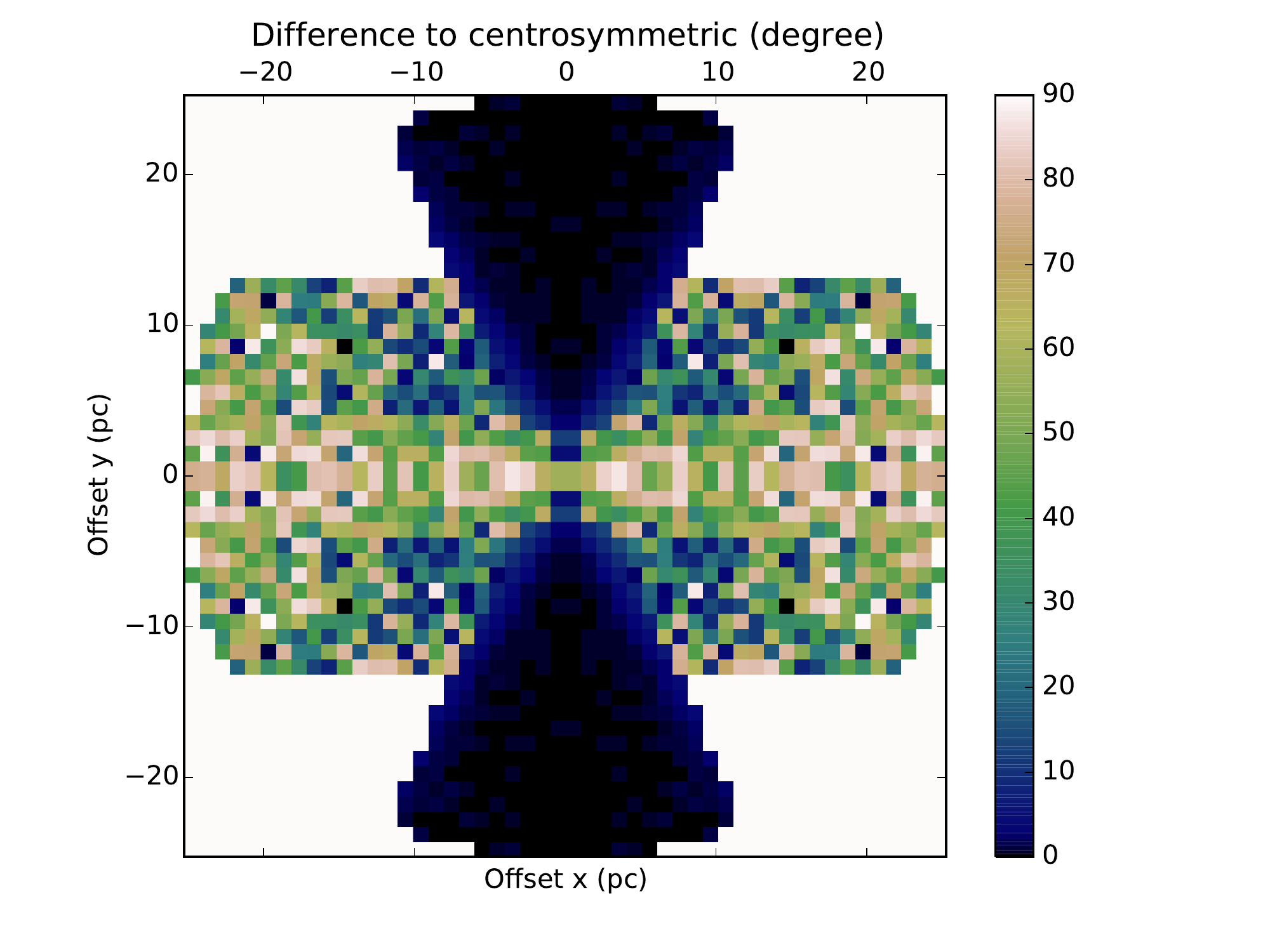}
   \caption{Maps of observed Q tangential (first, corresponds to the image of the centro-symmetric polarised intensity), and of difference angle to centro-symmetric pattern so as computed in \cite{Gratadour2015}, with an inclination angle of 90$^\circ$ at 1.6 $\mu$m.}
   \label{simu_1068_othertheta_maps}
\end{figure}

\section{MontAGN Pseudo-code}

\textbf{General MontAGN algorithm:}

\vspace{0.5cm}
\begin{itemize}
\item \textbf{MontAGN 01}

\hspace{0.5cm} Read of the input parameters and conversion into corresponding class objects. Parallelisation if asked.
\vspace{0.5cm}

\item \textbf{MontAGN 02}

\hspace{0.5cm} Read of the parameters for dust geometry (given as keyword, asked or from parameter file): dust densities, grains properties and structures associated.
\vspace{0.5cm}

\item \textbf{MontAGN 03}

\hspace{0.5cm} Creation of the 3D grid and filling with dust densities: compute of the density of each grain type for each cell. Temperature initialisation.
\vspace{0.5cm}

\item \textbf{MontAGN 04}

\hspace{0.5cm} Computation of the tables of all propagation elements: Mueller Matrices, albedo, Q$_{abs}$, Q$_{ext}$, phase functions, for each grain type, and different wavelength/grain radius.
\vspace{0.5cm}

\item \textbf{MontAGN 05}

\hspace{0.5cm} Start of the simulation:
\vspace{0.5cm}

\item \textbf{MontAGN 06}

\hspace{0.5cm} Selection of a source based on relative luminosity.
\vspace{0.5cm}

\item \textbf{MontAGN 07}

\hspace{0.5cm} Determination of the photons wavelength and initial directions of propagation and of polarisation. The wavelength is randomly determined from the source's spectral energy distribution. The Stokes vector and propagation vectors are initialised.

\vspace{0.5cm}

\item \textbf{MontAGN 08}

\hspace{0.5cm} Propagation of the packet until reaching a non empty cell:

- Determination of the cell density.

- If null, direct determination of the next cell, back to the beginning of MontAGN 08.

- If radius smaller than sublimation radius, the cell is considered empty.
\vspace{0.5cm}

\item \textbf{MontAGN 09}

\hspace{0.5cm} Determination of a $\tau$ that the packet will be able to penetrate, computed by inversion from optical depth penetration definition $P(\tau_x > \tau)=e^{-\tau}$
\vspace{0.5cm}

\item \textbf{MontAGN 10}

\hspace{0.5cm} Determination of the next event of the packet: 

- Interaction if $\tau$ is too low to allow the packet to exit the cell

- Or exit from the cell (in that case, decrease of $\tau$ and going to MontAGN 18).
\vspace{0.5cm}

\item \textbf{MontAGN 11}

\hspace{0.5cm} Determination of the properties of the encountered grain/electron: size and therefore albedo, Q$_{abs}$ and Q$_{ext}$.
\vspace{0.5cm}

\item \textbf{MontAGN 12}

\hspace{0.5cm} Case 1 (no re-emission): 

decrease of the energy of the packet according to the albedo (and going to MontAGN 16).

\hspace{0.5cm} Case 2 (re-emission): 

determination whether the packet is absorbed from albedo.

If no go to MontAGN 16.
\vspace{0.5cm}

\item \textbf{MontAGN 13}

\hspace{0.5cm} Determination of the new temperature of the cell, from the previous one by equalising incoming and emitted energy.
\vspace{0.5cm}

\item \textbf{MontAGN 14}

\hspace{0.5cm} If the new temperature is above the Sublimation temperature of a type of grain, update of the sublimation radius of this type of grains.
\vspace{0.5cm}

\item \textbf{MontAGN 15}

\hspace{0.5cm} Determination of the new wavelength and direction of propagation of the packet. The directions are determined randomly, the wavelength is determined from the difference between new and old temperatures of the cell.

Going to MontAGN 18.
\vspace{0.5cm}

\item \textbf{MontAGN 16}

\hspace{0.5cm} Determination of the angles of scattering and of the new directions of propagation and polarisation:

- From grain properties and wavelength, determination of the scattering angle $\alpha$ and of the Muller Matrix components $S_{1}$ and $S_{2}$.

- Determination of scattering angle $\beta$.

- Computation of the new propagation vectors.
\vspace{0.5cm}

\item \textbf{MontAGN 17}

\hspace{0.5cm} Construction of the matrices. From the angles, $S_1$ and $S_2$, construction of Mueller matrix and of the rotation matrix (as described in appendix \ref{Mueller}). Applied to Stokes vector of the packet.
\vspace{0.5cm}

\item \textbf{MontAGN 18}

\hspace{0.5cm} If the packet still is in the simulation box, going back to MontAGN 10, else the packets exits the simulation.
\vspace{0.5cm}

\item \textbf{MontAGN 19}

\hspace{0.5cm} Computation of the packet properties in the observer's frame. From the propagation vectors, determination of the orientation of the polarisation frame and correction of Stokes parameters using a rotation matrix.
\vspace{0.5cm}

\item \textbf{MontAGN 20}

\hspace{0.5cm} Record the packet's properties by writing it in the files specified.
\vspace{0.5cm}

\item \textbf{MontAGN 21}

\hspace{0.5cm} Is there is still others packets to launch, back to MontAGN 06.
\vspace{0.5cm}

\item \textbf{MontAGN 22}

\hspace{0.5cm} End of simulation.
\vspace{0.5cm}

\item \textbf{MontAGN 23}

\hspace{0.5cm} Computation of the displays if asked.

\end{itemize}

\end{appendix}

\end{document}